\documentclass[12pt,journal,onecolumn,draft]{IEEEtran}

\usepackage{xspace}
\usepackage{amsopn}
\usepackage{amsfonts}
\usepackage{color}
\usepackage{graphicx}
\usepackage{epsfig}
\usepackage{float}
\usepackage{cuted}
\usepackage[intlimits]{amsmath}
\usepackage{multicol}
\usepackage{hyperref}
\usepackage{subcaption}

\usepackage{tikz-cd,pgfplots,pgfgantt,pdflscape}
\usetikzlibrary{arrows}
\usepackage{tikz-3dplot}

\usepackage{amssymb,amsmath,bm}
\usepackage{amsthm}

\newtheorem{theorem}{Theorem}
\newtheorem{proposition}[theorem]{Proposition}

\newtheorem{corollary}[theorem]{Corollary}
\newtheorem{lemma}[theorem]{Lemma}

\newcommand{\boldkappa}{\boldsymbol{\kappa}}
\newcommand{\boldeta}{\boldsymbol{\eta}}

\newcommand{\boldtheta}{\boldsymbol{\theta}}
\newcommand{\boldupsilon}{\boldsymbol{\Upsilon}}
\newcommand{\boldlambda}{\boldsymbol{\lambda}}

\DeclareMathOperator*{\argmax}{argmax}
\DeclareMathOperator*{\argmin}{argmin}
\begin{document}
\setlength{\abovedisplayskip}{3pt}
\setlength{\belowdisplayskip}{3pt}
\allowdisplaybreaks

\title{Sequential Multi-hypothesis Testing in Multi-armed Bandit Problems:\\
An Approach for Asymptotic Optimality}

\author{Gayathri R. Prabhu, Srikrishna Bhashyam, Aditya Gopalan, Rajesh Sundaresan
\thanks{This work was supported by the Science and Engineering Research Board, Department of Science and Technology [grant no. EMR/2016/002503]. The authors acknowledge fruitful discussions with Aditya O. Deshmukh.}
\thanks{G. R. Prabhu and S. Bhashyam are with the Department of Electrical Engineering, IIT Madras, Chennai 600036, India.}
\thanks{A. Gopalan is with the Department of Electrical Communication Engineering and R. Sundaresan is with the Department of Electrical Communication Engineering and the Robert Bosch Centre for Cyber-Physical Systems, at the Indian Institute of Science, Bangalore 560012, India.}
} 

\maketitle
\begin{abstract}
We consider a multi-hypothesis testing problem involving a $K$-armed bandit. Each arm's signal follows a distribution from a vector exponential family. The actual parameters of the arms are unknown to the decision maker. The decision maker incurs a delay cost for delay until a decision and a switching cost whenever he switches from one arm to another. His goal is to minimise the overall cost until a decision  is reached on the true hypothesis. Of interest are policies that satisfy a given constraint on the probability of false detection. This is a sequential decision making problem where the decision maker gets only a limited view of the true state of nature at each stage, but can control his view by choosing the arm to observe at each stage. An information-theoretic lower bound on the total cost (expected time for a reliable decision plus total switching cost) is first identified, and a variation on a sequential policy based on the generalised likelihood ratio statistic is then studied. Due to the vector exponential family assumption, the signal processing at each stage is simple; the associated conjugate prior distribution on the unknown model parameters enables easy updates of the posterior distribution. The proposed policy, with a suitable threshold for stopping, is shown to satisfy the given constraint on the probability of false detection. Under a continuous selection assumption, the policy is also shown to be asymptotically optimal in terms of the total cost among all policies that satisfy the constraint on the probability of false detection.
\end{abstract}

\IEEEpeerreviewmaketitle

\section{Introduction}
We consider a multi-hypothesis testing problem involving a $K$-armed bandit.
The observations from each arm $i$, $1 \leq i \leq K$, follow a distribution from a {\em vector exponential family} parameterised by its {\em natural (vector) parameter} $\boldeta_i$. The parameters $\overline{\boldeta} = (\boldeta_1, \ldots, \boldeta_K)$ of the arms are unknown. The parameter $\overline{\boldeta}$ belongs to one of the sets $\Theta_1, \ldots, \Theta_M$. The goal is to identify the set $\Theta_l$ to which it belongs. At each successive stage or round, the decision maker chooses exactly one among the $K$ arms for observation. The decision maker therefore has only a limited view of the true state of nature at each stage. But the decision maker can control his view by choosing the arm to observe. The decision maker also incurs a cost whenever he switches from one arm to another. Specifically, the decision maker has to minimise the overall cost of expected time for a reliable decision plus total switching cost, subject to a constraint on the probability of false detection.

We can model the above problem as a sequential hypothesis testing problem with control \cite{chernoff} and unknown distributions \cite{ref:albert1961sequential} or parameters \cite{vaidhiyan15}. The control here is in the choice of arm for observation at each stage which is determined by the sampling strategy of the policy. Many problems fall within the aforementioned framework, for e.g., anomaly detection, best arm identification, A/B testing, etc.; see Section \ref{sec:problem_model} for several examples.

\subsection{Remarks on the model assumptions}

1. {\em Exponential families}. Our interest in exponential families is for three reasons.
\begin{itemize}
    \item It unifies most of the widely used statistical models such as the Gaussian, the Binomial, the Poisson, the Gamma distributions, among countless others.
    \item The generalisation forces us to rely on, and therefore bring out, the key properties of exponential families that make the analysis tractable. These include the usefulness of the convex conjugate (or convex dual) of the log partition function, the existence of easily amenable formulae for relative entropy, and the usefulness of the conjugate prior in the analysis.
    \item The conjugate prior enables extremely easy signal processing for posterior updates. This is of great value in practice.
\end{itemize}

2. {\em Switching cost}. Incorporation of switching cost is motivated by a number of applications.
\begin{itemize}
    \item In visual search tasks, where one is searching for say an odd object among many objects, a switching action implies a change in the location of visual focus via movement of the eyes. This is called a {\em saccade} and results in a delay cost \cite{vaidhiyan17}.
    \item In robotics applications, relocation of robots (or other autonomous decision makers such as unmanned aerial vehicles) incurs considerable cost in terms of energy or delay \cite{chen2019active}.
    \item In manufacturing industry applications, switching refers to reconfiguration of a production line and causes extra delays \cite{dekel2014bandits}.
\end{itemize}

\subsection{Results and Methodology}

{\em Converse:} We use the results from \cite{kaufmann16} to obtain an information-theoretic lower bound on the conditional expected total cost for any policy that satisfies an upper bound constraint on the probability of false detection, say $\alpha$. The lower bound suggests that the conditional expected total cost is asymptotically proportional to $\log (1/\alpha)$, i.e., it grows as $\log(1/\alpha)/D^*(\overline{\boldeta})$, where $D^*(\overline{\boldeta})$ is a relative-entropy based constant which we shall study in some detail in this paper. An examination of this lower bound reveals that it may be viewed as the best performance achievable when a more informed decision maker that knows the parameters to be either $\overline{\boldeta} \in \Theta_l$ or its `nearest alternative', a suitable $\overline{\boldeta}' \in \Theta_{-l} := \bigcup\limits_{m=1}^M \Theta_m\backslash \Theta_l$, is attempting to decide which of these is true, as quickly as possible in a sequential fashion and with a control on the false alarm probability.

{\em Achievability}: A commonly used test in problems with unknown parameters is the generalised likelihood ratio (GLR) test; see for example the text book \cite{Poor}. The basic idea of the policy, for stopping problems with control, dates back at least to Chernoff's \textit{Procedure A} \cite{chernoff}.
In our case, taking a cue from \cite{vaidhiyan15}, we use a modified GLR where the numerator of the generalised likelihood ratio is replaced by an averaged likelihood function. The average is computed with respect to an artificial prior on the unknown parameters. Each hypothesis is tested against its {\em nearest alternative} by taking the minimum, across the alternatives, of the modified GLRs. This yields a suitable statistic that quantifies the decision maker's confidence on each hypothesis.

At each stage, then, we choose the hypothesis with the largest statistic. If the statistic exceeds a pre-defined threshold, we declare this hypothesis as the one likely to be true and stop further sampling. Else, we decide randomly, based on a coin toss, whether to sample the current arm or choose another one according to the policy's sampling strategy. This slowed switching is to handle the switching costs. The bias of the coin determines the speed of switching thereby providing a control on the switching cost. The threshold for a decision in our policy, and therefore stoppage of further sampling, depends only on the tolerable probability of false detection ($\alpha$) and the number of hypotheses ($M$); in particular, the threshold is not time-varying. We show that such a policy meets the constraint on the probability of false detection (i.e., the policy is admissible). It is in proving this admissibility where the modification to the GLR comes in handy.

As remarked earlier, our approach involves the computation of generalised likelihoods. These provide best estimates of the unknown parameters obtained by (estimation-theoretic) constrained optimisation. We then adopt the principle of {\em certainty equivalence}, i.e., we assume that the latest estimated parameters are correct, solve an associated (decision-theoretic) optimisation problem for identifying the optimal sampling strategy, and then take actions according to this optimal prescription. The estimated parameters, at best, can approach the true parameters for, after all, the parameters take values in a continuum. This leads to two requirements. First, to enable the convergence to the true parameters, there should be {\em sufficient exploration}. Second, the arg-max of the decision-theoretic optimisation problem at each stage, assuming the estimated parameters at that stage, may have several solutions and therefore several sampling prescriptions; we then need a {\em continuous selection} of the arg-max. Otherwise, information will not be gained at the required rate $D^*(\overline{\boldeta})$ to meet the lower bound.

When a continuous selection exists, with just barely sufficient exploration, we show that the sampling strategy of our proposed policy has performance that is asymptotically close to the lower bound; the asymptotics is as the target probability of false detection $\alpha$ goes down to zero. We also show that, asymptotically, the total cost scales as $\log\left(1/\alpha\right)/D^*(\overline{\boldeta})$, where $D^*(\overline{\boldeta})$ is the optimal scaling factor suggested by the lower bound. We then show that the continuous selection assumption holds for some examples.

Under the vector exponential family assumption, the information processing at each stage is extremely simple. The decision maker maintains the parameters of the associated conjugate priors, corresponding to the posterior distributions of the model parameters, via easy-to-implement update rules.

\subsection{Closely related prior works}
Two special cases of great interest in literature are the cases of {\em best arm identification} and {\em odd arm identification}. Garivier et al. \cite{garivier16} have characterised the complexity of best arm identification in one-parameter bandit problems in the fixed confidence setting. Kaufmann et al. \cite{kaufmann16} have discussed the case of identifying $m$ best arms in a stochastic multi-armed bandit model for both fixed confidence and fixed budget settings. In \cite{vaidhiyan17}, the authors have considered the odd arm identification problem with switching costs, but the statistics of the observations were assumed to be known and Poisson-distributed. In \cite{vaidhiyan15}, the authors have considered a learning setting where the parameters of the Poisson distribution were not known but the switching costs were not taken into account. In earlier versions of this work, \cite{8755796} (a workshop paper) and \cite{gayathri} (technical report), under some restricting assumptions, we considered the odd arm identification problem with switching costs when the distributions are from a vector exponential family. This work substantially extends the results in \cite{8755796,gayathri} to much more general parameter structure classes.

A related problem  is discussed in \cite{deshmukh2018controlled,deshmukh2019sequential} where the authors have considered a multi-hypothesis problem with controlled sensing of the observations. The parameter set is partitioned into various subsets, one for each hypothesis. Each subset is further assumed to be a finite union of convex sets; then projections on closures of such sets exist (uniqueness holds in each closed convex part). In \cite{juneja2018sample}, the authors have considered the problem of identifying the partition to which a set of arms belong, given a finitely partitioned universe of such set of arms. In all these works, \cite{deshmukh2018controlled,deshmukh2019sequential,juneja2018sample}, the authors have assumed that the observations come from a single parameter exponential family of distributions. The important and practical issue of switching costs is also not taken into consideration. Further, their choice of the statistic requires the employment of a time-varying threshold.

Our work thus provides a significant generalisation of the results in \cite{deshmukh2018controlled,deshmukh2019sequential,juneja2018sample} to general vector exponential families, analyses the effect of switching cost on search complexity, all in the presence of learning. More importantly, we provide a fuller understanding of the subtleties associated with learning the true parameters: the use of forced exploration, the usefulness of the existence of a continuous selection, and the analysis methods that show convergence despite the adaptation based on the estimated parameters that change with time.

For connections to, and limitations of, the classical works of Chernoff \cite{chernoff} and Albert \cite{ref:albert1961sequential}, see a detailed summary in \cite[Sec. I-A]{vaidhiyan15}.


\subsection{Our contributions}

\begin{itemize}
\item We provide a significant generalisation of the odd arm identification problem in \cite{gayathri} to a much more general sequential hypothesis testing setting. At least six problems already studied in the literature are highlighted as special cases; see Section \ref{subsec:examples}.
\item We provide generalisations of the problems discussed in closely related papers \cite{deshmukh2018controlled,deshmukh2019sequential} in three aspects:
\begin{itemize}
    \item Observations come from a vector exponential family of distributions.
    \item We incorporate switching costs based on the idea of slowed switching; see \cite{vaidhiyan17}, \cite{vaidhiyan15_costs} and \cite{krishnaswamy2017augmenting}.
    \item The threshold for decision is time invariant.
\end{itemize}

\item We show that the proposed policy, which incorporates learning, is asymptotically optimal even with switching costs. This is in the sense that the growth rate of the total cost with switching, as the probability of false detection and the switching parameter that controls the speed of switching are both driven to zero, is the same as that without switching costs. Of course, optimality is only in terms of the growth rate (slope). As our simulations indicate, the slowed switching leads to an additional delay. However, the delay does not affect the growth rate since it does not show up in the slope.

\item We highlight why the continuous selection assumption for the arg-max over sampling strategies may be essential to meet, asymptotically, the lower bound.

\item We demonstrate the usefulness of forced exploration for learning the parameters, and suggest a range of exploration rates useful in our context.
\end{itemize}


\subsection{Organisation of the paper}

The paper is organised as follows. In Section~\ref{sec:preliminaries} we describe preliminaries related to exponential families. In Section~\ref{sec:problem_model}, we describe the problem model and discuss several examples that fall within our framework. We then preview the main result. In Section~\ref{sec:converse}, we discuss a lower bound on the expected search time for admissible policies. In Section~\ref{sec:achievability}, we describe the proposed policy that can be made to come arbitrarily close to meeting the lower bound. In Section~\ref{sec:simulations}, we provide insightful simulation results that corroborate the developed theory. The proofs and the verification of the assumption of continuous selection for some examples are all relegated to the appendices.

\section{Preliminaries: Exponential family basics}
\label{sec:preliminaries}

In this section we discuss formulae associated with the exponential family that will help in our analysis. Those familiar with exponential families may skip this section.

A probability distribution is a member of a vector exponential family if its probability density function (or probability  mass function) can be written as
\begin{equation}\label{gen_exp_family}
{f\left(x|\boldeta\right) = h\left(x\right)\exp\left(\boldeta^T\textbf{T}(x)-\mathcal{A}\left(\boldeta\right)\right)} \quad \forall x \in \mathbb{R},
\end{equation}
where $\boldeta$ is the vector parameter of the family, {{with  $\boldeta$ in some open convex subset $\Psi$ of $\mathbb{R}^d$}}, $\textbf{T}(x) \in \mathbb R^d$ is the sufficient statistic for the family, and $\mathcal{A}\left(\boldeta\right)$ is the log partition function given by
\begin{equation*}
\mathcal{A}\left(\boldeta\right) = \log \int_{\mathbb{R}^d}  h\left(x\right)\exp\left(\boldeta^T\textbf{T}(x)\right) dx.
\end{equation*}

The expression in (\ref{gen_exp_family}) gives the {\em canonical} parameterisation of the exponential family. We restrict ourselves to minimal representations \cite[p.~40]{wainwright2008graphical} which enables us to represent the distributions in the family using the {\em expectation} parameter defined as
\begin{equation}\label{exp_parameter}
{\boldkappa(\boldeta) := } E_{\boldeta}[\textbf{T}(x)]= \nabla_{\boldeta} \mathcal{A}\left(\boldeta\right)
\end{equation}

\noindent whenever $\mathcal{A}{(\cdot)}$ is {continuously} differentiable. We now continue with some additional observations on exponential families. The mapping $\boldeta \mapsto \mathcal{A}(\boldeta)$ is strictly convex \cite[Prop.~3.1]{wainwright2008graphical}, a fact that can be easily verified via the H\"{o}lder inequality. The strictness comes from the minimality of the representation. If $\mathcal{A}(\cdot)$ is twice differentiable, then the Hessian $\textsf{Hess}(\mathcal{A})(\boldeta)$ is just the covariance of $\mathbf{T}(x)$ when the canonical parameter is $\boldeta$. If $\mathcal{A}(\cdot)$ twice continuously differentiable, then the covariance matrix is a continuous function of its parameter.

The convex conjugate of $\mathcal{A}(\boldeta)$ evaluated at an arbitrary $\boldkappa$ and denoted $\mathcal{F}(\boldkappa)$ is given by
\begin{equation}\label{conv_conj_sup}
\mathcal{F}\left(\boldkappa\right) := \sup\limits_{\boldeta \in \mathbb{R}^d} \{\boldeta^T \boldkappa-\mathcal{A}\left(\boldeta\right)\};
\end{equation}
this is also a convex function. Since $\mathcal{A}(\cdot)$ is convex, we can recover $\mathcal{A}(\cdot)$ as the convex conjugate of $\mathcal{F}(\cdot)$, i.e.,
\begin{equation}\label{conv_conj_sup2}
\mathcal{A}\left(\boldeta\right) := \sup\limits_{\boldkappa \in \mathbb{R}^d} \{\boldeta^T \boldkappa-\mathcal{F}\left(\boldkappa\right)\}.
\end{equation}

We will assume henceforth that {$\mathcal{F}(\cdot)$ and $\mathcal{A}(\cdot)$} are strictly convex and $C^2$ functions (twice continuously differentiable) at all points in their domains of definition. Optimising (\ref{conv_conj_sup}) over $\boldeta$, recalling the strict convexity of $\mathcal{A}(\cdot)$, we get that the optimising $\boldeta$ is unique and satisfies $\boldkappa = \nabla_{\boldeta} \mathcal{A}(\boldeta)$ which is the expectation parameter (\ref{exp_parameter}) evaluated at the optimising $\boldeta$. Similarly, optimising (\ref{conv_conj_sup2}) over $\boldkappa$, we get an equation analogous to (\ref{exp_parameter}), i.e., $\boldeta = \nabla_{\boldkappa} \mathcal{F}(\boldkappa)$. Thus the optimising $\boldkappa$ and $\boldeta$ are dual to each other and are in one-to-one correspondence. Indeed, we can move from $\boldeta$ to its corresponding $\boldkappa$ and from $\boldkappa$ to its corresponding $\boldeta$ via
\begin{equation}\label{dual}\boldkappa \left(\boldeta\right) = \nabla_{\boldeta} \mathcal{A}\left(\boldeta\right) \text{ and } \boldeta\left(\boldkappa\right) = \nabla_{\boldkappa} \mathcal{F}\left(\boldkappa\right).
\end{equation}
From this one-to-one relation between $\boldeta$ and $\boldkappa$ in (\ref{dual}), we also have
\begin{eqnarray}\label{conv_conj_base}
\begin{array}{c}
\mathcal{F}\left(\boldkappa\right) = \boldeta(\boldkappa)^T \boldkappa-\mathcal{A}\left(\boldeta(\boldkappa)\right), \\
\mathcal{A}\left(\boldeta\right) = \boldeta^T \boldkappa(\boldeta)-\mathcal{F}\left(\boldkappa(\boldeta)\right).
\end{array}
\end{eqnarray}
When we know that a particular $\boldeta$ and a particular $\boldkappa$ are dual to each other, we simplify the notation in (\ref{conv_conj_base}) to
\begin{eqnarray}\label{conv_conj}
\mathcal{F}\left(\boldkappa\right) + \mathcal{A}(\boldeta) = \boldeta^T \boldkappa.
\end{eqnarray}
That the dual parameter $\boldkappa(\boldeta)$ (respectively, $\boldeta(\boldkappa)$) is involved should be clear from the context since, in (\ref{conv_conj}), the supremum that appears in (\ref{conv_conj_sup2}) (respectively, (\ref{conv_conj_sup})) is absent. See \cite[Section 3.3.2]{Boyd} for these basic properties on convex duals.

The expressions for Kullback-Leibler (KL) divergence or relative entropy in terms of the natural parameter and in terms of the expectation parameter (by (\ref{conv_conj})) are
\begin{eqnarray}
D\left(\boldeta_1 \parallel \boldeta_2\right) &:=& D\left(f(\cdot|\boldeta_1)  \parallel  f(\cdot|\boldeta_2) \right) \nonumber \\
\label{eqn:relativeentropy-natural}
&=& \left(\boldeta_1-\boldeta_2\right)^T \boldkappa_1-\mathcal{A}\left(\boldeta_1\right) + \mathcal{A}\left(\boldeta_2\right)\\
\label{eqn:relativeentropy-expectation}
&=& \left(\boldkappa_2-\boldkappa_1\right)^T\boldeta_2 + \mathcal{F}\left(\boldkappa_1\right) - \mathcal{F}\left(\boldkappa_2\right).
\end{eqnarray}
Note that we have used the duality relation between $\boldkappa_i$ and $\boldeta_i$, i.e., $\boldkappa_i = \boldkappa(\boldeta_i)$, $i=1,2$. Let $\textsf{Hess}(\mathcal{A})(\cdot)$ denote the Hessian associated with the function $\mathcal{A}(\cdot)$. Another useful formula is obtained by expanding \eqref{eqn:relativeentropy-natural} using the Taylor series centred at $\boldeta_1$:
\begin{eqnarray}
  D(\boldeta_1 \parallel \boldeta_2) & = &
  \mathcal{A}(\boldeta_2) - \mathcal{A}(\boldeta_1) - (\boldeta_2 - \boldeta_1)^T \nabla \mathcal{A}(\boldeta_1) \nonumber
  \\
  & = & \frac{1}{2}(\boldeta_2 - \boldeta_1)^T \left[ \int_0^1 (1-t) \textsf{Hess}(\mathcal{A})(\boldeta_1 + t( \boldeta_2 - \boldeta_1))~dt \right] (\boldeta_2 - \boldeta_1). \label{eqn:TaylorSeries}
\end{eqnarray}
These useful formulae will be exploited in later sections.

\section{Problem Model, Specific Examples, and Preview of the Main Result}\label{sec:problem_model}

In this section, we first discuss the model and explain the costs under consideration. We then provide several examples considered in the literature that are encompassed by our generalised framework. We end the section with a formal problem statement and a preview of the main result.

\subsection{Problem Model}
Let the set of arms be denoted as $\mathcal{K}:=\{1,2, \ldots, K\}$. The distribution of the observations from arm $i$ is a member of the vector exponential family with the natural (canonical) parameter $\boldeta_i \in \Psi_i$ (open, convex subset of a Euclidean space). Let
\begin{equation}
\Omega = \Psi_1 \times \Psi_2 \times \ldots \times \Psi_K.
\end{equation}
Let $\overline{\boldeta} \in \Omega$ denote the tuple of vector parameters associated with the set of arms:
\begin{equation}
\overline{\boldeta} = (\boldeta_1,\boldeta_2,\ldots, \boldeta_K ).
\end{equation}
Let $\mathcal{M} = \{1,2, \ldots, M\}$ denote the set of hypotheses. Under the hypothesis $m\in \mathcal{M}$, $\overline{\boldeta} \in \Theta_m$, where $\Theta_m \subset \Omega$. We assume that the sets $\Theta_m$, $m \in \mathcal{M}$ are disjoint. In addition, we also make an assumption that $\Theta_m$ is open in aff($\Theta_m$), i.e., $\Theta_m = \text{relint}(\Theta_m)$, where aff($\cdot$) is the affine hull and relint($\cdot$) is the relative interior. Recall that we also assume that $\mathcal{A}(\cdot)$ is strictly convex and $C^2$; so the second central moment exists for each $\overline{\boldeta} \in \Theta_m$ and for each $m$, namely, $E_{\boldeta}[(\textbf{T}(x)-\boldkappa(\boldeta))(\textbf{T}(x)-\boldkappa(\boldeta))^T]$ exists and is finite.

Let $\mathcal{P}\left(\mathcal{K}\right)$ be the set of probability distributions on $\mathcal{K}$. Let $a_n \in \mathcal{K}$ denote the index of the arm chosen for observation at the instant $n$, and let $x_n$ denote the value of the observation during instant $n$. We write $x^n$ for $(x_1, x_2, \ldots, x_n)$ and $a^n$ for $(a_1, a_2, \ldots, a_n)$. At any stage, say $n$, given the past observations and actions up to time $n-1$, a policy $\pi$ must choose an action ${\overline{A}_n}$ of the form:
\begin{itemize}
\item ${\overline{A}_n} = (stop, \delta)$ which is a decision to stop and decide the hypothesis as $\delta \in {\mathcal{M}}$, or
\item ${\overline{A}_n} = (continue, \boldlambda, \delta = \textsf{null})$ which is a decision to continue and sample the next arm to pull according to a probability measure $\boldlambda$ on the finite set of arms.
\end{itemize}

We define the stopping time of the policy $\pi$ as
\begin{equation}
\tau\left(\pi\right) := \inf\{n \geq 1 : {\overline{A}_n} = \left(stop,\cdot \right)\}.
\end{equation}

Given the false detection probability constraint $\alpha$, with $0<\alpha<1$, let $\Pi\left(\alpha\right)$ be the set of {\em admissible policies} that meet the following constraint on the probability of false detection:
\begin{equation}
\Pi\left(\alpha\right) ~ =  ~ \left\{\pi: P\left(\delta \neq l ~|~ \overline{\boldeta} \right)\leq \alpha, \quad \forall \overline{\boldeta} \in \Theta_l, \quad \forall l \right\},
\end{equation}
with $\delta$ being the decision made when the algorithm stops. It is important to note that a policy is in $\Pi(\alpha)$ only if it does well for each $\overline{\boldeta} \in \Theta_l$, for each $l$. Policies tuned to specific $\overline{\boldeta}$'s or specific $\Theta_l$ will likely fail when other hypotheses are in vogue.

\subsection{Examples}
\label{subsec:examples}
We discuss several examples already studied in the literature and show how they fit within our general framework. In each case, we first pose the problem and show how to embed it within our framework. The embedding sheds light on the structural aspects, associated with the parameter sets, of the specific problem under consideration.

\begin{itemize}
\item \textit{Generalised best arm identification in a multi-armed bandit setting}: We consider a set of $K$ arms, each following a distribution from the vector exponential family. Our objective is to identify the {\em best} arm $i \in \mathcal{K}$ such that
\begin{equation}
{\bf c}^T\boldeta_i > {\bf c}^T\boldeta_j, \forall j \in \mathcal{K} \backslash i,
\end{equation}
where ${\bf c} \in \mathbb{R}^d$.
We assume it is a priori known that there is exactly one such arm.

We cast this problem into our framework as follows. Let the number of hypotheses $M = K$. The parameter set under hypothesis $m$ can be taken to be
\begin{equation}
\Theta_m = \{\overline{\boldeta} \in \Omega: {\bf c}^T\boldeta_m > {\bf c}^T\boldeta_{m'} , \forall m' \neq m\}.
\end{equation}
For the scalar parameter case where the parameter is the mean and ${\bf c} = 1$, we have the best arm identification problem; see Fig. \ref{bestarm2}.
\begin{figure}[ht]
\begin{center}
\begin{subfigure}{.25\textwidth}
\begin{tikzpicture}[scale=3]
\draw[thick,->] (0,0) -- (0,1.2) node[anchor=north east]{$\eta_2$};
\draw[thick,->] (0,0) -- (1.2,0) node[anchor=north east]{$\eta_1$};
\def\x{1}
\filldraw[draw=black, fill=gray!20]
         (0,0)
            -- (1,0)
            -- (1,1)
            -- cycle;
\node[] at (0.6,0.25) {$\Theta_1$};
\end{tikzpicture}
\caption{}
\end{subfigure}
\quad
\begin{subfigure}{.25\textwidth}
\begin{tikzpicture}[scale=3]
\draw[thick,->] (0,0) -- (0,1.2) node[anchor=north east]{$\eta_2$};
\draw[thick,->] (0,0) -- (1.2,0) node[anchor=north east]{$\eta_1$};
\def\x{1}
\filldraw[draw=black, fill=gray!20]
         (0,0)
            -- (0,1)
            -- (1,1)
            -- cycle;
\node[] at (0.25,0.6) {$\Theta_2$};
\end{tikzpicture}
\caption{}
\end{subfigure}
\caption{Scalar best arm identification with $K=2$. Under hypothesis $1$, we have $\eta_1 > \eta_2$ and $\Theta_1$ is the shaded region indicated in (a). Under hypothesis $2$, $\Theta_2$ is the shaded region in (b). It is assumed that $\eta_1 \neq \eta_2$.}
\label{bestarm2}
\end{center}
\end{figure}
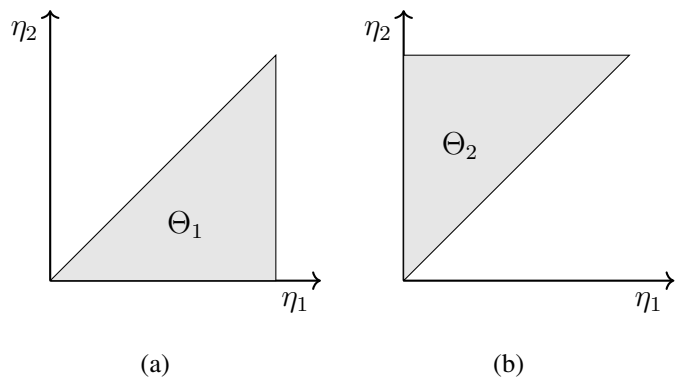

This problem was posed at least as early as Chernoff \cite{chernoff} and was subsequently studied by Albert \cite{ref:albert1961sequential}. For a more recent study with better performance for the best arm problem see Garivier and Kaufmann \cite{garivier16}. For results in the nonasymptotic regime see  \cite{kaufmann2013information,degenne2019nonasymptotic}. 

\item \textit{Multi-bandit best arm identification}: We consider a set of $K$ arms, each following a distribution from the vector exponential family. We assume that there are $b$ possibly overlapping group of arms denoted by the subsets $B_1, B_2, \ldots , B_b$ of $\{1,2, \ldots, K\}$. Each arm is present in at least one of these groups and each group has at least two arms and a unique best arm. Our objective is to find the best arm in each of these groups, i.e., to find $m=\{m_1,m_2, \ldots ,m_b\} \subset \mathcal{K}$ such that for $m_k \in B_k$, and  for each $k \in \{1,2, \ldots,b\}$
\begin{equation}
{\bf{c}}^T \boldeta_{m_k} > {\bf{c}}^T \boldeta_j, ~~ \forall j \in B_k, ~~j \neq m_k.
\end{equation}

We can embed this problem into our setting as follows. Let  the number of hypothesis $M=|B_1| \times |B_2| \times \cdots \times |B_b|$. Take $m = (m_1, m_2, \ldots, m_b) \in B_1 \times B_2 \times \cdots \times B_p$. The parameter set under the hypothesis $m$ can be defined as
\begin{equation}
\Theta_m = \{\overline{\boldeta} \in \Omega: {\bf{c}}^T \boldeta_{m_k} > {\bf{c}}^T \boldeta_j, ~ \forall j \in B_k \setminus \{m_k\}, ~~ \forall k  \}.
\end{equation}

A special case of this problem was studied in \cite{scarlett2019overlapping}, where $\boldeta$ is the mean parameter and the distributions are sub-Gaussian, i.e., $E[e^{sX}] \leq e^{\frac{\sigma^2 s^2}{2}}, \forall s \in \mathbb{R}$ with $\sigma \leq 1/2$.

\item \textit{Odd arm identification in multi-armed bandit setting}\label{eg:odd_arm}: We consider a set of $K$ arms, each following a distribution from the vector exponential family, in which all but one have the same distribution. The objective is to identify the odd arm, i.e., to find the $i \in \mathcal{K}$ such that
\begin{equation}
\boldeta_i=\boldtheta, \text{ and } \boldeta_j = \boldtheta', \forall j \in \mathcal{K} \backslash \{i\} \text{ for some } \boldtheta' \neq \boldtheta.
\end{equation}
The decision maker knows, a priori, that there is such an odd arm.

This problem too can be embedded in our setting as follows. Let the number of hypotheses $M=K$, and let the parameter set under hypothesis $m$ be defined as
\begin{equation}
\Theta_m = \{\overline{\boldeta}\in \Omega: \boldeta_m = \boldsymbol{\theta}, \text{ and } \boldeta_{m'} = \boldsymbol{\theta}', \forall m' \neq m, ~ \boldsymbol{\theta}' \neq \boldsymbol{\theta}\},
\end{equation}
where the hypothesis $m$ indicates that the arm $m$ is the odd one.
\tdplotsetmaincoords{70}{100}
\begin{figure}[ht]
\begin{center}
\begin{subfigure}{.25\textwidth}
\begin{tikzpicture}[scale=3,tdplot_main_coords]
    \draw[thick,->] (0,0,0) -- (1,0,0) node[anchor=north east]{$\eta_2$};
    \def\x{1}
    \draw[thin] (0,0,0) -- (1,1,0) node[above] {$\eta_2=\eta_3$};
    \filldraw[
        draw=black,opacity=0.5,
        fill=gray!20,%
    ]          (0,0,0)
            -- (1,1,0)
            -- (1,1,1)
            -- (0,0,1)
            -- cycle;
    \draw[thick,->] (0,0,0) -- (0,1,0) node[anchor=north west]{$\eta_3$};
    \draw[thick,->] (0,0,0) -- (0,0,1) node[anchor=south]{$\eta_1$};
\node[] at (0.5,0.5,0.6) {$\Theta_1$};
\end{tikzpicture}
\caption{}
\end{subfigure}
\quad
\begin{subfigure}{.3\textwidth}
\begin{tikzpicture}[scale=3,tdplot_main_coords]
    \draw[thick,->] (0,0,0) -- (1,0,0) node[anchor=north east]{$\eta_2$};
    \def\y{1}
    \draw[thin] (0,0,0) -- (0,1,1) node[above] {$\eta_3=\eta_1$};
    \filldraw[
        draw=black,opacity=0.5,
        fill=gray!20,%
    ]          (0,0,0)
            -- (0,1,1)
            -- (1,1,1)
            -- (1,0,0)
            -- cycle;
    \draw[thick,->] (0,0,0) -- (0,1,0) node[anchor=north west]{$\eta_3$};
    \draw[thick,->] (0,0,0) -- (0,0,1) node[anchor=south]{$\eta_1$};
\node[] at (0.5,0.5,0.5) {$\Theta_2$};
\end{tikzpicture}
\caption{}
\end{subfigure}
\quad
\begin{subfigure}{.3\textwidth}
\begin{tikzpicture}[scale=3,tdplot_main_coords]
    \draw[thick,->] (0,0,0) -- (1,0,0) node[anchor=north east]{$\eta_2$};
    \def\z{1}
    \draw[thin] (0,0,0) -- (1,0,1) node[above] {$\eta_1=\eta_2$};
    \filldraw[
        draw=black,opacity=0.5,
        fill=gray!20,%
    ]          (0,0,0)
            -- (1,0,1)
            -- (1,1,1)
            -- (0,1,0)
            -- cycle;
    \draw[thick,->] (0,0,0) -- (0,1,0) node[anchor=north west]{$\eta_3$};
    \draw[thick,->] (0,0,0) -- (0,0,1) node[anchor=south]{$\eta_1$};
\node[] at (0.5,0.5,0.6) {$\Theta_3$};
\end{tikzpicture}
\caption{}
\end{subfigure}
\end{center}
\caption{Odd arm identification with $K=3$. Under hypothesis $1$, $\eta_2=\eta_3\neq\eta_1$ and $\Theta_1$ is the shaded portion indicated in (a) excluding the line $\eta_1=\eta_2=\eta_3$. In (b) and (c), the shaded portions excluding the line $\eta_1=\eta_2=\eta_3$ indicate the parameter sets $\Theta_2$ and $\Theta_3$, respectively.}
\end{figure}
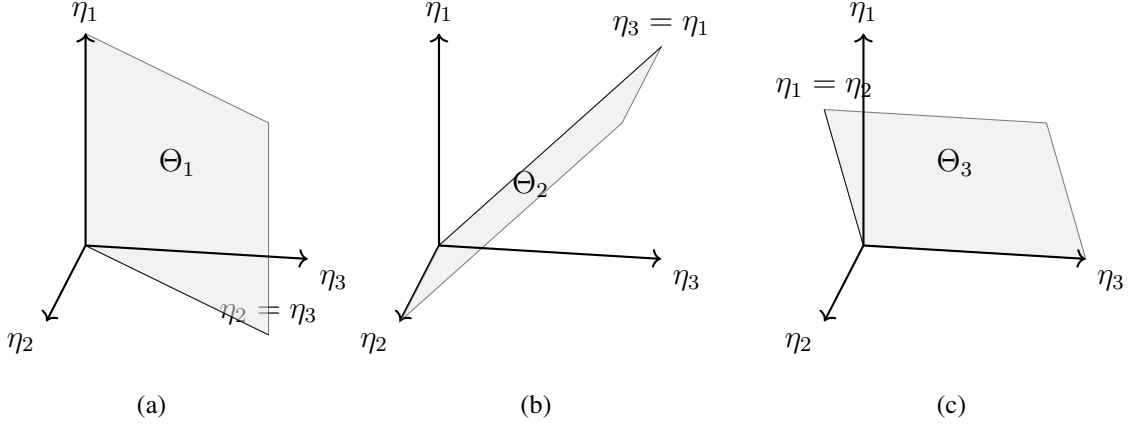

This problem was considered in the conference version \cite{8755796} and its accompanying technical report \cite{gayathri} as the exponential family generalisation of \cite{vaidhiyan17,vaidhiyan15} where the observations were restricted to be Poisson random variables. Note that the odd arm detection problem is a particular case of structured best arm identification \cite{huang2017structured}.

\item \textit{L-anomalous arms identification in the multi-armed bandit setting}: Here, we consider the case when we have multiple anomalous arms, i.e., out of the $K$ arms, $L$ arms have a distribution different from the rest.

We can embed this problem into our setting as follows. Let the number of hypothesis $M= {K \choose L}$ and enumerate the subsets as $S_1, S_2, \ldots, S_M$, $|S_m|=L$, $1 \leq m \leq M$. The parameter set under hypothesis $m$ associated with $S_m$ is defined as
\begin{equation}
\Theta_m = \{\overline{\boldeta}\in \Omega: \boldeta_i = \boldsymbol{\theta}, \forall i \in S_m \text{ and } \boldeta_j = \boldsymbol{\theta}', \forall j \notin S_m, \boldtheta' \neq \boldtheta\}.
\end{equation}
The hypothesis $m$ indicates that the arms in $S_m$ are anomalous.

This has been considered in \cite{li2017universal}.
For a summary of various other kinds of anomaly detection problems, see \cite{tajer2014outlying}.

\item \textit{High reward outlier detection in the multi-armed bandit setting}: Consider the problem of identifying outlier arms with extremely high expected reward compared to the other arms. An arm is defined as an outlier if the expectation parameter is greater than the mean plus $k$ times standard deviation of the expectation parameters of all the arms, i.e., arm $i$ is an outlier when
\begin{equation}
\kappa_i > \mu + k \sigma=\theta,
\end{equation}
where $\kappa_i$ is the expectation parameter of arm $i$, $\mu$ and $\sigma$ are calculated as
\begin{equation}
\mu = \frac{1}{K} \sum\limits_{i=1}^K \kappa_i \text{ and } \sigma = \sqrt{\frac{1}{K}\sum\limits_{i=1}^K (\kappa_i-\mu)^2},
\end{equation}
respectively. It is a priori known that this set is nonempty.

We can embed this problem too into our setting as follows. Consider a set of $K$ arms, each following a distribution from the exponential family.  Let the number of hypothesis $M = 2^K-1$ and enumerate the subsets $P_m$ of $2^\mathcal{K}\backslash\emptyset$, $m = 1,2, \ldots, M$ . Let the parameter set under hypothesis $m$ be defined as, with $\overline{\boldeta} = (\eta_1, \ldots, \eta_K)$,
\begin{equation}
\Theta_{m} = \{\overline{\boldeta} \in \Omega: \kappa_i > \theta, \forall \eta_i \in P_m\}.
\end{equation}
The hypothesis $m$ indicates that the arms in the set $P_m$ are outliers.

This problem was studied in \cite{zhuang2017identifying}.

\item \textit{Partition identification problem in the multi-armed bandit setting}: In this setting, the parameter space is {\em partitioned} into $M$ sets. The goal is to identify the subset of the partition in which the parameter belongs. The general problem addressed in this paper does not require the $\Theta_m, ~m \in \mathcal{M}, $ to be a partition of $\Omega$. Two such problems (with $M = 2$) and their embedding in our framework are described below. We take $\overline{\boldeta} = (\eta_1, \ldots, \eta_K)$, where the $\eta_i$ are (in this set of examples) scalars.


\begin{itemize}
\item \textit{Threshold crossing problem} \cite{juneja2018sample}: In this setting, our objective is to check if there is at least one arm whose parameter is above a given threshold value $u$. Define the parameter set $\Theta_1$ to be
\begin{equation}
\Theta_1 = \left\{{\overline{\boldeta}} \in \Omega: \max\limits_{i \leq K} \eta_i > u \right\},
\end{equation}
and  $\Theta_2 = \text{relint}(\Theta_1^c)$.
\item \textit{Half-space identification problem} \cite{juneja2018sample}: In this setting, our objective is to identify the half-space that contains the parameters. Fix constants $(a_1,a_2,\ldots, a_k,b)$. Define the parameter set $\Theta_1$ to be
\begin{equation}
\Theta_1 = \left\{\overline{\boldeta} \in \Omega: \sum\limits_{i=1}^Ka_i\eta_i >b \right\},
\end{equation}
and $\Theta_2 = \text{relint}(\Theta_1^c)$.
\end{itemize}
\end{itemize}

As one can see from the rich set of examples above, our framework is sufficiently general to cover all these examples considered in the literature.

\subsection{Costs}
The total cost is the sum of the switching cost and the delay cost in arriving at a decision, as in \cite{vaidhiyan15_costs}. We now make this precise.

\subsubsection{Switching cost}
Let $g\left(a,a'\right)$ denote the cost of switching from arm $a$ to arm $a'$. This is incurred every time a switch of arms is executed. We assume
\begin{equation*}
g\left(a,a'\right) \geq 0 \text{ } \forall a, a' \in \mathcal{K} \text{ and } g\left(a,a\right) = 0 \quad \forall a \in \mathcal{K}.
\end{equation*}
The assumption $g(a,a) = 0$ says there is no switching cost if the controller does not switch arms.
Define $g_{\max}$ as follows and assume that it is finite:
$$g_{\max} := \max\limits_{a,a' \in \mathcal K} g\left(a,a'\right) < \infty.$$
\subsubsection{Total cost}
For a policy $\pi \in \Pi\left(\alpha\right)$, the total cost $C\left(\pi\right)$ is the sum of the stopping time (delay) and the net switching cost:
\begin{equation*}
C\left(\pi\right) := \tau\left(\pi\right) + \sum\limits_{l=1}^{\tau\left(\pi\right)-1} g\left(a_l,a_{l+1}\right).
\end{equation*}

\subsection{Problem statement and a preview of main result}

{\em Problem Statement}: Our goal is to identify, for each $l$ and for each $\overline{\boldeta} \in \Theta_l$, the asymptotic growth rate of the cost $\inf_{ \pi \in \Pi(\alpha)} E[C(\pi) ~|~ \overline{\boldeta}]$ with respect to $\log (1/\alpha)$ as the constraint on the probability of false detection $\alpha$ vanishes. More precisely, we wish to identify

\begin{equation*}
  \lim_{\alpha \downarrow 0} \inf_{\pi \in \Pi(\alpha)} \frac{E[ C(\pi) ~|~ \overline{\boldeta}] }{\log (1/\alpha)}.
\end{equation*}

\vspace*{5mm}

{\em A preview of the main result}: We will argue that

\begin{equation*}
  \lim_{\alpha \downarrow 0} \inf_{\pi \in \Pi(\alpha)} \frac{E[ C(\pi) ~|~ \overline{\boldeta} ] }{\log (1/\alpha)} = \frac{1}{D^*(\overline{\boldeta})},
\end{equation*}
where $D^*(\overline{\boldeta})$ is the solution to a max-min problem to be defined later in (\ref{eqn:D*}). The converse, as usual, follows from a smart application of the data-processing inequality, and involves the stumbling block of a sequential hypothesis test between $\overline{\boldeta} \in \Theta_l$ and its nearest alternative in $\Theta_{-l}$.

The achievability result, however, requires us to address several nontrivial and nuanced issues which we now highlight.
\begin{itemize}
  \item We need the cumulant generating function $\mathcal{A}(\cdot)$ to be strictly convex, a consequence of the minimality of the representation, and further in $C^1$ (continuously differentiable). The former ensures 1-to-1 correspondence between the $\boldeta$ and the $\boldkappa$ parameters. The latter ensures that the relative entropy is continuous in the parameters of the problem.
  \item To ensure that the estimated parameters approach the true parameters, one needs sufficient exploration. We use an $O(n^{\beta})$ exploration scheme, where $1/2 < \beta < 1$. See equation (\ref{eqn:D-tracking}) in Section \ref{sec:policy}.
  \item For certain concentration results to hold, we need finite second central moments and a regularity condition on relative entropy -- that it diverges as the separation between the associated canonical parameters grows without bound. If $\mathcal{A}(\cdot)$ is twice-differentiable, the former corresponds to the positive definiteness of the Hessian matrix and the latter corresponds to its condition number not vanishing too quickly.
  \item The optimisation problem leads to a set of maximising actions. However these may not be singletons. As a consequence, as we show later, we only get upper semicontinuity of the maximising set-valued correspondence. On the other hand, the use of the certainty equivalence principle involves actions based on maximisers associated with the estimated parameters. One therefore needs a continuous selection for the action mapping.
  \item As the estimated parameters approach the true parameters, the policies used also vary with time. One needs to show convergence in this complex regime which has time-varying estimates and certainty equivalence based actions.
  \item On account of zero switching cost for no switching and on account of $g_{\max} < \infty$, the asymptotic growth rate of $E[ C(\pi) ~|~ \overline{\boldeta}]$ can be made as close to the asymptotic growth rate without switching cost, i.e., the growth rate of $E[ \tau(\pi) ~|~ \overline{\boldeta}]$,  as one wishes. This involves the use of a {\em sluggish} policy that switches at a very slow rate, yet mimics the stationary distribution associated with the asymptotically optimal policy for no switching costs.
\end{itemize}

\section{The Converse (Lower bound on delay)}
\label{sec:converse}
The following proposition gives an information theoretic lower bound on the expected conditional stopping time for any policy that belongs to $\Pi\left(\alpha\right)$, given the true configuration is $\overline{\boldeta} \in \Theta_l$. The decision maker a priori does not know either $\overline{\boldeta}$ or $l$.

\begin{proposition}\label{prop:lower_bound}
Fix $0 < \alpha<1$. Let $\overline{\boldeta} \in \Theta_l$ be the true configuration. For any $\pi \in \Pi(\alpha)$, we have
\begin{equation}
E[\tau|\overline{\boldeta} ] \geq \frac{d_b(\alpha \parallel 1-\alpha)}{D^*(\overline\boldeta)}
\end{equation}
where $d_b(\cdot \parallel \cdot)$ is the binary relative entropy function, and $D^*(\overline{\boldeta})$ is defined as
\begin{equation}\label{eqn:D*}
D^*(\overline{\boldeta}) = \sup_{\boldlambda \in \mathcal{P}({\mathcal{K}})} ~~ \inf_{\overline{\boldeta}' \in \Theta_{-l}} ~~ \sum\limits_{i=1}^K \lambda_i D(\boldeta_i \parallel \boldeta_i'),
\end{equation}
where $\overline{\boldeta}' = (\boldeta_1',\boldeta_2',\ldots, \boldeta_K').$
\end{proposition}

\begin{IEEEproof}
The proof follows straightforwardly from \cite[Lem.~1]{kaufmann2016complexity} and involves an application of the data processing inequality and Wald's lemma. See also \cite{garivier2019explore}. We omit the details.
\end{IEEEproof}

\vspace*{.25cm}

The binary relative entropy function is given by the familiar expression:
\[
  d_b(\alpha \parallel 1-\alpha) = \alpha \log \frac{\alpha}{1-\alpha} + (1-\alpha) \log \frac{1-\alpha}{\alpha}.
\]
As the constraint on the probability of false detection $\alpha \rightarrow 0$, we have
$$d_b\left(\alpha  \parallel  1-\alpha\right)/(\log\left(1/\alpha\right))\rightarrow 1.$$ Hence, we get that the conditional expected stopping time of the optimal policy scales at least as $(\log\left(1/\alpha\right))/D^*\left(\overline{\boldeta}\right)$.

\begin{corollary}
\label{cor:lowerbound}
Fix $0 < \alpha<1$. Let $\overline{\boldeta} \in \Theta_l$ be the true configuration. For any $\pi \in \Pi(\alpha)$, we have
\begin{equation}
E[C\left(\pi\right)|\overline{\boldeta}] \geq \frac{d_b\left(\alpha  \parallel  1-\alpha\right)}{D^*\left(\overline{\boldeta}\right)}.
\end{equation}
\end{corollary}

\begin{IEEEproof}
With the switching costs added, we have $C\left(\pi\right) \geq \tau\left(\pi\right)$. Hence the corollary follows from Proposition \ref{prop:lower_bound}.
\end{IEEEproof}

{\em Interpretation of the sup-inf optimisation problem in \eqref{eqn:D*}}: We can interpret $D^*(\overline{\boldeta})$ as follows. Consider the simpler hypothesis testing problem where the decision maker has to decide between the given $\overline{\boldeta} \in \Theta_l$ and any alternative chosen from $\Theta_{-l}$ by an adversary. The decision maker may choose a sampling strategy $\boldlambda \in \mathcal{P}(\mathcal{K})$. Knowing this, the adversary may pick, from $\Theta_{-l}$, the nearest alternative to $\overline{\boldeta}$ that minimises the separation as measured by the $\boldlambda$-weighted average of the relative entropies of the arms. Realising this, the decision maker will ensure that his chosen $\boldlambda$ maximises the minimum separation. This is the decision maker's best guarding policy against the (more informed) adversary's strategy of picking the nearest alternative to $\overline{\boldeta}$ from outside $\Theta_{l}$, i.e., $\Theta_{-l}$.

In the next section, we discuss how to convert the above intuitive interpretation into a policy that achieves this lower bound.

\section{A Sluggish and Modified GLR Test with Forced exploration}
\label{sec:achievability}

In this section, we describe a suitable policy that can achieve the growth rate in the lower bound in Proposition \ref{prop:lower_bound}, as the constraint on the probability of false detection is driven to zero. The proposed policy is a modification of the policy $\pi_{SM}$ discussed in \cite{gayathri} where we replace the random sampling strategy by a forced exploration technique\footnote{The subscript SM in $\pi_{SM}$ stands for ``sluggish'' and ``modified''. We shall use $\pi_{SMF}$ where the added letter F stands for ``forced exploration''.} as in \cite{garivier16}.

Let us denote by $\mathcal{A}_i(\cdot)$ the cumulant generating function associated with arm $i$. Recall the assumption that $\mathcal{A}_i(\cdot)$ is $C^2$. This ensures $\boldkappa_i(\cdot)$ is continuous and furthermore that the relative entropy $D(\cdot \parallel \boldeta'_i)$ is continuous for each fixed $\overline{\boldeta}'$. As a consequence, for any $i$ and any $\boldeta_i,\boldeta_i' \in \Psi_i$ such that $\boldeta_i \neq \boldeta_i'$, we have $D(\boldeta_i \parallel \boldeta_i') < \infty$. Furthermore, on account of the $C^2$ condition, the observations have finite second central moments. Indeed, as already highlighted, the Hessian matrix of $\mathcal{A}_i(\boldeta_i)$ is just the covariance of $\mathbf{T}(x)$ when the parameter is $\boldeta_i$, the strict convexity of $\mathcal{A}_i(\boldeta_i)$ at all $\boldeta_i$ is the same as positive definiteness of the associated covariance matrix, and the $C^2$ condition on $\mathcal{A}_i(\boldeta_i)$ is the same as saying that the covariance matrix of $\mathbf{T}(x)$ has entries that are continuous in the parameter $\boldeta_i$.

\subsection{Continuous selection of the optimal sampling strategy}
\label{subsec:regularityassumptions}


In this section, we will highlight the usefulness of a continuous selection of the sampling strategy. To set the stage, we first prove the following property about the attainment of the supremum in the definition of $D^*(\overline{\boldeta})$ in (\ref{eqn:D*}).

\begin{proposition}\label{prop:lambda_continuity}
The supremum in (\ref{eqn:D*}) is a maximum, i.e., for each $l$ and for each $\overline{\boldeta} \in \Theta_l$, we can write
\begin{equation}
\label{eqn:lambda-opt}
\boldlambda^*(\overline{\boldeta}) = \argmax\limits_{\boldlambda \in \mathcal{P}({\mathcal{K}})} \inf\limits_{\overline{\boldeta}'\in \Theta_{-l}} \sum\limits_{i=1}^K \lambda_i D(\boldeta_i \parallel \boldeta_i').
\end{equation}
Furthermore  the mapping $\overline{\boldeta} \mapsto \boldlambda^*(\overline{\boldeta})$ is an upper semi-continuous convex-valued correspondence.
\end{proposition}
\begin{IEEEproof}
See Appendix \ref{app:lambda_continuity}.
\end{IEEEproof}

In any problem instance of the type considered in this paper, the learner does not know the true parameters, and must learn these parameters along the way. Our strategy to attain the lower bound is to estimate the parameters $\overline{\boldeta}$, say via $\widehat{\overline{\boldeta}}(n)$ at time instant $n$, to apply the certainty equivalence principle by taking the estimated $\widehat{\overline{\boldeta}}(n)$ to be true parameter, and then to apply a sampling strategy from the set $\boldlambda^*(\widehat{\overline{\boldeta}}(n))$. Since the estimated parameter can at best approach the true parameter as time $n \rightarrow \infty$, for our scheme to work, continuity in the sampling strategy will prove beneficial. If the desired continuity does not hold, the rate at which information and therefore confidence is gathered, based on $\boldlambda^*(\widehat{\overline{\boldeta}}(n))$ may not match the rate at which information should be gathered, as per $\boldlambda^*(\overline{\boldeta})$, to meet the lower bound. Observe however that the mapping $\boldeta \mapsto \boldlambda^*(\boldeta)$ is only an upper semicontinuous correspondence, and may not possess, in general, a continuous selection \cite[Sec.~9.2]{aubin2009set}. We shall therefore make the following assumption on the existence of a continuous selection. Let
\begin{equation}
  \label{eqn:F-definition}
  F(\boldlambda, \overline{\boldeta}) := \inf\limits_{\overline{\boldeta}'\in \Theta_{-l}} \sum\limits_{i=1}^K \lambda_i D(\boldeta_i \parallel \boldeta_i').
\end{equation}
Then $\boldlambda^*(\overline{\boldeta})$ in~(\ref{eqn:lambda-opt}) optimises $F(\cdot, \overline{\boldeta})$.

\vspace*{.25cm}

\noindent {\bf Assumption A}: The correspondence $\overline{\boldeta} \mapsto \boldlambda^*(\overline{\boldeta})$ admits a continuous selection.

\vspace*{.25cm}

This assumption holds for example when
\begin{description}
  \item{(i)} $\boldlambda^*(\cdot)$ is single-valued, a condition that holds when $F(\cdot, \overline{\boldeta})$ is strictly concave for each $\boldeta \in \Theta_{l}$, see \cite[Th.~9.17]{sundaram1996first}; or
  \item{(ii)} $\boldlambda^*(\cdot)$ is lower semicontinuous, see \cite[Sec.~9.1]{aubin2009set}.
\end{description}
We shall verify in Appendix \ref{app:verifyCS} that Assumption A holds in the examples of odd arm and best arm identifications. There are interesting situations where we have not yet been able to establish that Assumption A holds, and they involve nonconvex sets that are not necessarily finite unions of convex sets, e.g., two-dimensional independent Gaussians with variances 1 but unknown means $\eta_1$ and $\eta_2$, with $\Theta_1 = \{ \overline{\boldeta} = (\eta_1, \eta_2) ~:~ \eta_1^2 + \eta_2^2 < 1 \}$, and $\Theta_2 = \{ \overline{\boldeta} = (\eta_1, \eta_2) ~:~ 1 < \eta_1^2 + \eta_2^2 < 2 \}$. Whether a continuous selection exists for this setting is still open.

\subsection{Additional notations}
Let $N_i^n$ denote the number of times the arm $i$ was chosen for observation up to time $n$, i.e.,
\begin{equation}
  N_i^n = \sum\limits_{t=1}^n1_{\{a_t=i\}},\label{eqn:num-samples}
\end{equation}
{where $a_t$ is the arm chosen at time $t$. Clearly}  $n=\sum_{i=1}^K N_i^n$. Let $\textbf{Y}_i^n$ denote the sum of sufficient statistic of arm $i$ up to time $n$, i.e.,
\begin{equation}
  \textbf{Y}_i^n = \sum\limits_{t=1}^{n}\textbf{T(}x_t\textbf{)}1_{\{a_t=i\}}.\label{eqn:num-points}
\end{equation}

We will use the letter $f(\cdot)$ to denote all probability density functions. Conditional densities will be denoted by $f(\cdot|\cdot)$. The argument(s) will help identify the appropriate random variable(s) whose density (conditional density) is being represented. We also use it to denote {\em likelihoods} and {\em conditional likelihoods} without the normalisation needed to make them probability densities or conditional probability densities.

\subsection{Likelihood function}

Let $f(x^n|a^n,\overline{\boldeta})$ be the likelihood function of the observations upto time $n$, conditioned on the actions and the parameters $\overline{\boldeta}$, i.e.,
\begin{eqnarray}
f(x^n|a^n,\overline{\boldeta}) &=&\prod\limits_{t=1}^n f(x_t|a_t, \boldeta_{a_t})\\
&=& \prod\limits_{t=1}^n h(x_t)\exp\left\{\boldeta_{a_t}^T {\bf{T}}(x_t)-\mathcal{A}_{a_t}(\boldeta_{a_t})\right\}\\
							   &=& \left(\prod\limits_{t=1}^n h(x_t)\right)\prod\limits_{i=1}^{K}\exp\left\{\boldeta_i^T \sum\limits_{t=1}^n{\bf{T}}(x_t)1_{\{a_t=i\}}-N_i^n\mathcal{A}_i(\boldeta_i)\right\}\\
                               &=&  \left(\prod\limits_{t=1}^n h(x_t)\right) \prod\limits_{i=1}^{K}\exp\left\{\boldeta_i^T{\bf{Y}}_i^n-N_i^n\mathcal{A}_i(\boldeta_i)\right\}\label{eqn:likelihood}.
\end{eqnarray}
Then the log likelihood function is
\begin{eqnarray}
\log f(x^n|a^n,\overline{\boldeta}) &=& \sum\limits_{t=1}^{n}\log h(x_t) + n\sum\limits_{i=1}^{K} \frac{N_i^n}{n}\left\{\boldeta_i^T\frac{{\bf{Y}}_i^n}{N_i^n}-\mathcal{A}_i(\boldeta_i)\right\}\\
                                    &=& \sum\limits_{t=1}^{n}\log h(x_t) + n\sum\limits_{i=1}^{K} w_i\left\{\boldeta_i^T\hat{\boldkappa}_i-\mathcal{A}_i(\boldeta_i)\right\},
\end{eqnarray}
where $w_i := N_i^n/n$ and $\hat{\boldkappa}_i := {\bf{Y}}_i^n/N_i^n$ .
\subsection{Maximum likelihood function}
Consider a sequence $\delta_n \rightarrow 0$. Let the maximum likelihood estimates of the natural parameters under the hypothesis $m$ be defined as
\begin{equation}\label{eqn:ml_argument}
{\overline{\boldeta}}^*(m):= (\boldeta_1^*(m), \ldots, \boldeta_K^*(m)) \in \argmax\limits_{\overline{\boldeta}\in \Theta_m} \sum\limits_{i=1}^{K} w_i\left\{\boldeta_i^T\hat{\boldkappa}_i-\mathcal{A}_i(\boldeta_i)\right\}
\end{equation}
if the maximum is attained, and as some $\overline{\boldeta}^*(m) \in \Theta_m$ so that
\begin{equation}
\left|\sup\limits_{\overline{\boldeta} \in \Theta_m}\sum\limits_{i=1}^K w_i\left[\boldeta_i^T\hat{\boldkappa}_i-\mathcal{A}_i(\boldeta_i)\right]-\sum\limits_{i=1}^K w_i\left[\boldeta^*_i(m)^T\hat{\boldkappa}_i-\mathcal{A}_i(\boldeta^*_i(m))\right]\right| \leq \delta_n \label{eqn:nearest_ml}
\end{equation}
if the maximum is not attained.
When the maximum exists, the expression for relative entropy and some algebraic manipulation shows that
\begin{equation}\label{eqn:ml_argument_alt}
{\overline{\boldeta}}^*(m) \in \argmin\limits_{\overline{\boldeta}\in \Theta_m} \sum\limits_{i=1}^{K} w_i D(\hat{\boldeta}_i \parallel \boldeta_i).
\end{equation}
We differ here from Deshmukh et al. \cite{deshmukh2018controlled} in that our estimate is an ML estimate that recognises that $\overline{\boldeta} \in \Theta_m$ while Deshmukh et al. \cite{deshmukh2018controlled} first optimise over all $\overline{\boldeta}$ and then project this value on to $\Theta_m$. So our approach is more direct, but requires the existence of a continuous selection (Assumption A). In this context, note that Deshmukh et al. \cite{deshmukh2018controlled} also assume continuous selection by asking for the $\boldlambda^*(\cdot)$ to be single-valued (sufficient condition (i) for Assumption A to hold).

The log ML function is obtained as
\begin{eqnarray}
\log \hat{f}(x^n|a^n,\overline{\boldeta} \in \Theta_m) & = & \sum\limits_{t=1}^{n}\log h(x_t)+n\sum\limits_{i=1}^{K}w_i\left\{{\boldeta}_i^{*T}(m)\hat{\boldkappa}_i-\mathcal{A}_i({\boldeta}_i^{*}(m))\right\}.\label{eqn:ml_function}
\end{eqnarray}

\subsection{Average likelihood function}
When the parameters are unknown, a natural conjugate prior on the parameter $\boldeta_i$ enables easy updates of the posterior distribution based on observations. The conjugate prior, also denoted $f(\overline{\boldeta}| \overline{\boldeta}\in \Theta_l)$, is taken to be a product distribution over $i=1, \ldots, K$, with each marginal coming from an exponential family of the same form characterised by the hyper-parameters $\overline{\boldupsilon} = (\boldupsilon_1,\boldupsilon_2, \ldots, \boldupsilon_K)$ and ${\bf{n_0}} = (n_{01}, \ldots, n_{0K})$, i.e.,
\begin{equation}\label{eqn:conj_prior}
f(\overline{\boldeta}| \overline{\boldeta}\in \Theta_l) = \begin{cases}
                                              \mathcal{H}_l(\overline{\boldupsilon},{\bf{n_0}}) \prod\limits_{i=1}^K  \exp\left\{\boldeta^T_i\boldupsilon_i-n_{0i}\mathcal{A}_i(\boldeta_i)\right\}, &\text{ if } \overline{\boldeta} \in \Theta_l\\
                                                0, &\text{ otherwise}
                                                \end{cases}
\end{equation}
with
\begin{equation}\label{eqn:norm_factor}
\mathcal{H}_l(\overline{\boldupsilon},{\bf{n_0}}) = \left[~~\int\limits_{\overline{\boldeta}\in \Theta_l}\prod\limits_{i=1}^K\exp\left\{\boldeta^T_i\boldupsilon_i-n_{0i}\mathcal{A}_i(\boldeta_i)\right\}d\overline{\boldeta}\right]^{-1}
\end{equation}
as the normalising factor. We remark that the conjugate prior is an {\em artificial prior}, used mainly as an analytical artifice for easy posterior updates.

The average likelihood function at time $n$, averaged according to the artificial prior in (\ref{eqn:conj_prior}), is
\begin{eqnarray}
\tilde{f}_l(x^n|a^n) & := & \int\limits_{\overline{\boldeta} \in \Theta_l} f(x^n|a^n,\overline{\boldeta}) \cdot f(\overline{\boldeta}|\overline{\boldeta}\in \Theta_l) d\overline{\boldeta}\label{eqn:avg_likelihood_def}\\
										 &= & \left(\prod\limits_{t=1}^{n}h(x_t)\right) \mathcal{H}_l(\overline{\boldupsilon},{\bf{n_0}})\int\limits_{\overline{\boldeta} \in \Theta_l}\exp\left\{\sum\limits_{i=1}^K\boldeta_i^T({\bf{Y}}_i^n+\boldupsilon_i)-(N_i^n+n_{0i})\mathcal{A}_i(\boldeta_i)\right\}d\overline{\boldeta} \nonumber\\
										 & & \label{eqn:avg_likelihood_int}\\
										 &= & \left(\prod\limits_{t=1}^{n}h(x_t)\right)\frac{ \mathcal{H}_l(\overline{\boldupsilon},{\bf{n_0}})}{\mathcal{H}_l({\bf{Y}}+\overline{\boldupsilon},{\bf{N}}+{\bf{n_0}})},\label{eqn:avg_likelihood}
\end{eqnarray}
where ${\bf{Y}} = ({\bf{Y}}_1^n, \ldots, {\bf{Y}}_K^n)$ and ${\bf{N}}=(N_1^n,\ldots, N_K^n)$ (with the dependence of $\bf{Y}$ and $\bf{N}$ on $n$ understood and suppressed). The equality in (\ref{eqn:avg_likelihood_int}) is obtained by substituting (\ref{eqn:likelihood}) and (\ref{eqn:conj_prior}) in (\ref{eqn:avg_likelihood_def}). We then use (\ref{eqn:norm_factor}) in (\ref{eqn:avg_likelihood_int}) to get the final expression in (\ref{eqn:avg_likelihood}). Taking the $\log$, we get
\begin{eqnarray}
\log \tilde{f}_l(x^n|a^n) &= & \sum\limits_{t=1}^{n}\log h(x_t)+\log {\mathcal{H}_l(\overline{\boldupsilon},n_0)}-\log{\mathcal{H}_l({\bf{Y}}+\overline{\boldupsilon},{\bf{N}}+{\bf{n_0}})}.
\label{eqn:log_AL_H}
\end{eqnarray}

\subsection{Modified GLR statistic}
We define the modified GLR of hypothesis $l$ against hypothesis $m$ as
\begin{eqnarray}
\lefteqn{Z_{lm}(n) = \log \frac{\tilde{f}_l(x^n|a^n)}{\hat{f}(x^n|a^n,\overline{\boldeta} \in \Theta_m)}} \\
          &=& \log {\mathcal{H}_l(\overline{\boldupsilon},{\bf{n_0}})}-\log{\mathcal{H}_l({\bf{Y}}+\overline{\boldupsilon},{\bf{N}}+{\bf{n_0}})}-n\sum\limits_{i=1}^{K} w_i\left\{{\boldeta}_i^{*T}(m) \hat{\boldkappa}_i-\mathcal{A}_i({\boldeta}_i^{*}(m))\right\},\label{eqn:MGLR_statistic}
\end{eqnarray}
which is obtained using (\ref{eqn:ml_function}) and (\ref{eqn:log_AL_H}). The modification to the standard GLR is that the numerator contains an averaged likelihood function, averaged with respect to the artificial prior, rather than the maximum likelihood function. As we shall soon see, it is this modification that enables us to make the resulting policy admissible (i.e., a policy in $\Pi(\alpha)$).

Now let
\begin{equation}
Z_l(n) = \min\limits_{m \neq l} Z_{lm}(n)
\end{equation}
denote the modified GLR of hypothesis $l$ against its nearest alternative. The value of $Z_l(n)$ is a measure of the decision maker's confidence in the hypothesis $l$.

\subsection{Policy}
\label{sec:policy}
Let us denote our policy as  $\pi_{SMF}(L,\gamma,\beta)$ with SMF standing for Sluggish, Modified GLR based test with Forced exploration, where $L$ is a threshold parameter, $\gamma$ is a switching parameter, and $\beta$ is a forced exploration parameter, to be explained soon. The policy will involve some new variables: $n^a$ is the number of instants when the decision maker {\em actively} samples using (\ref{eqn:D-tracking}) below, and $N_i^{n,a}$ is the number of times the arm $i$ is {\em actively} sampled. These will be clear from the pseudo-code.

\vspace*{5mm}
\hrule
\vspace*{.25cm}
\noindent{\bf Policy $\pi_{SMF}(L,\gamma,\beta)$}:
\vspace*{.25cm}
\hrule
\vspace*{.25cm}

\noindent Fix $L \geq 1$, $0< \gamma \leq 1$, $1/2 < \beta < 1$.\\
Initialize: Sample the first arm $a_1 = 1$, set $n^a = 1$, $N_1^{n,a} = 1$, $N_i^{n,a} = 0 \, \,\forall i \ne 1$, $N_1^{n} = 1$, $N_i^{n} = 0\, \,\forall i \ne 1$.\\
For $n = 1, 2, \hdots$, do:
\begin{itemize}
\item $l^*(n) = \argmax_{l} Z_l(n)$. Resolve ties uniformly at random.
\item If $Z_{l^*(n)} < \log((M-1)L)$ then choose $a_{n+1}$ via the following, and make the associated updates:
\begin{itemize}
\item Generate $U_{n+1} \sim Bern(\gamma)$, independent of all other random variables.
\item If $U_{n+1}=0$, $a_{n+1} = a_n$.
\item If $U_{n+1}=1$, then update $n^a = n^a+1$ and choose $a_{n+1}$ according to
\begin{equation}\label{eqn:D-tracking}
a_{n+1} \in \begin{cases}
		   \argmin\limits_{i} N_i^{n,a} & \text{ if } \ni i : N_i^{n,a} < (n^a)^{\beta}-(\beta K)^{\beta/(1-\beta)}\\
		   \argmax\limits_{i} \{n^a\lambda_i^*(\overline{{\boldeta}}^*(l^*(n)) - N_i^{n,a}\} & \text{ otherwise}.
           \end{cases}
\end{equation}
Resolve ties uniformly at random.\\
Update $N_i^{n,a}$ as $N_i^{n,a} = N_i^{n,a}+1$, whenever $a_{n+1} = i$.
\item $N_i^{n} = N_i^{n}+1$, whenever $a_{n+1} = i$.
\end{itemize}

\item If $Z_{l^*(n)} \geq \log((M-1)L)$, then stop and declare $\delta = l^*(n)$ as the decision.
\end{itemize}
\vspace*{.5cm}
\hrule
\vspace*{.5cm}

We now explain the sampling rule in words. When $U_{n+1} = 0$, the sampling arm is not changed for the next instant, i.e., there is no switching. Our policy is {\em sluggish} because of this possibility of no switching. When $U_{n+1} = 1$,  we actively sample based on the sampling rule in (\ref{eqn:D-tracking}). This is  a variation on the \textit{D-tracking} sampling rule of \cite{garivier16} that includes a forced exploration component. In the above policy, as already described, $n^a$ is the number of instants when the decision maker {\em actively} samples using (\ref{eqn:D-tracking}) and $N_i^{n,a}$ is the number of times the arm $i$ is {\em actively} sampled. $N_i^n$ is the number of times arm $i$ is sampled, actively or otherwise, up to time $n$. The threshold to stop is $\log((M-1)L)$ which depends on the threshold parameter $L$. Observe that the threshold is fixed upfront and does not change over time. It is important to note that the switching parameter $\gamma$ cannot be chosen to be 0 (for ergodicity considerations).
%

\subsection{Main Result}
\label{subsec:mainresult}

We can now state and prove the main result.
\begin{theorem}
\label{thm:maintheorem}
Let Assumption A hold. Fix $l$. Consider $K$ arms with the configuration $\overline{\boldeta} \in \Theta_l$. Let $\left(\alpha_n\right)_{n\geq 1}$ be a sequence of tolerances such that $\lim\limits_{n \rightarrow \infty} \alpha_n = 0$. Then, for each $n$ and for each $\gamma > 0$, the policy $\pi_{SMF}\left(L_n,\gamma,\beta\right)$ with $L_n = 1/ \alpha_n$ belongs to $\Pi\left(\alpha_n\right)$. Furthermore,
\begin{eqnarray}
\lefteqn{\liminf\limits_{n \rightarrow \infty} \inf\limits_{\pi \in \Pi\left(\alpha_n\right)} \frac{E[C\left(\pi\right)|\overline{\boldeta}]}{\log\left(L_n\right)}}\\
			 &=& \lim\limits_{\gamma \downarrow 0} \lim\limits_{n \rightarrow \infty} \frac{E[C\left(\pi_{SMF}\left(L_n,\gamma,\beta\right)\right)|\overline{\boldeta}]}{\log\left(L_n\right)}= \frac{1}{D^*\left(\overline{\boldeta}\right)}.\nonumber
\end{eqnarray}
\end{theorem}

\begin{IEEEproof}
The main steps of the proof are to verify that the $n$-indexed sequence of policies $\pi_{SMF}\left(L_n,\gamma,\beta\right)$ satisfies the following.
\begin{enumerate}
  \item For each $n$, $\pi_{SMF}\left(L_n,\gamma,\beta\right)$ stops in finite time.
  \item For each $n$, $\pi_{SMF}\left(L_n,\gamma,\beta\right) \in \Pi(\alpha_n)$, i.e., it is admissible with error tolerance $\alpha_n$.
  \item As $n \rightarrow \infty$, the sequence of policies $\pi_{SMF}\left(L_n,\gamma,\beta\right)$ indexed by $n$ can be made arbitrarily close to being asymptotically optimal by a suitable choice of $\gamma$.
\end{enumerate}

We proceed to show these in the Propositions \ref{prop:finite_stopping_time}, \ref{prop:admissibility}, and \ref{prop:achievability} next.

Let us begin with the assertion that the proposed policy almost surely (a.s.) stops in finite time.

\begin{proposition}[Probability of stopping in finite time]
\label{prop:finite_stopping_time}
Fix the threshold parameter $L>1$ and switching parameter $0 < \gamma \leq 1$. Fix $l \in \{1, \cdots, M\}$. Let $\overline{\boldeta} \in \Theta_l$ be the true configuration. Then, the policy $\pi_{SMF}\left(L,\gamma,\beta\right)$ stops in finite time with probability 1, i.e., $$P\left(\tau\left(\pi_{SMF}\left(L,\gamma,\beta\right)\right) <\infty |\overline{\boldeta} \right) = 1.$$
\end{proposition}

\begin{IEEEproof}
To prove this, we show that under the true hypothesis, almost surely, the test statistic $Z_l\left(n\right)$ grows as $\Omega(n^{\beta})$ and, therefore, crosses the threshold $\log\left(\left(M-1\right)L\right)$ in finite time. See Appendix \ref{app:finite_stopping_time} for details.
\end{IEEEproof}

We next assert the admissibility of the proposed policy.

\begin{proposition}[Admissibility]
\label{prop:admissibility}
Fix $\alpha > 0$, $\gamma >0$, and let $L=1/\alpha$. We then have $\pi_{SMF}(L,\gamma,\beta) \in \Pi(\alpha)$.
\end{proposition}

\begin{IEEEproof}
The proof exploits the properties of the modified GLR and involves a change of measure argument. See Appendix \ref{app:admissibility}.
\end{IEEEproof}

We next assert that our policy is not only admissible, but is also asymptotically arbitrarily close to the lower bound.

\begin{proposition}[Achievability]
\label{prop:achievability}
Fix $\gamma > 0$. Consider the policy $\pi_{SMF}(L,\gamma,\beta)$. Let $\overline{\boldeta} \in \Theta_l$ be the true configuration. Under Assumption A, we have
\begin{eqnarray}
\label{eqn:upper_bound_tau}
\limsup\limits_{L \rightarrow \infty} \frac{\tau(\pi_{SMF}(L,\gamma,\beta))}{\log(L)}
& \leq & \frac{1}{D^*(\overline{\boldeta})} \quad \text{a.s.}, \\
\label{eqn:upper_bound_Etau}
\limsup\limits_{L \rightarrow \infty} \frac{E[\tau(\pi_{SMF}(L,\gamma,\beta))~|~ \overline{\boldeta}]}{\log(L)}
& \leq & \frac{1}{D^*(\overline{\boldeta})},
\end{eqnarray}
and furthermore,
\begin{equation}\label{eqn:upper_bound_cost}
\limsup\limits_{L \rightarrow \infty} \frac{E\left[C(\pi_{SMF}(L,\gamma,\beta)) ~|~ \overline{\boldeta} \right]}{\log(L)} \leq \frac{1}{D^*(\overline{\boldeta})} + \frac{g_{\max}\gamma}{D^*(\overline{\boldeta})}
\end{equation}
\end{proposition}

\begin{IEEEproof}
The main ideas are as follows.

The key to showing \eqref{eqn:upper_bound_tau} is that, as the target probability of false alarm goes to 0, the policy must wait longer and longer to decide. But this, along with the chosen sampling strategy and forced exploration, ensures that the estimated parameters approach the true parameters. By Assumption A, the continuously evolving sampling strategy approaches the desired sampling strategy that guards the correct hypothesis against its nearest alternative. Since relative entropy is continuous in its parameters, the logarithm of the GLR grows at the correct rate, almost surely, and reaches the threshold for a decision within the desired time duration.

The main idea behind \eqref{eqn:upper_bound_Etau} is to leverage the second moment condition for uniform integrability. This then helps us turn an almost-sure bound into a bound on the expectation. For the second moment condition itself, concentration inequalities play a crucial role.

The proof of \eqref{eqn:upper_bound_cost} leverages the fact that the total cost is upper bounded by $(1 + g_{\max} \gamma)$ times the delay cost. Since the $g_{\max}$ is finite and $\gamma>0$ is arbitrary, we obtain this inequality as well.

See Appendix \ref{app:achievability} for the proof of each of these.
\end{IEEEproof}

Propositions \ref{prop:finite_stopping_time}, \ref{prop:admissibility}, and \ref{prop:achievability} combined with Corollary \ref{cor:lowerbound} establish Theorem \ref{thm:maintheorem}.
\end{IEEEproof}

\section{Simulation Results}
\label{sec:simulations}

We end the main body of the paper with some simulation results.

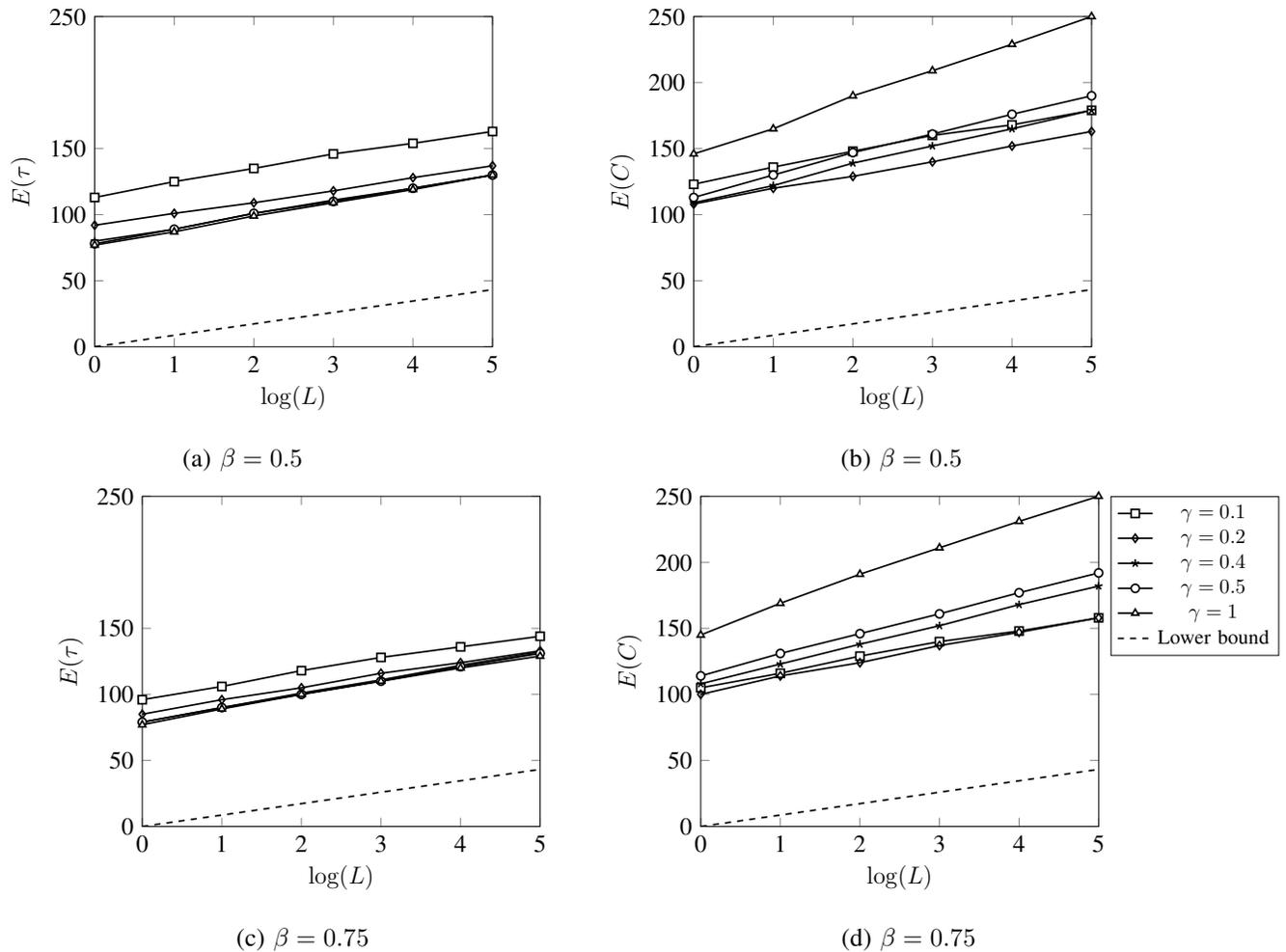
\begin{figure*}
\begin{subfigure}[h]{.4\textwidth}
\centering
\begin{tikzpicture}[scale=.8]
\begin{axis}
[xlabel = $\log(L)$, ylabel = $E(\tau)$, xmin =0, xmax=5,ymin=0, ymax=250,
xtick={0,1,2,3,4,5},
xticklabels={0,1,2,3,4,5},
ytick={0,50,100,150,250},
yticklabels={0,50,100,150,250},
]

\addplot[mark = square*,black,thick, mark options={fill=white}] coordinates{
(0,113)
(1,125)
(2,135)
(3,146)
(4,154)
(5,163)
};

\addplot[mark = diamond,black,thick, mark options={fill=white}] coordinates{
(0,92)
(1,101)
(2,109)
(3,118)
(4,128)
(5,137)
};

\addplot[mark = star,black,thick, mark options={fill=white}] coordinates{
(0,80)
(1,89)
(2,101)
(3,111)
(4,120)
(5,130)
};
%
\addplot[mark = *,black,thick, mark options={fill=white}] coordinates{
(0,78)
(1,89)
(2,101)
(3,110)
(4,120)
(5,130)
};
%
\addplot[mark = triangle*,black,thick, mark options={fill=white}] coordinates{
(0,77)
(1,87)
(2,99)
(3,109)
(4,119)
(5,130)
};

\addplot[mark = none,black,dashed,thick] coordinates{
(0,0)
(1,8.654)
(2,17.31)
(3,25.96)
(4,34.62)
(5,43.26)
};

\end{axis}
\end{tikzpicture}
\caption{$\beta = 0.5$}
\end{subfigure}
\begin{subfigure}{.5\textwidth}
\begin{tikzpicture}[scale=.8]
\begin{axis}
[xlabel = $\log(L)$, ylabel = $E(C)$, xmin =0, xmax=5,ymin=0, ymax=250,
xtick={0,1,2,3,4,5},
xticklabels={0,1,2,3,4,5},
ytick={0,50,100,150,200,250,300},
yticklabels={0,50,100,150,200,250},
]

\addplot[mark = square*,black,thick, mark options={fill=white}] coordinates{
(0,123)
(1,136)
(2,148)
(3,160)
(4,168)
(5,179)
};

\addplot[mark = diamond,black,thick, mark options={fill=white}] coordinates{
(0,108)
(1,120)
(2,129)
(3,140)
(4,152)
(5,163)
};

\addplot[mark = star,black,thick, mark options={fill=white}] coordinates{
(0,109)
(1,122)
(2,139)
(3,152)
(4,165)
(5,179)
};
%
\addplot[mark = *,black,thick, mark options={fill=white}] coordinates{
(0,113)
(1,130)
(2,147)
(3,161)
(4,176)
(5,190)
};
%
\addplot[mark = triangle*,black,thick, mark options={fill=white}] coordinates{
(0,146)
(1,165)
(2,190)
(3,209)
(4,229)
(5,250)
};

\addplot[mark = none,black,dashed,thick] coordinates{
(0,0)
(1,8.654)
(2,17.31)
(3,25.96)
(4,34.62)
(5,43.26)
};

\end{axis}
\end{tikzpicture}
\caption{$\beta = 0.5$}
\end{subfigure}
\begin{subfigure}{.5\textwidth}
\centering
\begin{tikzpicture}[scale=.8]
\begin{axis}
[xlabel = $\log(L)$, ylabel = $E(\tau)$, xmin =0, xmax=5,ymin=0, ymax=250,
xtick={0,1,2,3,4,5},
xticklabels={0,1,2,3,4,5},
ytick={0,50,100,150,250},
yticklabels={0,50,100,150,250},
]

\addplot[mark = square*,black,thick, mark options={fill=white}] coordinates{
(0,96)
(1,106)
(2,118)
(3,128)
(4,136)
(5,144)
};

\addplot[mark = diamond,black,thick, mark options={fill=white}] coordinates{
(0,85)
(1,96)
(2,105)
(3,116)
(4,124)
(5,133)
};

\addplot[mark = star,black,thick, mark options={fill=white}] coordinates{
(0,79)
(1,90)
(2,101)
(3,111)
(4,122)
(5,132)
};
%
\addplot[mark = *,black,thick, mark options={fill=white}] coordinates{
(0,79)
(1,90)
(2,100)
(3,110)
(4,121)
(5,131)
};
%
\addplot[mark = triangle*,black,thick, mark options={fill=white}] coordinates{
(0,77)
(1,89)
(2,100)
(3,110)
(4,120)
(5,129)
};

\addplot[mark = none,black,dashed,thick] coordinates{
(0,0)
(1,8.654)
(2,17.31)
(3,25.96)
(4,34.62)
(5,43.26)
};

\end{axis}
\end{tikzpicture}
\caption{$\beta = 0.75$}
\end{subfigure}
\begin{subfigure}{.5\textwidth}
\begin{tikzpicture}[scale=.8]
\begin{axis}
[xlabel = $\log(L)$, ylabel = $E(C)$, xmin =0, xmax=5,ymin=0, ymax=250,
xtick={0,1,2,3,4,5},
xticklabels={0,1,2,3,4,5},
ytick={0,50,100,150,200,250,300},
yticklabels={0,50,100,150,200,250},
legend style={legend columns=1,font=\small,legend pos=outer north east,fill=none}
]

\addplot[mark = square*,black,thick, mark options={fill=white}] coordinates{
(0,105)
(1,116)
(2,129)
(3,140)
(4,148)
(5,158)
};
\addlegendentry{$\gamma=0.1$}

\addplot[mark = diamond,black,thick, mark options={fill=white}] coordinates{
(0,100)
(1,114)
(2,124)
(3,137)
(4,147)
(5,158)
};
\addlegendentry{$\gamma=0.2$}

\addplot[mark = star,black,thick, mark options={fill=white}] coordinates{
(0,108)
(1,123)
(2,138)
(3,152)
(4,168)
(5,182)
};
\addlegendentry{$\gamma=0.4$}
\addplot[mark = *,black,thick, mark options={fill=white}] coordinates{
(0,114)
(1,131)
(2,146)
(3,161)
(4,177)
(5,192)
};
\addlegendentry{$\gamma=0.5$}
\addplot[mark = triangle*,black,thick, mark options={fill=white}] coordinates{
(0,145)
(1,169)
(2,191)
(3,211)
(4,231)
(5,250)
};
\addlegendentry{$\gamma=1$}

\addplot[mark = none,black,dashed,thick] coordinates{
(0,0)
(1,8.654)
(2,17.31)
(3,25.96)
(4,34.62)
(5,43.26)
};
\addlegendentry{Lower bound}

\end{axis}
\end{tikzpicture}
\caption{$\beta = 0.75$}
\end{subfigure}
\caption{Gaussian distribution with unknown means and known variances. The true parameters are $\mu_1 = 0$, $\sigma_1^2=1$, $\mu_2=1$, $\sigma_2^2 = 1$, $K=8$ , $g_{\max}=1$ and $D^*=0.1156$.}
\label{fig:gaussian_um}
\end{figure*}

Figures~\ref{fig:gaussian_um}-\ref{fig:gaussian_umv} provide our simulation results for the average delay and average cost as a function of $\log(L)$, where $\alpha = 1/L$, for three versions of the odd arm problem.
The following describe the setting for each plot, and describe the subplots within each plot.

\begin{itemize}
  \item All observations are taken to have the Gaussian distribution.

  \item Figure \ref{fig:gaussian_um} is for the odd arm problem with unknown means but known variance, among 8 arms. Figure \ref{fig:gaussian_uv} is for the odd arm identification problem, among 8 arms, with unknown variance but known mean.

  \item Figure \ref{fig:gaussian_umv} is for the odd arm identification problem with both mean and variance unknown. Again, there are a total of $K=8$ arms. This is truly a vector parameter for the exponential family.


  \item Subplots with subindices (a) and (c) refer to the expected delay.

  \item Subplots with subindices (b) and (d) refer to the expected cost.

  \item Subplots (a) and (b) are for the forced exploration parameter $\beta = 0.5$ while subplots (c) and (d) are for the forced exploration parameter $\beta = 0.75$. 

  \item Each plot contains the asymptotic lower bound (dashed curve). Each plot also contains five other curves (Figures~\ref{fig:gaussian_um}-\ref{fig:gaussian_umv}) 
      showing the average stopping times and average cost as the sluggishness parameter $\gamma$ varies.

  \item In Figures~\ref{fig:gaussian_um}-\ref{fig:gaussian_umv}, $\gamma$ is taken to be either 0.1, 0.2, 0.4, 0.5, or 1.0. 

  \item In Figures~\ref{fig:gaussian_um}-\ref{fig:gaussian_umv} each point involves an averaging over 5000 sample points. 
\end{itemize}

We make the following observations that corroborate the theory.
\begin{itemize}
\item We observe that, in all the cases, the slopes of the empirical average delays match the slope of the lower bound, thereby validating the asymptotic optimality of the policy.
\item The slopes of the empirical average values of cost roughly match the slopes of the lower bounds, as expected, for small $\gamma$.

\item For smaller values of $\gamma$, the average delay in arriving at a decision increases (due to limited exploration).

\item As $\gamma$ decreases, the total cost first decreases  due to reduced switching, but then increases due to increased decision delay because of sluggishness and limited exploration. A value of $\gamma$ of around $0.2$ seems to be the best choice in Figures~\ref{fig:gaussian_um}-\ref{fig:gaussian_umv}, for the chosen settings. 
\end{itemize}

\begin{figure*}
\begin{subfigure}[h]{.4\textwidth}
\centering
\begin{tikzpicture}[scale=.8]
\begin{axis}
[xlabel = $\log(L)$, ylabel = $E(\tau)$, xmin =0, xmax=5,ymin=0, ymax=250,
xtick={0,1,2,3,4,5},
xticklabels={0,1,2,3,4,5},
ytick={0,50,100,150,200,250},
yticklabels={0,50,100,150,200,250},
enlargelimits=false
]

\addplot[mark = square*,black,thick, mark options={fill=white}] coordinates{
(0,87)
(1,98)
(2,108)
(3,116)
(4,125)
(5,133)
};

\addplot[mark = diamond,black,thick, mark options={fill=white}] coordinates{
(0,71)
(1,83)
(2,92)
(3,97)
(4,106)
(5,115)
};

\addplot[mark = star,black,thick, mark options={fill=white}] coordinates{
(0,66)
(1,74)
(2,86)
(3,92)
(4,101)
(5,110)
};

\addplot[mark = *,black,thick, mark options={fill=white}] coordinates{
(0,65)
(1,75)
(2,83)
(3,92)
(4,102)
(5,111)
};

\addplot[mark = triangle*,black,thick, mark options={fill=white}] coordinates{
(0,67)
(1,77)
(2,86)
(3,93)
(4,102)
(5,111)
};

\addplot[mark = none,black,dashed,thick] coordinates{
(0,0)
(1,7.18)
(2,14.36)
(3,21.5)
(4,28.7)
(5,35.9)
};

\end{axis}
\end{tikzpicture}
\caption{$\beta = 0.5$}
\end{subfigure}
\begin{subfigure}{.5\textwidth}
\begin{tikzpicture}[scale=.8]
\begin{axis}
[xlabel = $\log(L)$, ylabel = $E(C)$, xmin =0, xmax=5,ymin=0, ymax=250,
xtick={0,1,2,3,4,5},
xticklabels={0,1,2,3,4,5},
ytick={0,50,100,150,200,250},
yticklabels={0,50,100,150,200,250},
legend pos=outer north east,
legend style={legend columns=2,font=\small,fill=none}
]

\addplot[mark = square*,black,thick, mark options={fill=white}] coordinates{
(0,94)
(1,107)
(2,118)
(3,127)
(4,137)
(5,146)
};

\addplot[mark = diamond,black,thick, mark options={fill=white}] coordinates{
(0,84)
(1,98)
(2,108)
(3,115)
(4,125)
(5,137)
};

\addplot[mark = star,black,thick, mark options={fill=white}] coordinates{
(0,89)
(1,100)
(2,117)
(3,126)
(4,138)
(5,150)
};

\addplot[mark = *,black,thick, mark options={fill=white}] coordinates{
(0,94)
(1,109)
(2,121)
(3,134)
(4,149)
(5,162)
};

\addplot[mark = triangle*,black,thick, mark options={fill=white}] coordinates{
(0,126)
(1,146)
(2,164)
(3,177)
(4,195)
(5,213)
};

\addplot[mark = none,black,dashed,thick] coordinates{
(0,0)
(1,7.18)
(2,14.36)
(3,21.5)
(4,28.7)
(5,35.9)
};

\end{axis}
\end{tikzpicture}
\caption{$\beta=0.5$}
\end{subfigure}
\begin{subfigure}{.5\textwidth}
\begin{tikzpicture}[scale=.8]
\begin{axis}
[xlabel = $\log(L)$, ylabel = $E(C)$, xmin =0, xmax=5,ymin=0, ymax=250,
xtick={0,1,2,3,4,5},
xticklabels={0,1,2,3,4,5},
ytick={0,50,100,150,200,250},
yticklabels={0,50,100,150,200,250},
legend pos=outer north east,
legend style={legend columns=2,font=\small,fill=none}
]

\addplot[mark = square*,black,thick, mark options={fill=white}] coordinates{
(0,88)
(1,94)
(2,105)
(3,113)
(4,123)
(5,129)
};

\addplot[mark = diamond,black,thick, mark options={fill=white}] coordinates{
(0,75)
(1,83)
(2,93)
(3,101)
(4,109)
(5,114)
};

\addplot[mark = star,black,thick, mark options={fill=white}] coordinates{
(0,68)
(1,77)
(2,87)
(3,97)
(4,106)
(5,112)
};

\addplot[mark = *,black,thick, mark options={fill=white}] coordinates{
(0,69)
(1,78)
(2,87)
(3,95)
(4,104)
(5,112)
};

\addplot[mark = triangle*,black,thick, mark options={fill=white}] coordinates{
(0,67)
(1,76)
(2,87)
(3,94)
(4,104)
(5,112)
};

\addplot[mark = none,black,dashed,thick] coordinates{
(0,0)
(1,7.18)
(2,14.36)
(3,21.5)
(4,28.7)
(5,35.9)
};

\end{axis}
\end{tikzpicture}
\caption{$\beta=0.75$}
\end{subfigure}
\begin{subfigure}{.5\textwidth}
\begin{tikzpicture}[scale=.8]
\begin{axis}
[xlabel = $\log(L)$, ylabel = $E(C)$, xmin =0, xmax=5,ymin=0, ymax=250,
xtick={0,1,2,3,4,5},
xticklabels={0,1,2,3,4,5},
ytick={0,50,100,150,200,250},
yticklabels={0,50,100,150,200,250},
legend pos=outer north east,
legend style={legend columns=1,font=\small,fill=none}
]

\addplot[mark = square*,black,thick, mark options={fill=white}] coordinates{
(0,96)
(1,102)
(2,115)
(3,124)
(4,134)
(5,142)
};
\addlegendentry{$\gamma=0.1$}

\addplot[mark = diamond,black,thick, mark options={fill=white}] coordinates{
(0,89)
(1,98)
(2,110)
(3,119)
(4,129)
(5,136)
};
\addlegendentry{$\gamma=0.2$}

\addplot[mark = star,black,thick, mark options={fill=white}] coordinates{
(0,92)
(1,104)
(2,119)
(3,132)
(4,145)
(5,153)
};
\addlegendentry{$\gamma=0.4$}

\addplot[mark = *,black,thick, mark options={fill=white}] coordinates{
(0,99)
(1,113)
(2,127)
(3,139)
(4,152)
(5,164)
};
\addlegendentry{$\gamma=0.5$}

\addplot[mark = triangle*,black,thick, mark options={fill=white}] coordinates{
(0,124)
(1,142)
(2,165)
(3,180)
(4,200)
(5,216)
};
\addlegendentry{$\gamma=1$}

\addplot[mark = none,black,dashed,thick] coordinates{
(0,0)
(1,7.18)
(2,14.36)
(3,21.5)
(4,28.7)
(5,35.9)
};
\addlegendentry{Lower bound}

\end{axis}
\end{tikzpicture}
\caption{$\beta=0.75$}
\end{subfigure}
\caption{Gaussian distribution with known means and unknown variances. The true parameters are $\mu_1 = 0$, $\sigma_1^2=5$, $\mu_2=0$, $\sigma_2^2 = 1$, $K=8$ , $g_{\max}=1$ and $D^*=0.1392$.}
\label{fig:gaussian_uv}
\end{figure*}
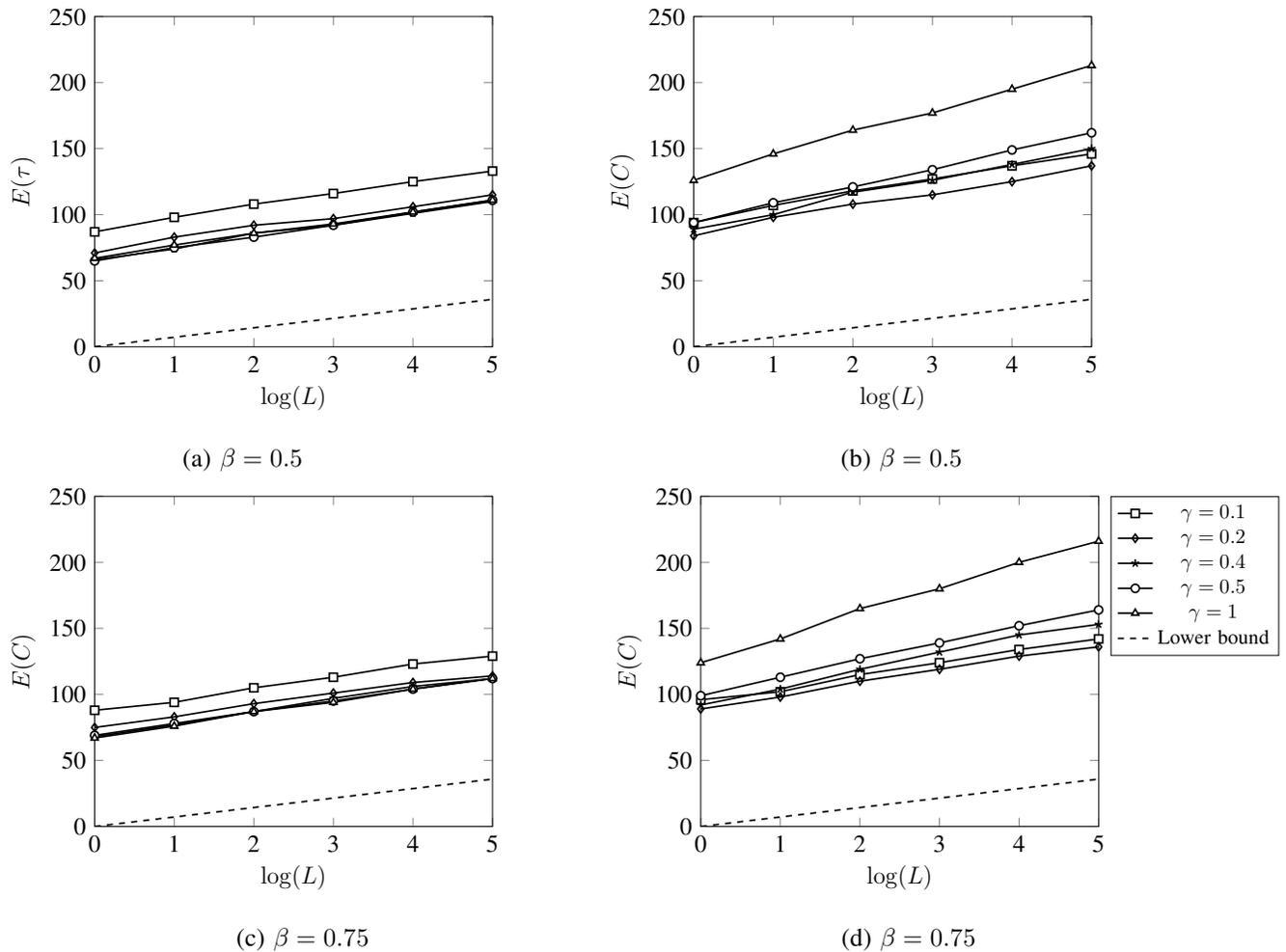
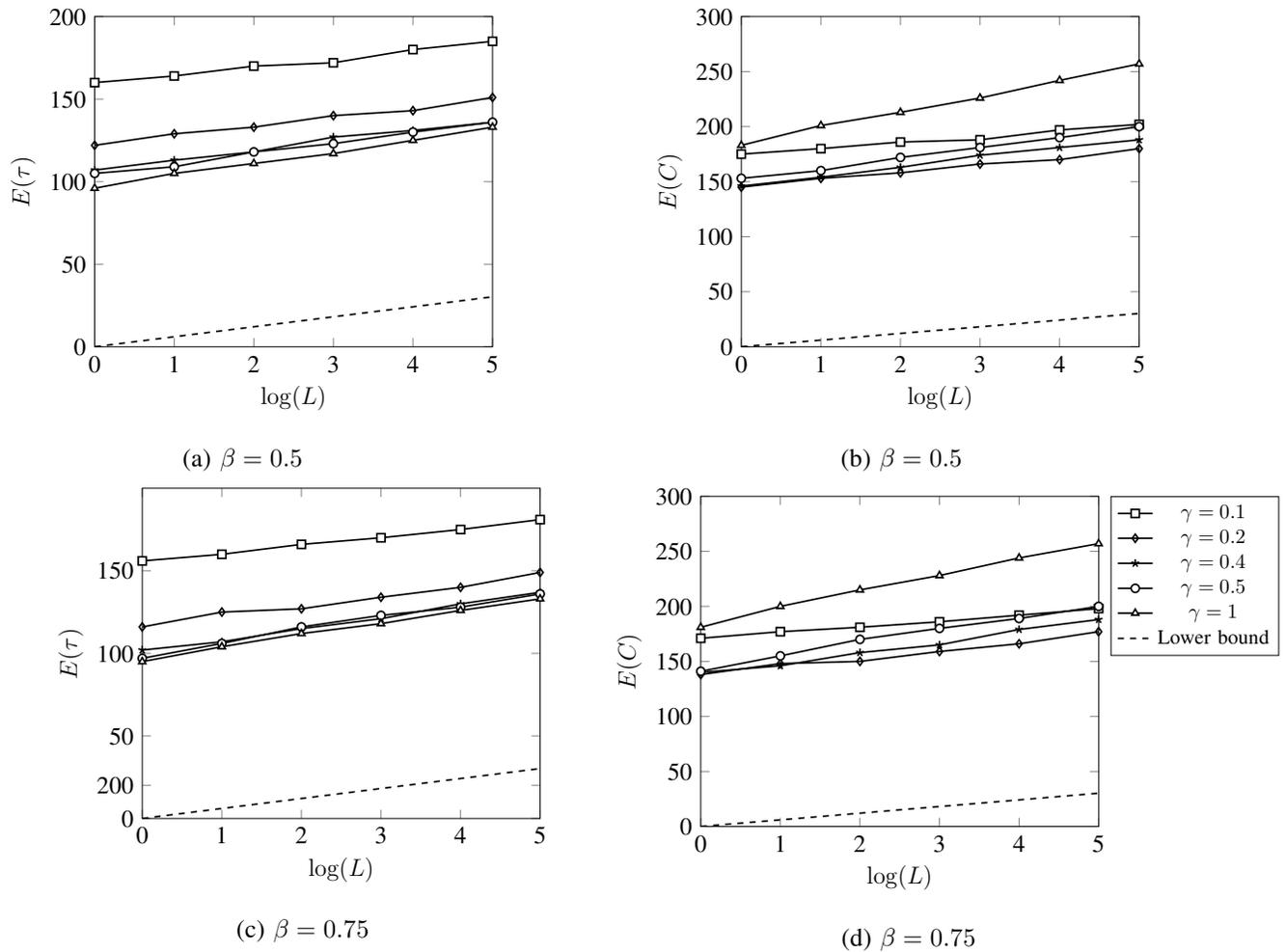
\begin{figure*}
\begin{subfigure}[h]{.4\textwidth}
\centering
\begin{tikzpicture}[scale=.8]
\begin{axis}
[xlabel = $\log(L)$, ylabel = $E(\tau)$, xmin =0, xmax=5,ymin=0, ymax=200,
xtick={0,1,2,3,4,5},
xticklabels={0,1,2,3,4,5},
ytick={0,50,100,150,200},
yticklabels={0,50,100,150,200},
]

\addplot[mark = square*,black,thick, mark options={fill=white}] coordinates{
(0,160)
(1,164)
(2,170)
(3,172)
(4,180)
(5,185)
};

\addplot[mark = diamond,black,thick, mark options={fill=white}] coordinates{
(0,122)
(1,129)
(2,133)
(3,140)
(4,143)
(5,151)
};

\addplot[mark = star,black,thick, mark options={fill=white}] coordinates{
(0,107)
(1,113)
(2,118)
(3,127)
(4,131)
(5,136)
};

\addplot[mark = *,black,thick, mark options={fill=white}] coordinates{
(0,105)
(1,109)
(2,118)
(3,123)
(4,130)
(5,136)
};

\addplot[mark = triangle*,black,thick, mark options={fill=white}] coordinates{
(0,96)
(1,105)
(2,111)
(3,117)
(4,125)
(5,133)
};

\addplot[mark = none,black,dashed,thick] coordinates{
(0,0)
(1,6.048)
(2,12.09)
(3,18.145)
(4,24.19)
(5,30.24)
};

\end{axis}
\end{tikzpicture}
\caption{$\beta = 0.5$}
\end{subfigure}
\begin{subfigure}{.5\textwidth}
\centering
\begin{tikzpicture}[scale=.8]
\begin{axis}
[xlabel = $\log(L)$, ylabel = $E(C)$, xmin =0, xmax=5,ymin=0, ymax=300,
xtick={0,1,2,3,4,5},
xticklabels={0,1,2,3,4,5},
ytick={0,50,100,150,200,250,300},
yticklabels={0,50,100,150,200,250,300}
]

\addplot[mark = square*,black,thick, mark options={fill=white}] coordinates{
(0,175)
(1,180)
(2,186)
(3,188)
(4,197)
(5,202)
};

\addplot[mark = diamond,black,thick, mark options={fill=white}] coordinates{
(0,145)
(1,153)
(2,158)
(3,166)
(4,170)
(5,180)
};

\addplot[mark = star,black,thick, mark options={fill=white}] coordinates{
(0,146)
(1,154)
(2,163)
(3,174)
(4,181)
(5,188)
};

\addplot[mark = *,black,thick, mark options={fill=white}] coordinates{
(0,153)
(1,160)
(2,172)
(3,181)
(4,190)
(5,200)
};

\addplot[mark = triangle*,black,thick, mark options={fill=white}] coordinates{
(0,183)
(1,201)
(2,213)
(3,226)
(4,242)
(5,257)
};

\addplot[mark = none,black,dashed,thick] coordinates{
(0,0)
(1,6.048)
(2,12.09)
(3,18.145)
(4,24.19)
(5,30.24)
};

\end{axis}
\end{tikzpicture}
\caption{$\beta = 0.5$}
\end{subfigure}
\begin{subfigure}{.5\textwidth}
\centering
\begin{tikzpicture}[scale=.8]
\begin{axis}
[xlabel = $\log(L)$, ylabel = $E(\tau)$, xmin =0, xmax=5,ymin=0, ymax=200,
xtick={0,1,2,3,4,5},
xticklabels={0,1,2,3,4,5},
ytick={0,50,100,150,20},
yticklabels={0,50,100,150,200},
]

\addplot[mark = square*,black,thick, mark options={fill=white}] coordinates{
(0,156)
(1,160)
(2,166)
(3,170)
(4,175)
(5,181)
};

\addplot[mark = diamond,black,thick, mark options={fill=white}] coordinates{
(0,116)
(1,125)
(2,127)
(3,134)
(4,140)
(5,149)
};

\addplot[mark = star,black,thick, mark options={fill=white}] coordinates{
(0,102)
(1,107)
(2,115)
(3,121)
(4,130)
(5,137)
};

\addplot[mark = *,black,thick, mark options={fill=white}] coordinates{
(0,97)
(1,106)
(2,116)
(3,123)
(4,128)
(5,136)
};

\addplot[mark = triangle*,black,thick, mark options={fill=white}] coordinates{
(0,95)
(1,104)
(2,112)
(3,118)
(4,126)
(5,133)
};

\addplot[mark = none,black,dashed,thick] coordinates{
(0,0)
(1,6.048)
(2,12.09)
(3,18.145)
(4,24.19)
(5,30.24)
};

\end{axis}
\end{tikzpicture}
\caption{$\beta = 0.75$}
\end{subfigure}
\begin{subfigure}{.5\textwidth}
\centering
\begin{tikzpicture}[scale=.8]
\begin{axis}
[xlabel = $\log(L)$, ylabel = $E(C)$, xmin =0, xmax=5,ymin=0, ymax=300,
xtick={0,1,2,3,4,5},
xticklabels={0,1,2,3,4,5},
ytick={0,50,100,150,200,250,300},
yticklabels={0,50,100,150,200,250,300},
legend style={legend pos = outer north east,legend columns=1,font=\small,fill=none}
]

\addplot[mark = square*,black,thick, mark options={fill=white}] coordinates{
(0,171)
(1,177)
(2,181)
(3,186)
(4,192)
(5,198)
};
\addlegendentry{$\gamma=0.1$}

\addplot[mark = diamond,black,thick, mark options={fill=white}] coordinates{
(0,138)
(1,148)
(2,150)
(3,159)
(4,166)
(5,177)
};
\addlegendentry{$\gamma=0.2$}

\addplot[mark = star,black,thick, mark options={fill=white}] coordinates{
(0,140)
(1,146)
(2,158)
(3,165)
(4,179)
(5,188)
};
\addlegendentry{$\gamma=0.4$}

\addplot[mark = *,black,thick, mark options={fill=white}] coordinates{
(0,141)
(1,155)
(2,170)
(3,180)
(4,189)
(5,200)
};
\addlegendentry{$\gamma=0.5$}

\addplot[mark = triangle*,black,thick, mark options={fill=white}] coordinates{
(0,181)
(1,200)
(2,215)
(3,228)
(4,244)
(5,257)
};
\addlegendentry{$\gamma=1$}

\addplot[mark = none,black,dashed,thick] coordinates{
(0,0)
(1,6.048)
(2,12.09)
(3,18.145)
(4,24.19)
(5,30.24)
};
\addlegendentry{Lower bound}

\end{axis}
\end{tikzpicture}
\caption{$\beta = 0.75$}
\end{subfigure}
\caption{Gaussian distribution with unknown means and unknown variances. The true parameters are $\mu_1 = 0$, $\sigma_1^2=2$, $\mu_2=1$, $\sigma_2^2 = 10$, $K=8$ , $g_{\max}=1$ and $D^*=0.1653$.}
\label{fig:gaussian_umv}
\end{figure*}

\clearpage
\newpage
\appendices

\section{A regularity lemma}
\label{app:regularity lemma}

In this appendix, we prove the following regularity lemma.

\begin{lemma}
\label{lem:regularity}
For any $i$, for any compact set $C$,
\[
  \inf_{\boldeta_i \in C}\| \boldeta_i' - \boldeta_i\| \rightarrow \infty ~ \Rightarrow ~ \inf_{\boldeta_i \in C} D(\boldeta_i \parallel \boldeta_i') \rightarrow \infty.
\]
\end{lemma}

In other words, the relative entropy of (the distribution associated with) a confined parameter $\boldeta_i$ with respect to (the distribution associated with) a parameter $\boldeta'_i$ approaches infinity as the parameter separation between $\boldeta'_i$ and $\boldeta_i$ grows without bound.

\begin{IEEEproof}
The lemma holds vacuously when the parameter set is bounded. Let $\mathbf{d} := \boldeta'_i - \boldeta_i$ and assume that $ ||\mathbf{d}|| > 1$. Then, from the formula \eqref{eqn:TaylorSeries} for relative entropy, we get
\begin{eqnarray}
  D(\boldeta_i \parallel \boldeta'_i)
  & = & \frac{1}{2}(\boldeta'_i - \boldeta_i)^T \left[ \int_0^1 (1-t) \textsf{Hess}(\mathcal{A}_i)(\boldeta_i + t( \boldeta'_i - \boldeta_i))~dt \right] (\boldeta'_i - \boldeta_i) \\
  & \geq & \frac{1}{2}\mathbf{d}^T \left[ \int_0^{1/||\mathbf{d}||} (1-t) \textsf{Hess}(\mathcal{A}_i)(\boldeta_i + t( \boldeta'_i - \boldeta_i))~dt \right] \mathbf{d} \label{eqn:d2hessian}\\
  & \geq & \frac{\| \mathbf{d} \|^2}{2} \left(1 - \frac{1}{\| \mathbf{d} \|} \right) \int_0^{1/||\mathbf{d}||} \lambda_{\min}\left( \textsf{Hess}(\mathcal{A}_i)(\boldeta_i + t( \boldeta'_i - \boldeta_i)) \right)~dt \label{eqn:hessianbound}\\
  & \geq & \frac{1}{2}\left(\| \mathbf{d} \| - 1 \right) \times \min \{ \lambda_{\min}\left( \textsf{Hess}(\mathcal{A}_i)(\boldeta_i + s {\mathbf{d}}/\| \mathbf{d} \|) \right): s \in [0,1] \} \\
  & \rightarrow & \infty \mbox{ as } \| \mathbf{d} \| \rightarrow \infty. \label{eqn:hessianbound2}
\end{eqnarray}
In the above sequence of inequalities, \eqref{eqn:d2hessian} follows because of the positive definiteness of the Hessian of $\mathcal{A}_i$. Next, \eqref{eqn:hessianbound} follows from $1 - t \geq 1 - 1/\|\mathbf{d}\|$ in the interval under consideration and the fact that $\mathbf{d}^T H \mathbf{d} \geq \lambda_{\min}(H) \|\mathbf{d}\|^2$ where $\lambda_{\min}(H)$ is the smallest eigenvalue of any positive definite matrix $H$. Finally, \eqref{eqn:hessianbound2} follows because $\lambda_{\min}(\textsf{Hess}(\mathcal{A}_i)(\cdot)$ is strictly positive in the unit neighbourhood around $\boldeta_i$; indeed, $\lambda_{\min}(\textsf{Hess}(\mathcal{A}_i)(\cdot))$ is a continuous function of $\tilde{\boldeta}_i$ and therefore cannot attain the value zero on account of the strict convexity of $\mathcal{A}(\cdot)$ leading to $\lambda_{\min}(\mathcal{A}(\cdot))$ being strictly positive in the unit neighbourhood around $\boldeta_i$.
\end{IEEEproof}

\section{Proof of Proposition \ref{prop:lambda_continuity}}\label{app:lambda_continuity}

Define
\[
  F(\boldlambda , \overline{\boldeta}) := \inf\limits_{\overline{\boldeta}'\in \Theta_{-l}} \sum\limits_{i=1}^K \lambda_i D(\boldeta_i \parallel \boldeta_i').
\]
It suffices show that $(\boldlambda, \overline{\boldeta}) \mapsto F(\boldlambda,\overline{\boldeta})$ is continuous everywhere on its domain and that $\boldlambda \mapsto F(\boldlambda , \overline{\boldeta})$ is concave for each $\overline{\boldeta}$. Then by Berge's maximum theorem \cite[Theorems 9.14 and 9.17(2)]{sundaram1996first} the set $\boldlambda^*(\overline{\boldeta})$ where the maximum is attained is nonempty, and the set-valued map $\overline{\boldeta} \mapsto \boldlambda^*(\overline{\boldeta})$ is upper semi-continuous, compact-valued and convex-valued.

Fix $\overline{\boldeta} \in \Theta_l$. Since $\boldlambda \mapsto F(\boldlambda , \overline{\boldeta})$ is an infimum of linear functions parameterised by $\overline{\boldeta}'$, concavity immediately follows.

We now proceed to show that $(\boldlambda, \overline{\boldeta}) \mapsto F(\boldlambda,\overline{\boldeta})$ is continuous. On account of concavity, the issue of continuity arises only at a boundary. Nevertheless, we give a general proof.

\begin{description}
\item[(i)] Fix an arbitrary $\varepsilon > 0$. Fix $\overline{\boldeta}' \in \Theta_{-l}$ so that
\begin{equation}
  \label{eqn:F-approx}
  \sum_i \lambda_i D(\boldeta_i \parallel \boldeta'_i) \leq F(\boldlambda, \overline{\boldeta}) + \varepsilon.
\end{equation}
Observe that, for each $i$,
\[
  \overline{\boldeta} \mapsto D(\boldeta_i \parallel \boldeta'_i) = (\boldeta_i - \boldeta'_i)^T \boldkappa_i(\boldeta_i) - \mathcal{A}_i(\boldeta_i) + \mathcal{A}_i(\boldeta'_i)
\]
is a continuous function of $\overline{\boldeta} \in \Theta_l$ by virtue of the continuity of $\boldkappa_i(\cdot)$ and the continuously differentiable property of $\mathcal{A}_i(\cdot)$. (Indeed, the continuity of $\boldkappa_i(\cdot)$ itself follows from the continuous differentiability of $\mathcal{A}_i(\cdot)$.) Now consider a sequence $(\boldlambda(n), \overline{\boldeta}(n)) \rightarrow (\boldlambda, \overline{\boldeta}) \in \mathcal{P}(K) \times \Theta_l$ as $n \rightarrow \infty$.
We then have, for all sufficiently large $n$,
\begin{equation}
  \label{eqn:lambda-D-approx-up}
  \boldlambda(n) \leq \boldlambda + \varepsilon {\bf 1} \mbox{ and } D(\boldeta_i(n) \parallel \boldeta'_i) \leq D(\boldeta_i \parallel \boldeta'_i)  + \varepsilon ~ \forall i,
\end{equation}
where the first inequality is to be taken component-wise with ${\bf 1}$ being the all-1 vector. Hence, for all sufficiently large $n$, we have
\begin{eqnarray*}
  F(\boldlambda(n), \overline{\boldeta}(n))
  & \leq &
  \sum_i \lambda_i(n) D(\boldeta_i(n) \parallel \boldeta'_i) \quad (\mbox{left-side is an infimum, for each fixed $n$})
  \\
  & \leq & \sum_i (\lambda_i + \varepsilon)(D(\boldeta_i \parallel \boldeta'_i) + \varepsilon)\\
  & = & \sum_i \lambda_i D(\boldeta_i \parallel \boldeta'_i) + \varepsilon \left(1 + \sum_i D(\boldeta_i \parallel \boldeta'_i) \right) + \varepsilon^2 K \\
  & \leq & F(\boldlambda, \overline{\boldeta}) + \varepsilon \left(2 + \sum_i D(\boldeta_i \parallel \boldeta'_i) + \varepsilon K \right),
\end{eqnarray*}
where the last inequality follows from (\ref{eqn:F-approx}). Since $\sum_i D(\boldeta_i \parallel \boldeta'_i)$ is finite for any $\overline{\boldeta} \in \Theta_l$ and $\overline{\boldeta}' \in \Theta_{-l}$, and since $\varepsilon$ was arbitrary, we get
\begin{equation}
\label{eqn:limsupF}
\limsup_{n \rightarrow \infty} F(\boldlambda(n), \overline{\boldeta}(n)) \leq F(\boldlambda, \overline{\boldeta}).
\end{equation}

\item[(ii)] Once again fix $\varepsilon > 0$, and consider a sequence $(\boldlambda(n), \overline{\boldeta}(n)) \rightarrow (\boldlambda, \overline{\boldeta}) \in \mathcal{P}(K) \times \Theta_l$ as $n \rightarrow \infty$, but this time choose a convergent sequence\footnote{\label{footnote1}Observe that if, for some $i$, $\boldeta_i'(n) \rightarrow \infty$, since $\boldeta_i(n)$ is confined to a compact neighbourhood of $\boldeta_i$, by Lemma \ref{lem:regularity}, we must have $D(\boldeta_i(n) \parallel \boldeta_i'(n)) \rightarrow \infty$. On account of (\ref{eqn:F-lower}), we can replace the offending $\boldeta_i'(n)$ with another bounded quantity yielding a bounded $D(\boldeta_i(n) \parallel \boldeta_i'(n))$, without affecting inequality (\ref{eqn:F-lower}). Hence, without loss of generality, we may take that $\overline{\boldeta}'(n)$ is bounded. Furthermore, by passing to a subsequence if necessary, we may further take $\overline{\boldeta}'(n) \rightarrow \overline{\boldeta}'$ for some limit $\overline{\boldeta}'$.}
    $\overline{\boldeta}'(n) \in \Theta_{-l}$, such that for every $n$, we have:
\begin{equation}
  \label{eqn:F-lower}
  F(\boldlambda(n), \overline{\boldeta}(n))
  \geq \sum_i \lambda_i(n) D(\boldeta_i(n) \parallel \boldeta'_i(n)) - \varepsilon,
\end{equation}
and $\overline{\boldeta}'(n) \rightarrow \overline{\boldeta}'$. Analogous to (\ref{eqn:lambda-D-approx-up}), using the boundedness of the sequence $\boldeta'(n)$, for all sufficiently large $n$, we have
\begin{equation}
  \label{eqn:lambda-D-approx-down}
  \boldlambda(n) \geq \boldlambda -\varepsilon {\bf 1} \mbox{ and } D(\boldeta_i(n) \parallel \boldeta'_i(n)) \geq D(\boldeta_i \parallel \boldeta'_i(n)) - \varepsilon ~ \forall i.
\end{equation}
Using (\ref{eqn:lambda-D-approx-down}) in (\ref{eqn:F-lower}), we get
\begin{eqnarray*}
  F(\boldlambda(n), \overline{\boldeta}(n))
  & \geq & \sum_i \lambda_i D(\boldeta_i \parallel \boldeta'_i(n)) - \varepsilon \left( 2 + \sum_i D(\boldeta_i \parallel \boldeta'_i(n)) - \varepsilon K \right)\\
  & \geq & F(\boldlambda, \overline{\boldeta}) - \varepsilon \left( 2 + \sum_i D(\boldeta_i \parallel \boldeta'_i(n)) - \varepsilon K \right),
\end{eqnarray*}
where the last inequality follows from the observation that $\overline{\boldeta}'(n) \in \Theta_{-l}$ for all $n$ and by then taking the infimum over all $\overline{\boldeta}' \in \Theta_{-l}$. Since $\overline{\boldeta}'(n) \rightarrow \overline{\boldeta}'$, the quantity $\sum_i D(\boldeta_i \parallel \boldeta'_i(n))$ converges to $\sum_i D(\boldeta_i \parallel \boldeta'_i)$ and is therefore bounded. Since $\varepsilon$ was arbitrary, we obtain
\begin{equation}
\label{eqn:liminfF}
\liminf_{n \rightarrow \infty} F(\boldlambda(n), \overline{\boldeta}(n)) \geq F(\boldlambda, \overline{\boldeta}).
\end{equation}
%
%
%
%
%
%
%
%
%
%
\end{description}
From (\ref{eqn:limsupF}) and (\ref{eqn:liminfF}), we have the continuity of $F(\boldlambda, \overline{\boldeta})$.

\section{Finite stopping time}\label{app:finite_stopping_time}
We show in a series of steps that the proposed policy stops in finite time. We begin with the proof of ML estimates of the parameters converging to the true parameter values and then show that the statistic grows as $\Omega(n^{\beta})$ as time $n \rightarrow \infty$. We then show that, with this growth, the statistic crosses any fixed threshold in finite time and, hence, the stopping time is finite almost surely.

Before we begin with the proof, as done in \cite{vaidhiyan15}, we also consider two variants of $\pi_{SMF}\left(L,\gamma,\beta\right)$ which are useful in the analysis.

\begin{enumerate}
\item \textit{Policy} $\pi_{SMF}^l\left(L,\gamma,\beta\right)$ is like policy $\pi_{SMF}\left(L,\gamma,\beta\right)$ but stops only at decision $l$, when $Z_l\left(n\right)\geq \log\left(\left(M-1\right)L\right)$.
\item \textit{Policy} $\tilde{\pi}_{SMF}\left(\gamma,\beta\right)$ is also like $\pi_{SMF}\left(L,\gamma,\beta\right)$ but never stops. So the policy does not depend on the policy parameter $L$.
\end{enumerate}

The policy $\tilde{\pi}_{SMF}\left(\gamma,\beta\right)$ will be used in Proposition \ref{prop:ml_convergences} and the policy $\pi_{SMF}^l\left(L,\gamma,\beta\right)$ will be used in the proof of Proposition \ref{prop:finite_stopping_time}.

\subsection{Convergence results}

Let us begin with a preliminary result about the forced exploration rule.

\begin{lemma}\label{lemma:lambda_conv}
Let $\overline{\boldeta} \in \Theta_l$ be the true configuration. The proposed sampling rule ensures that
$$N_i^{n,a} \geq [(n^a)^{\beta}-(\beta (K+1))^{\beta/(1-\beta)}]_+-1.$$
Furthermore, for all $\epsilon>0$ and for all $n_0$, setting $n_\epsilon = \max\{n_0, \epsilon^{-1}\}/(3\epsilon)$, we have the implication:
\begin{equation}
\sup\limits_{n^a\geq n_0}\max\limits_i |\boldlambda^*({\overline{\boldeta}^*(l)})-\boldlambda^*(\overline{\boldeta})| \leq \epsilon \Rightarrow \sup\limits_{n^a \geq n_\epsilon} \max\limits_i \left|\frac{N_i^{n,a}}{n^a}-\lambda_i^*(\overline{\boldeta})\right| \leq 3(K-1)\epsilon.
\end{equation}
\end{lemma}
\begin{IEEEproof}
The proof follows from \cite[Lemmas~17]{garivier16}. Specifically, from \cite[Appendix~B.2]{garivier16}, we need to check that, if $g(n) = [n^{\beta} - (\beta (K+1))^{\beta/(1-\beta)}]_+$, then $g(0) = 0$, $g(n)/n \rightarrow 0$ as $n \rightarrow \infty$, and for every $m \geq 1$
\[
  \inf \{ k \in \mathbb{N}: g(k) \geq m \} > \inf \{ k \in \mathbb{N}: g(k) \geq m-1 \} + K.
\]
The first two conditions are straightforward. To check the third condition, observe that
$\inf \{ k \in \mathbb{N}: g(k) \geq m \} = \lceil [m + (\beta (K+1))^{\beta/(1-\beta)}]^{1/\beta} \rceil$. Thus, with $u = m - 1 + (\beta (K+1))^{\beta/(1-\beta)}$, we have
\begin{eqnarray*}
  \inf \{ k \in \mathbb{N}: g(k) \geq m \} - \inf \{ k \in \mathbb{N}: g(k) \geq m-1 \}
  & \geq & (u+1)^{1/\beta} - \lceil u^{1/\beta} \rceil \\
  & > & \frac{1}{\beta} u^{(1-\beta)/\beta} - 1 \\
  & \geq & \frac{1}{\beta} u_{\min}^{(1-\beta)/\beta} -1\\
  & = & K
\end{eqnarray*}
where the strict inequality follows from strict convexity of the function $u^{1/\beta}$, the following inequality follows from monotonicity of $u^{(1-\beta)/\beta}$, and the last equality follows from the observation that $u_{\min} = (\beta (K+1))^{\beta/(1-\beta)}$ (obtained when $m=1$). Since the conditions for \cite[Lemma~17]{garivier16} hold, the rest of the proof follows \cite[Appendix~B.2]{garivier16}, whose examination indicates that we may take $n_{\epsilon} = \max\{ n_0, \epsilon^{-1}\}/(3 \epsilon)$.
\end{IEEEproof}

We next establish a concentration lemma. The notation ${\bf a} \succ {\bf b}$ stands for component-wise strict inequality.

\begin{lemma}
\label{lem:concentration}
Let ${\boldsymbol{\epsilon}} \in \mathbb{R}^d$ be made of strictly positive entries. Then there exists a finite positive constant $C$ such that
\[
  P \left( \left| \frac{{\bf{Y}}_i^n}{N_i^n} - {\boldkappa_i} \right| \succ {\boldsymbol{\epsilon}} \right) \leq  \frac{C}{n^4}.
\]
\end{lemma}

\begin{IEEEproof}
Thanks to the union bound, it suffices to focus on one component, say $l$ where $1\leq l \leq d$. Let the $l$th components be represented as $Y_i^n(l)$, $\kappa_i(l)$, and $\epsilon_l$. We will show that
\[
  P \left( \left| \frac{{Y}_i^n(l)}{N_i^n} - \kappa_i(l) \right| > \epsilon_l \right) \leq \frac{C_l}{n^4}
\]
for some finite positive constant $C_l$.

Fix $l$. Observe that $M_n := {Y}_i^n(l) - \kappa_i(l) N_i^n$ is a martingale whose quadratic variation process $\langle M \rangle_n$ has the following property:
\begin{eqnarray*}
  \langle M \rangle_n
  & := & \sum_{t=1}^n E\left[ \left((Y_{i,t}(l) - \kappa_i(l)) 1\{a_t = i\} \right)^2 \mid {\bf{Y}}_i^{t-1}, A^{t-1} \right] \\
  & \leq & n \sigma^2(l),
\end{eqnarray*}
a consequence of the existence of the second central moment for the observations. Hence, we have the following inequalities for all sufficiently $n$:
\begin{eqnarray*}
  \lefteqn{ P \left( \left| \frac{{Y}_i^n(l)}{N_i^n} - \kappa_i(l) \right| > \epsilon_l \right)} \\
  & \leq & P \Big( \left| {Y}_i^n(l) - \kappa_i(l) N_i^n \right| > N_i^n \epsilon_l  \Big) \\
  & \leq & P \Big( \left| {Y}_i^n(l) - \kappa_i(l) N_i^n \right|  > (n^a)^{\beta} \epsilon_l/2  \Big) \quad (\mbox{since $N_i^n \geq N_i^{n,a} \geq (n^a)^{\beta}/2$ for all suff. large $n$})\\
  & \leq & P \Big( \left| {Y}_i^n(l) - \kappa_i(l) N_i^n \right|  > (\gamma n/2)^{\beta} \epsilon_l/2, \quad n^a \geq \gamma n/2  \Big) + P \Big( n^a < \gamma n / 2 \Big) \\
  & \leq & P \Big( \sup_{1 \leq t \leq n} |{Y}_i^t(l) - \kappa_i(l) N_i^t| > n^{\beta} (\gamma/2)^{\beta} \epsilon_l/2  \Big) + P \Big( n^a - \gamma n < -\gamma n /2 \Big) \\
  & \leq & \frac{E[\left( \sup_{1 \leq t \leq n} |{Y}_i^t(l) - \kappa_i(l) N_i^t| \right)^p]}{(\epsilon_l/2)^p (\gamma/2)^{\beta p} n^{\beta p}} \quad (\mbox{Markov inequality}) \\
  & & \quad \quad \quad \quad \quad +~ e^{-n\gamma^2 /16} \quad (\mbox{Bernstein inequality}) \\
  & \leq &  \frac{c_p}{(\epsilon_l/2)^p (\gamma/2)^{\beta p} n^{\beta p}} E [|\langle M \rangle_n|^{p/2}] + e^{-n\gamma^2 /16} \quad (\mbox{Burkholder inequality}) \\
  & \leq & \frac{c_p}{(\epsilon_l/2)^p (\gamma/2)^{\beta p}n^{\beta p}} \cdot \sigma^p(l) n^{p/2} + e^{-n\gamma^2 /16} \quad (\mbox{from quadratic variation bound}) \\
  & \leq & \frac{C_l}{n^4} \quad (\mbox{taking $p = 4/(\beta - 1/2)$ and choosing $C_l$ suitably});
\end{eqnarray*}
recall that in the inequality above where the Burkholder inequality \cite[p.414]{chow2012probability} is employed with $p = 4/(\beta - 1/2)$, we made use of finiteness of the variances of the observations. This establishes the lemma. Choosing a suitably larger $C_l$ if needed, we can make the probability inequality be upper bounded by $C_l/n^4$ for all $n$.
\end{IEEEproof}

\begin{proposition}\label{prop:ml_convergences}
Let $\overline{\boldeta} \in \Theta_l$ be the true configuration. Consider the non-stopping policy $\tilde{\pi}_{SMF}(\gamma,\beta)$. Then, the following convergences hold almost surely as $n \rightarrow \infty$:
\begin{equation}\label{eqn:ml_convergence_k}
\hat{{\boldkappa}_i} :=\frac{{\bf{Y}}_i^n}{N_i^n} \rightarrow {\boldkappa_i} \text{ for all } i,
\end{equation}
\begin{equation}\label{eqn:ml_convergence_eta}
\hat{{\boldeta}_i} \rightarrow {\boldeta_i} \text{ for all } i,
\end{equation}
\begin{equation}\label{eqn:nearest_ml_conv}
{\boldeta}_i^*(l) \rightarrow {\boldeta}_i \text{ for all } i,
\end{equation}
\begin{equation}
\label{eqn:positive drift}
\liminf\limits_{n \rightarrow \infty} \frac{Z_{lm}(n)}{n^{\beta}} > 0.
\end{equation}
\end{proposition}
\begin{IEEEproof}
We prove the statements one after another.

(i) Proof of (\ref{eqn:ml_convergence_k}): This follows from Lemma \ref{lem:concentration} and the Borel-Cantelli lemma, since the series involving the upper bound in Lemma \ref{lem:concentration} is summable.

(ii) Proof for (\ref{eqn:ml_convergence_eta}): follows from (\ref{eqn:ml_convergence_k}) and the continuity of the function $ \boldkappa \mapsto {{\boldeta(\boldkappa)}}$.

(iii) Proof for (\ref{eqn:nearest_ml_conv}): Since $\overline{\boldeta} \in \Theta_l$, an open set, by (\ref{eqn:ml_convergence_eta}), $\hat{\overline{\boldeta}} \in \Theta_l$ for all sufficiently large $n$. From \eqref{eqn:ml_argument_alt}, it follows $\overline{\boldeta}^*(l) = \hat{\overline{\boldeta}}$ for all sufficiently large $n$, and hence (\ref{eqn:ml_convergence_eta}) implies (\ref{eqn:nearest_ml_conv}).

(iv) Proof for (\ref{eqn:positive drift}): We have from (\ref{eqn:MGLR_statistic}) that
\begin{equation}
\frac{Z_{lm}(n)}{n^{\beta}} = \frac{1}{n^{\beta}}\left( \log {\mathcal{H}_l(\overline{\boldupsilon},{\bf{n_0}})}-\log{\mathcal{H}_l({\bf{Y}}+\overline{\boldupsilon},{\bf{N}}+{\bf{n_0}})}-n\sum\limits_{i=1}^{K} w_i\left\{{\boldeta}_i^{*T}(m) \hat{\boldkappa}_i-\mathcal{A}_i({\boldeta}_i^{*}(m))\right\}\right).
\end{equation}
We then observe that
\begin{eqnarray}
\lefteqn{\liminf\limits_{n \rightarrow \infty} \frac{Z_{lm}(n)}{n^{\beta}}}\nonumber\\
 &=& \liminf\limits_{n \rightarrow \infty} \Bigg(\frac{1}{n^{\beta}} \log {\mathcal{H}_l(\overline{\boldupsilon},{\bf{n_0}})} + \frac{1}{n^{\beta}} \log \left\{\int\limits_{\overline{\boldeta}' \in \Theta_l}\exp\left\{\sum\limits_{i=1}^K \left[ ({\boldeta}_i')^{T}({\bf{Y}}_i^n+\boldupsilon_i)-(N_i^n+n_{0i})\mathcal{A}_i(\boldeta'_i) \right] \right\} d\overline{\boldeta}'\right\}\nonumber\\
     & & \quad \quad \quad \quad -\frac{1}{n^{\beta}} \cdot n \sum\limits_{i=1}^{K} w_i\left\{{\boldeta}_i^{*T}(m) \hat{\boldkappa}_i-\mathcal{A}_i({\boldeta}_i^{*}(m))\right\}\Bigg) \label{eqn:replace_H}\\
     &\geq& \liminf\limits_{n \rightarrow \infty} \Bigg(\frac{1}{n^{\beta}} \log \left\{\int\limits_{\overline{\boldeta}' \in \Theta_l} \exp\left\{\sum\limits_{i=1}^K \left[ (\boldeta_i')^T({\bf{Y}}_i^n+\boldupsilon_i)-(N_i^n+n_{0i})\mathcal{A}_i(\boldeta_i') \right] \right\}d\overline{\boldeta}'\right\}\nonumber\\
     & & \quad \quad \quad \quad - \frac{1}{n^{\beta}} \cdot n \sum\limits_{i=1}^{K} w_i\left\{{\boldeta}_i^{*T}(m) \hat{\boldkappa}_i-\mathcal{A}_i({\boldeta}_i^{*}(m))\right\}\Bigg) \label{eqn:super_additivity}\\
     & & \quad \quad \quad \quad \mbox{(since the first term in (\ref{eqn:replace_H}) is inconsequential)} \nonumber \\
     &=& \liminf\limits_{n \rightarrow \infty} \Bigg(\frac{1}{n^{\beta}} \log \int\limits_{\overline{\boldeta}' \in \Theta_l}
     \exp\left\{n\sum\limits_{i=1}^K \left[ \frac{N_i^n}{n}(\boldeta_i')^T\left(\frac{{\bf{Y}}_i^n}{N_i^n}+\frac{\boldupsilon_i}{N_i^n}\right)-\frac{(N_i^n+n_{0i})}{n}\mathcal{A}_i(\boldeta_i') \right] \right\}d\overline{\boldeta}'\nonumber\\
     & & \quad \quad \quad \quad - \frac{1}{n^{\beta}} \log \exp \left\{n\sum\limits_{i=1}^{K} \frac{N_i^n}{n}\left\{({\boldeta}^*_i(m))^{T} \hat{\boldkappa}_i-\mathcal{A}_i({\boldeta}_i^{*}(m))\right\}\right\}\Bigg)\label{eqn:re-write_ml}\\
     &=& \liminf\limits_{n \rightarrow \infty} \frac{1}{n^{\beta}} \log\int\limits_{\overline{\boldeta}' \in \Theta_l} \exp\left\{n \sum\limits_{i=1}^K\frac{N_i^n}{n}\left(\left(\boldeta_i'-\boldeta^*_i(m)\right)^T{\boldkappa}_i-\mathcal{A}_i(\boldeta_i')+\mathcal{A}_i(\boldeta_i^*(m))\right) \right. \nonumber
     \\ & & \left. \quad \quad \quad \quad \quad \quad \quad \quad \quad \quad \quad + n
     \sum\limits_{i=1}^K \left[ \frac{N_i^n}{n}\left(\left(\boldeta_i'-\boldeta^*_i(m)\right)^T (\hat{\boldkappa}_i - \boldkappa_i)
     + (\boldeta_i')^T \frac{\boldupsilon_i}{N_i^n} \right)
     - \frac{n_{0i}}{n} \mathcal{A}_i(\boldeta_i') \right\} \right] d\overline{\boldeta}'. \nonumber \\
     \label{eqn:hyper_0}
\end{eqnarray}
Note that ${\boldeta}_i(\boldkappa_i)$  optimises the function $\boldeta'_i \mapsto (\boldeta')^T {\boldkappa}_i - \mathcal{A}_i(\boldeta'_i)$, for each $i = 1, \ldots, K$, over the set $\Theta_l$, since $\overline{\boldeta} \in \Theta_l$. We now leverage this.

Write $B_r(\overline{\boldeta}(\overline{\boldkappa}))$ for the open Euclidean ball of radius $r$ around $\overline{\boldeta}(\overline{\boldkappa})$. Fix $\epsilon > 0$. There is then an $\delta > 0$ and a $C_{\delta} > 0$ such that, almost surely, for all sufficiently large $n$ and all $\overline{\boldeta}' \in B_{\delta}(\overline{\boldeta}(\overline{\boldkappa}))$, we have:
\begin{eqnarray*}
|| \hat{\boldkappa}_i - \boldkappa_i||_{\infty} & \leq &  \varepsilon \\
\left|\left(\left(\boldeta_i'-\boldeta^*_i(m)\right)^T (\hat{\boldkappa}_i - \boldkappa_i)
     + (\boldeta_i')^T \frac{\boldupsilon_i}{N_i^n} \right)\right| & < & C_{\delta} \varepsilon \quad (\mbox{since $\boldeta^*_i(m)$ is bounded, see footnote \ref{footnote1}}) \\
\left| n_{0i} \mathcal{A}_i(\boldeta_i') \right| & \leq & C_{\delta}  \\
\left| (\boldeta'_i)^T {\boldkappa}_i - \mathcal{A}_i(\boldeta'_i) - (\boldeta_i(\boldkappa_i))^T {\boldkappa}_i + \mathcal{A}_i(\boldeta_i(\boldkappa_i))  \right| & \leq & \varepsilon.
\end{eqnarray*}
Furthermore, we can lower bound the integral in (\ref{eqn:hyper_0}) by restricting the integral to the set $\Theta_l \cap B_{\delta}(\overline{\boldeta}(\overline{\boldkappa}))$. Putting these ideas together, we get that (\ref{eqn:hyper_0}) is lower bounded by:
\begin{eqnarray}
  & \geq & \liminf\limits_{n \rightarrow \infty} \frac{1}{n^{\beta}} \log \int_{\overline{\boldeta}' \in \Theta_l \cap B_{\delta}(\overline{\boldeta}(\overline{\boldkappa}))}
  \hspace*{-.2cm}
  \exp\left\{n \sum\limits_{i=1}^K\frac{N_i^n}{n}\left(\left(\boldeta_i'-\boldeta^*_i(m)\right)^T{\boldkappa}_i-\mathcal{A}_i(\boldeta_i')+\mathcal{A}_i(\boldeta_i^*(m))\right) \right. \nonumber
     \\ & & \left. \quad \quad \quad \quad \quad \quad \quad \quad \quad  + n
     \sum\limits_{i=1}^K \left[ \frac{N_i^n}{n}\left(\left(\boldeta_i'-\boldeta^*_i(m)\right)^T (\hat{\boldkappa}_i - \boldkappa_i)
     + (\boldeta_i')^T \frac{\boldupsilon_i}{N_i^n} \right)
     - \frac{n_{0i}}{n} \mathcal{A}_i(\boldeta_i') \right] \right\}d\overline{\boldeta}' \nonumber \\
     & \geq & \liminf\limits_{n \rightarrow \infty} \frac{1}{n^{\beta}} \log \exp \left\{ n \sum\limits_{i=1}^K \frac{N_i^n}{n}\Big(  (\boldeta_i(\boldkappa_i) - \boldeta_i^*(m))^T {\boldkappa_i} - \mathcal{A}_i(\boldeta_i(\boldkappa_i)) + \mathcal{A}_i(\boldeta_i^*(m)) \Big)
     \right\} \\
     & & + \liminf\limits_{n \rightarrow \infty} \frac{1}{n^{\beta}} \log
     \int_{\overline{\boldeta}' \in \Theta_l \cap B_{\delta}(\overline{\boldeta}(\overline{\boldkappa}))}
     \exp \left\{ n \sum\limits_{i=1}^K \frac{N_i^n}{n} (-\varepsilon) + n \sum\limits_{i=1}^K \frac{N_i^n}{n} (-C_{\delta} \varepsilon) - C_{\delta} \right\} \, d{\overline{\boldeta'}} \nonumber \\
     & \geq & \liminf\limits_{n \rightarrow \infty} \sum\limits_{i=1}^K \frac{N_i^n}{n^{\beta}} D(\boldeta_i(\boldkappa_i)  \parallel  \boldeta_i^*(m)) \\
     & & + \liminf\limits_{n \rightarrow \infty} \frac{1}{n^{\beta}} \log \left( \textsf{Leb}\left(\overline{\boldeta}' \in \Theta_l \cap B_{\delta}(\overline{\boldeta}(\overline{\boldkappa})) \right) \right)
     - \limsup_{n \rightarrow \infty} \left( \sum\limits_{i=1}^K \frac{N_i^n}{n^{\beta}} (\varepsilon) + \sum\limits_{i=1}^K \frac{N_i^n}{n^{\beta}} (C_{\delta}\varepsilon) + \frac{C_{\delta}}{n^{\beta}}
     \right) \label{eqn:pre-laplace}\\
     & \geq & \liminf\limits_{n \rightarrow \infty} \sum\limits_{i=1}^K \frac{N_i^n}{n^{\beta}} \left( D(\boldeta_i(\boldkappa_i)  \parallel  \boldeta_i^*(m)) - (1 + C_{\delta}) \varepsilon \right), \label{eqn:laplace}
\end{eqnarray}
where the last inequality holds because the Lebesgue measure $\textsf{Leb}\left(\overline{\boldeta}' \in \Theta_l \cap B_{\delta}(\overline{\boldeta}(\overline{\boldkappa})) \right) > 0$. Continuing with the inequality in (\ref{eqn:laplace}), by virtue of our sampling rule and by choosing $\varepsilon$ sufficiently small, almost surely, for some constant $a > 0$, the lower bound becomes
\begin{eqnarray}
  & \geq & a \left( \inf_{\boldeta' \in \Theta_m} \sum_{i=1}^K D(\boldeta_i  \parallel  \boldeta'_i) - K(1 + C_{\delta})\varepsilon \right) \\
  & > & 0,
\end{eqnarray}
where the last strict inequality holds because
$$\inf_{\overline{\boldeta}' \in \Theta_m} \sum_{i=1}^K D(\boldeta_i \parallel  \boldeta'_i) > 0, \quad \forall m \neq l,$$
a fact that comes from the assumptions that  $\Theta_l$ and $\Theta_m$ are disjoint and open.
This completes the proof of the Proposition.
\end{IEEEproof}

\subsection{Proof of Proposition \ref{prop:finite_stopping_time}}
\begin{IEEEproof} The following inequalities hold almost surely:
\begin{eqnarray}
\tau(\pi_{SMF}(L,\gamma,\beta)) &\leq& \tau(\pi^l_{SMF}(L,\gamma,\beta))\\
                                &=& \inf\{n \geq 1| Z_l(n) > \log((M-1)L)\}\\
                                &\leq & \inf\{n \geq 1|Z_{lm}(n')>\log((M-1)L), \forall n'\geq n, \forall m\neq l\}\\
                                &<&\infty,
\end{eqnarray}
where the last inequality follows because of (\ref{eqn:positive drift}) in Proposition \ref{prop:ml_convergences}.
\end{IEEEproof}

\section{Proof of Proposition \ref{prop:admissibility} on Admissibility}
\label{app:admissibility}

\begin{IEEEproof}
Fix $\overline{\boldeta} \in \Theta_l$ as the true configuration. We begin with
\begin{eqnarray}
P(\delta \neq l| \overline{\boldeta}) &=& \sum\limits_{m \neq l}P(\delta =m|\overline{\boldeta} ) + P(\tau(\pi_{SMF}(L,\gamma,\beta))=\infty|\overline{\boldeta})\\
													&=&	\sum\limits_{m \neq l}P(\delta =m|\overline{\boldeta}),\label{eqn:Pe}		
\end{eqnarray}
where (\ref{eqn:Pe}) follows from Proposition \ref{prop:finite_stopping_time}. Let
\begin{equation}
\Delta_m^n = \{(x^n, a^n) : \tau(\pi_{SMF}(L,\gamma,\beta)(x^n, a^n) = n, \delta(x^n,a^n) = m\}
\end{equation}
denote the sample paths for which the decision maker stops sampling after $n$ time slots and decides in favour of $H=m$. The decision region in favour of $m$ is denoted $\Delta_m:=\bigcup\limits_{n \geq 1} \Delta_m^n$. Note that
\begin{equation}
\Delta_m^n \cap \Delta_m^k = \emptyset \text{ for all } k\neq n.
\end{equation}
We then have
\begin{eqnarray}
P(\delta \neq l| \overline{\boldeta}) &=& \sum\limits_{m \neq l}P(\delta =m|\overline{\boldeta})\\
                                                   &=& \sum\limits_{m \neq l}\sum\limits_{n \geq 1} \int\limits_{(x^n,a^n) \in \Delta_m^n} dP((x^n,a^n)|\overline{\boldeta})\\
                                                   &=& \sum\limits_{m \neq l}\sum\limits_{n \geq 1} \int\limits_{(x^n,a^n) \in \Delta_m^n} \prod\limits_{t=1}^n\left[P\left(a_t|a^{t-1},x^{t-1}\right) \cdot f(x_t|a_t,\boldeta_{a_t}) \right] d(x^n,a^n)\label{eqn:change}\\
                                                   &=& \sum\limits_{m \neq l}\sum\limits_{n \geq 1} \int\limits_{(x^n,a^n) \in \Delta_m^n} \prod\limits_{t=1}^n f(x_t|a_t,\boldeta_{a_t})\cdot\left[\prod\limits_{t=1}^n P\left(a_t|a^{t-1},x^{t-1}\right)\right] d(x^n,a^n)\label{eqn:change1}\\
                                                   &\leq & \sum\limits_{m \neq l}\sum\limits_{n \geq 1} \int\limits_{(x^n,a^n) \in \Delta_m^n} \frac{\hat{f}(x^n|a^n,\tilde{\overline{\boldeta}} \in \Theta_l)}{\tilde{f}_m(x^n|a^n)} \tilde{f}_m(x^n|a^n)\cdot\left[\prod\limits_{t=1}^n  P\left(a_t|a^{t-1},x^{t-1}\right)\right] d(x^n,a^n)\nonumber\\
                                                   & & \label{eqn:ML_bound}\\
                                                   &\leq& \sum\limits_{m \neq l}\frac{1}{(M-1)L}\sum\limits_{n \geq 1} \int\limits_{(x^n,a^n) \in \Delta_m^n} \tilde{f}_m(x^n|a^n) \prod\limits_{t=1}^n  P\left(a_t|a^{t-1},x^{t-1}\right) d(x^n,a^n) \label{eqn:thresholdbound}\\
                                                   &\leq& \sum\limits_{m \neq l} \frac{1}{(M-1)L} \cdot \tilde{P}(\delta = m | \tilde{\overline{\boldeta}} \in \Theta_m) \label{eqn:ptilde}\\
                                                   &\leq& \frac{1}{L}. \label{eqn:alpha}
\end{eqnarray}
In (\ref{eqn:change}), the term $P\left(a_t|a^{t-1},x^{t-1}\right)$ indicates the probability of choosing arm $a_t$, with the convention that at time $t=1$, the term represents $P(a_1)$. Inequality (\ref{eqn:ML_bound}) follows from the definition of maximum likelihood function, in particular $\prod\limits_{t=1}^n f(x_t|a_t,\tilde{\boldeta}_{a_t}) = f(x^n|a^n,\tilde{\overline{\boldeta}}) \leq \hat{f}(x^n|a^n,\tilde{\overline{\boldeta}} \in \Theta_l)$. In (\ref{eqn:thresholdbound}), we have used $$\frac{\hat{f}(x^n|a^n,\tilde{\overline{\boldeta}} \in \Theta_l)}{\tilde{f}_m(x^n|a^n)} \leq \frac{1}{(M-1)L}$$
for $(x^n,a^n) \in \Delta_m^n$. In (\ref{eqn:ptilde}), $\tilde{P}$ is the probability under $\Theta_m$ when the prior on $\tilde{\overline{\boldeta}}$ is $f(\tilde{\overline{\boldeta}}|\tilde{\overline{\boldeta}} \in \Theta_m)$.  Inequality (\ref{eqn:alpha}) follows from $\tilde{P}(\delta = m |\tilde{\overline{\boldeta}} \in \Theta_m) \leq 1$ and the union bound. Choosing $L=1/\alpha$ completes the proof.
\end{IEEEproof}

\section{Achievability}\label{app:achievability}

We first show some preliminary results before we get to achievability. In the following Proposition, we assert that several statements hold almost surely. We show that the hypothesis $l^*(n)$ chosen by the policy is eventually the correct one. In addition, we show that the parameters $\boldeta_i^*(l^*(n))$ chosen by the policy converge to the true or the actual parameters. Furthermore, we will strengthen (\ref{eqn:positive drift}) to show that $Z_{lm}(n)$ is linear in $n$ and that the drift is at least $D^*(\overline{\boldeta})$.

\begin{proposition}\label{prop:conv_results}
Let Assumption A hold. Let $\overline{\boldeta} \in \Theta_l$ be the true configuration. Consider the non-stopping policy $\tilde{\pi}_{SMF}(\gamma)$. Then, the following convergences hold almost surely:
\begin{equation}\label{eqn:conv_l}
l^*(n) \rightarrow l,
\end{equation}
\begin{equation} \label{eqn:conv_l_nearest_ml}
{{\boldeta^*_i(l^*(n)) \rightarrow \boldeta_i \text{ for all } i}},
\end{equation}
\begin{equation}\label{eqn:conv_l_lambda}
\lambda_i^*(\overline{\boldeta}^*(l^*(n)) \rightarrow \lambda_i^*(\overline{\boldeta}) \text{ for all } i,
\end{equation}
\begin{equation}\label{eqn:conv_l_Ni_a}
\frac{N_i^{n,a}}{n^a} \rightarrow \lambda_i^*(\overline{\boldeta}) \text{ for all } i
\end{equation}
\begin{equation}\label{eqn:conv_l_Ni}
\frac{N_i^n}{n} \rightarrow \lambda_i^*(\overline{\boldeta}) \text{ for all } i
\end{equation}
\begin{equation}\label{eqn:constant_drift}
\liminf\limits_{n \rightarrow \infty} \frac{Z_l(n)}{n} \geq  D^*(\overline{\boldeta}).
\end{equation}
\end{proposition}
\begin{IEEEproof}
From (\ref{eqn:positive drift}), we have
\begin{equation}
\liminf\limits_{n \rightarrow \infty} Z_l(n) = \liminf\limits_{n \rightarrow \infty} Z_{lm}(n) >0 \text{ almost surely}.
\end{equation}
Fix $m \neq l$. Then, the following inequalites hold almost surely:
\begin{eqnarray}
\limsup\limits_{n \rightarrow \infty} Z_m(n) &=& \limsup\limits_{n \rightarrow \infty} \min\limits_{p \neq m} Z_{mp}(n)\\
											 &\leq& \limsup\limits_{n \rightarrow \infty} Z_{ml}(n)\\
											 &\leq& \limsup\limits_{n \rightarrow \infty} -Z_{lm}(n) \quad \mbox{(a property of the modified GLR)}\\
											 &=& -\liminf\limits_{n \rightarrow \infty} Z_{lm}(n)\\
											 &\leq& -\liminf\limits_{n \rightarrow \infty} \min\limits_{p \neq l} Z_{lp}(n)\\
											 &=& -\liminf\limits_{n \rightarrow \infty} Z_l(n)\\
											 &<& 0.
\end{eqnarray}
This further implies that a.s. $l^*(n) = \max\limits_{p}Z_p(n) = l$, for all sufficiently large $n$. This completes the proof for (\ref{eqn:conv_l}).

For (\ref{eqn:conv_l_nearest_ml}), we use (\ref{eqn:conv_l}) to get, a.s.,
\begin{equation}
\boldeta^*_i(l^*(n)) = \boldeta^*_i(l)
\end{equation}
for all sufficiently large $n$, and then Proposition \ref{prop:ml_convergences} to get $\boldeta^*_i(l) \rightarrow \boldeta_i$, which then yields $\boldeta^*_i(l^*(n)) \rightarrow \boldeta_i$.

The convergence in (\ref{eqn:conv_l_lambda}) follows from (\ref{eqn:conv_l_nearest_ml}) and Assumption A.

The convergence in (\ref{eqn:conv_l_Ni_a}) follows from  (\ref{eqn:conv_l_lambda}) and Lemma \ref{lemma:lambda_conv}.

Proof of (\ref{eqn:conv_l_Ni}): Let $\{V_1,V_2, \ldots, V_{n^a}\}$ be such that $V_k$ is the number of sluggish instants plus one active instance corresponding to the $k$th active instance, $k = 1,2, \ldots, n^a$. Then $V_t$'s are independent and identical random variables with the geometric distribution of parameter $\gamma$. Additionally, to make the total of $n$ arm pulls at time instant $n$, the last `sluggish run' should also be accounted. We do this by re-writing the expression in (\ref{eqn:num-samples}) as
\begin{equation}
N_i^n = \sum\limits_{t=1}^{n^a} V_t1_{\{a_t=i\}} + \overline{V}_i
\end{equation}
where $\overline{V}_i$ is nonzero for at most for one $i$ and corresponds to the latest sluggish run at time instant $n$. To study the limit of $N_i^n/n$, it suffices to study
\begin{equation}\label{eqn:split_frac}
\frac{1}{n} \sum\limits_{t=1}^{n^a} V_t1_{\{a_t=i\}} = \frac{n^a}{n} \cdot \frac{N_i^{n,a}}{n^a} \cdot \frac{1}{N_i^{n,a}}\sum\limits_{t=1}^{n^a} V_t1_{\{a_t=i\}}.
\end{equation}
We consider each term on the right-hand side of (\ref{eqn:split_frac}) in detail. Note that $n^a/n \rightarrow \gamma$ and from (\ref{eqn:conv_l_Ni_a}) we get $N_i^{n,a}/n^a \rightarrow \lambda_i^*(\overline{\boldeta})$. Also by Lemma \ref{lemma:lambda_conv} we have $N_i^{n,a} \rightarrow \infty$ as $n \rightarrow \infty$.  Note that the summation in (\ref{eqn:split_frac}) has $N_i^{n,a}$ terms, and hence the sample mean converges to the expected value of $V_t$ which is $1/\gamma$. Hence,we get, almost surely,
\begin{equation}\label{eqn:exp_frac}
\lim\limits_{n \rightarrow \infty} \frac{N_i^n}{n} = \gamma \cdot \lambda_i^*(\overline{\boldeta}) \cdot \frac{1}{\gamma} = \lambda_i^*(\overline{\boldeta}).
\end{equation}
This concludes the proof of (\ref{eqn:conv_l_Ni}).

Proof of (\ref{eqn:constant_drift}): In the proof of (\ref{eqn:positive drift}), instead of scaling by $1/n^{\beta}$, rescale by $1/n$, and arrive at
\[
  \liminf_{n \rightarrow \infty} \frac{Z_l(n)}{n} \geq \inf_{\overline{\boldeta'} \in \Theta_m}  \liminf_{n \rightarrow \infty} \sum_{i=1}^K \frac{N_i^n}{n} D(\boldeta_i \parallel \boldeta'_i),
\]
which is the equivalent of \eqref{eqn:laplace} with an additional infimum over all $\overline{\boldeta}' \in \Theta_m$. The result in (\ref{eqn:constant_drift}) then follows from (\ref{eqn:conv_l_Ni}).
\end{IEEEproof}

In the next three subsections we prove each of the three claims in Proposition \ref{prop:achievability}.

\subsection{Proof of (\ref{eqn:upper_bound_tau})}
\begin{IEEEproof}
We begin by proving that, as the probability of false detection constraint goes to zero, the stopping time of the policy goes to infinity (Lemma \ref{lemma:tau_L}). We then combine this result with Proposition \ref{prop:conv_results} to complete the required proof.
\begin{lemma}\label{lemma:tau_L}
Let $\overline{\boldeta} \in \Theta_l$ be the true configuration. Consider the policy $\pi_{SMF}(L,\gamma,\beta)$. Then,
\begin{equation}
\label{eqn:taugoestoinfinity}
\liminf\limits_{L \rightarrow \infty} \tau(\pi_{SMF}(L,\gamma,\beta)) \rightarrow \infty \text{ a.s.}
\end{equation}
\end{lemma}

\begin{IEEEproof}
It suffices to show that, as $L \rightarrow \infty$,
\begin{equation}
P(\tau(\pi_{SMF}(L,\gamma,\beta)) < n) \rightarrow 0 \text{ for all } n.
\end{equation}
Fix some $\boldeta'(m) \in \Theta_{-m}$, for each $m$. We begin with
\begin{eqnarray}
\lefteqn{\limsup\limits_{L \rightarrow \infty} P(\tau(\pi_{SMF}(L,\gamma,\beta)) < n)}\nonumber\\
&=& \limsup\limits_{L \rightarrow \infty} P\left(\max\limits_{1 \leq t \leq n} Z_m(t) > \log((M-1)L) \text{ for some $m$}\right)\\
&\leq& \limsup\limits_{L \rightarrow \infty} \sum\limits_{m=1}^M \sum\limits_{t=1}^n P\left(Z_m(t) > \log((M-1)L)\right)\label{eqn:UB1_union_bound}\\
& \leq & \limsup\limits_{L \rightarrow \infty} \frac{1}{\log((M-1)L)}\sum\limits_{m=1}^M \sum\limits_{t=1}^n E \left[\sum\limits_{i=1}^K N_i^tD(\hat{\boldeta}_i \parallel \boldeta_i'(m))\right]\label{eqn:UB1_markov}\\
&\leq& \limsup\limits_{L \rightarrow \infty} \frac{1}{\log((M-1)L)}\sum\limits_{m=1}^M \sum\limits_{t=1}^n \sum\limits_{i=1}^K E\left[t \left(\hat{\boldeta}_i^T \hat{\boldkappa}_i-\mathcal{A}_i(\hat{\boldeta}_i)-\boldeta_i'(m)^T\hat{\boldkappa}_i+\mathcal{A}_i(\boldeta_i'(m))\right)\right] \quad \quad \label{eqn:UB1_D}\\
&\leq & \limsup\limits_{L \rightarrow \infty} \frac{1}{\log((M-1)L)}\sum\limits_{m=1}^M \sum\limits_{t=1}^n \sum\limits_{i=1}^K t\left(E\left[\hat{\boldeta}_i^T\hat{\boldkappa}_i\right] - \mathcal{A}_i(\boldeta_i) - \boldeta_i'(m)^T{\boldkappa}_i +\mathcal{A}_i(\boldeta'_i(m))\right) \quad \quad \label{eqn:jen}\\
&=&0\label{eqn:UB1_Z}.
\end{eqnarray}
Inequality in (\ref{eqn:UB1_union_bound}) follows from union bound. Inequality (\ref{eqn:UB1_markov}) is discussed below. Equality in (\ref{eqn:UB1_D}) is obtained using the expression for relative entropy and the fact that $N_i^t \leq t$. The inequality in (\ref{eqn:jen}) is obtained from the observation that $\mathcal{A}_i(\cdot)$ is convex and by an application of Jensen's inequality. The final equality in (\ref{eqn:UB1_Z}) follows because $E\left[\hat{\boldeta}_i^T\hat{\boldkappa}_i\right]$ is finite. Inequality in (\ref{eqn:UB1_markov}) is obtained using the following result
\begin{eqnarray*}
Z_m(t) &=& \min\limits_{p \neq m} \log \frac{\tilde{f}_m(x^t|a^t)}{\hat{f}(x^t|x^t,\overline{\boldeta}\in \Theta_p)}\label{eqn:Zm_redn1}
\\
       &=& \min\limits_{p \neq m}\inf\limits_{\overline{\boldeta}'\in \Theta_p} \log\frac{\tilde{f}_m(x^t|a^t)}{f(x^t|a^t,\overline{\boldeta}')}\label{eqn:Zm_redn2}\\
       &\leq& \log\frac{\tilde{f}_m(x^t|a^t)}{f(x^t|a^t,\overline{\boldeta}'(m))}, \quad \mbox{ using the chosen } \overline{\boldeta}'(m) \in \Theta_{-m}\label{eqn:Zm_redn3}\\
       &\leq& \log\frac{\sup\limits_{\tilde{\overline{\boldeta}}\in \Omega}f(x^t|a^t,\tilde{\overline{\boldeta}})}{f(x^t|a^t,\overline{\boldeta}'(m))}\label{eqn:Zm_redn4}\\
       &=& \sum\limits_{i=1}^K N_i^t \left[\hat{\boldeta}_i^T \hat{\boldkappa}_i-\mathcal{A}_i(\hat{\boldeta}_i)-\boldeta_i'(m)^T\hat{\boldkappa}_i+\mathcal{A}_i(\boldeta_i'(m))\right]\label{eqn:Zm_redn5}\\
       &=& \sum\limits_{i=1}^K N_i^t D(\hat{\boldeta}_i \parallel \boldeta_i'(m)).\label{eqn:Zm_redn6}
\end{eqnarray*}
The last quantity being positive, we can now apply the Markov inequality and (\ref{eqn:UB1_markov}) follows.
This finishes the proof of the lemma.
\end{IEEEproof}
\begin{lemma}
Let Assumption A hold. Let $\overline{\boldeta} \in \Theta_l$ be the true configuration. Consider the policy $\pi_{SMF}(L,\gamma,\beta)$. We then have
\begin{eqnarray}\label{eqn:Zl_tau}
\liminf\limits_{L \rightarrow\infty} \frac{Z_l(\tau(\pi_{SMF}(L,\gamma,\beta)))}{\tau(\pi_{SMF}(L,\gamma,\beta))} & \geq & D^*(\overline{\boldeta}) \text{ a.s.}, \\
\label{eqn:Zl_tau-2}
\liminf\limits_{L \rightarrow\infty} \frac{Z_l(\tau(\pi_{SMF}(L,\gamma,\beta))-1)}
{\tau(\pi_{SMF}(L,\gamma,\beta))-1} & \geq & D^*(\overline{\boldeta}) \text{ a.s.}
\label{eqn:Zl_tau-1}
\end{eqnarray}
\end{lemma}
\begin{IEEEproof}
The proofs of the two statements follow by focusing on sample paths that satisfy (\ref{eqn:constant_drift}) of Proposition~\ref{prop:conv_results} and \eqref{eqn:taugoestoinfinity} of Lemma \ref{lemma:tau_L}. The argument goes as follows. For any such sample path $\omega$ and any $\epsilon > 0$, there is an $N(\omega, \epsilon)$, independent of $L$, such that $Z_l(n)/n \geq D^*(\overline{\boldeta}) - \epsilon$ for all $n \geq N(\omega,\epsilon)$. Now take $L$ to infinity and employ Lemma \ref{lemma:tau_L} to get that $\tau(\pi_{SMF}(L,\gamma,\beta))$ is eventually bigger than $N(\omega, \epsilon) + 1$, and so $\tau(\pi_{SMF}(L,\gamma,\beta))-1 \geq N(\omega, \epsilon)$. So both \eqref{eqn:Zl_tau} and \eqref{eqn:Zl_tau-2} hold.
\end{IEEEproof}

\vspace*{.3cm}

We now begin the proof for (\ref{eqn:upper_bound_tau}). Using the definition for $\tau(\pi_{SMF}(L,\gamma,\beta))$, at the time slot prior to stoppage, we must have $Z_l(\tau(\pi_{SMF}(L,\gamma,\beta))-1)<\log((M-1)L)$. So,
\begin{equation}\label{eqn:Zl_L}
\limsup\limits_{L\rightarrow \infty} \frac{Z_l(\tau(\pi_{SMF}(L,\gamma,\beta))-1)}{\log(L)} \leq \limsup\limits_{L\rightarrow \infty} \frac{\log((M-1)L)}{\log(L)}=1.
\end{equation}
Thus
\begin{eqnarray}
1 & \geq & \limsup\limits_{L\rightarrow \infty} \frac{Z_l(\tau(\pi_{SMF}(L,\gamma,\beta))-1)}{\log(L)} \\
& \geq & \liminf\limits_{L \rightarrow \infty} \frac{Z_l(\tau(\pi_{SMF}(L,\gamma,\beta))-1)}{\tau(\pi_{SMF}(L,\gamma,\beta))-1} \cdot \limsup\limits_{L \rightarrow \infty} \frac{\tau(\pi_{SMF}(L,\gamma,\beta))-1}{\log L}\\
& \geq & D^*(\overline{\boldeta}) \cdot \limsup\limits_{L \rightarrow \infty} \frac{\tau(\pi_{SMF}(L,\gamma,\beta))-1}{\log L},
\end{eqnarray}
where in the last inequality, we have used (\ref{eqn:Zl_tau}). Finally,
\begin{eqnarray}
\limsup\limits_{L \rightarrow \infty} \frac{\tau(\pi_{SMF}(L,\gamma,\beta))}{\log(L)} &=& \limsup\limits_{L \rightarrow \infty} \frac{\tau(\pi_{SMF}(L,\gamma,\beta))-1}{\log(L)}\\
                 & \leq& \frac{1}{D^*(\overline{\boldeta})} \text{ a.s.}.
\end{eqnarray}
This completes the proof of (\ref{eqn:upper_bound_tau}).
\end{IEEEproof}

\subsection{Proof of (\ref{eqn:upper_bound_Etau})}

We begin with a couple of lemmas.

\begin{lemma}
\label{lem:Fl-continuous}
For every $l$, for every $i$, the mapping
\[
  \Theta_l \ni \boldeta_i \mapsto \inf\limits_{\overline{\boldeta}' \in \Theta_{-l}} D(\boldeta_i \parallel \boldeta_i')
\]
is continuous.
\end{lemma}

\begin{IEEEproof}
Write $\mathcal{G}_l (\boldeta_i) := \inf\limits_{\overline{\boldeta}' \in \Theta_{-l}} D(\boldeta_i \parallel \boldeta_i')$. Observe that $\mathcal{G}_l \geq 0$. Fix $\epsilon > 0$.

\noindent (i) Consider a sequence $\boldeta_i(n) \rightarrow \boldeta_i$ with all of them being in $\Theta_l$. By the definition of $\mathcal{G}_l (\boldeta_i)$, there exists $\boldeta_i' \in
\Theta_{-l}$ such that $D(\boldeta_i \parallel \boldeta_i') \leq \mathcal{G}_l(\boldeta_i) + \epsilon$. We then have
\begin{eqnarray*}
  \mathcal{G}_l(\boldeta_i(n)) & \leq & D(\boldeta_i(n) \parallel \boldeta_i') \quad \mbox{(since the left-side is an infimum)} \\
  & \leq & D(\boldeta_i \parallel \boldeta_i') + \epsilon ~ \mbox{(for all sufficiently large $n$, using continuity of relative entropy)} \\
  & \leq & \mathcal{G}_l(\boldeta_i) + 2\epsilon \quad \mbox{(by the choice of $\boldeta_i'$)}.
\end{eqnarray*}
Thus $\limsup_{n \rightarrow \infty } \mathcal{G}_l(\boldeta_i(n)) \leq \mathcal{G}_l(\boldeta_i)$.

\noindent (ii) Since $\mathcal{G}_l(\boldeta_i(n))$ is bounded, by Lemma~\ref{lem:regularity} (see footnote~\ref{footnote1}) there exists a convergent sequence of $\boldeta_i'(n) \rightarrow \boldeta_i'$. By the same argument that led to (\ref{eqn:liminfF}), using the bounded of the $\boldeta'_i(n)$ sequence, for all sufficiently large $n$, we have
\[
 \mathcal{G}_l(\boldeta_i(n)) \geq D(\boldeta_i \parallel \boldeta_i'(n)) - 2\epsilon \geq \mathcal{G}_l(\boldeta_i) - 2 \epsilon.
\]
This establishes that $\liminf_{n \rightarrow \infty } \mathcal{G}_l(\boldeta_i(n)) \geq \mathcal{G}_l(\boldeta_i)$.

Together, (i) and (ii) establish the continuity of $\mathcal{G}_l(\cdot)$.
\end{IEEEproof}

The next lemma provides an estimate of the probability that the likelihood for the correct hypothesis is small.

\begin{lemma}\label{lemma:exp_bounds_Zl}
Let Assumption A hold. Fix $L>1$. Let $\overline{\boldeta} \in \Theta_l$ be the true configuration. Then there exists a constant $0<B<\infty$ and a constant $N_0$, both independent of $L$, such that for all $n \geq \max\{ 2 \log((M-1)L) / D^*(\boldeta), \mathcal{N}_0 \}$, we have
\begin{equation}
P(Z_l(n) < \log((M-1)L)) < \frac{B}{n^3}.
\end{equation}
\end{lemma}
\begin{IEEEproof}
{\color{black}{Before we start with the proof, let us note that
\begin{eqnarray}
D^*(\overline{\boldeta}) &=& \inf\limits_{\overline{\boldeta}' \in \Theta_{-l}} \sum\limits_{i=1}^K \lambda_i^* D(\boldeta_i \parallel \boldeta_i')\\
                         &=& \inf\limits_{\overline{\boldeta}' \in \Theta_{-l}} \sum\limits_{i=1}^K \lambda_i^* \left[(\boldeta_i-\boldeta_i')^T\boldkappa_i-\mathcal{A}_i(\boldeta_i)+\mathcal{A}_i(\boldeta_i')\right]\\
                         &=&  \sum\limits_{i=1}^K \lambda_i^* \left(\boldeta_i^T\boldkappa_i-\mathcal{A}_i(\boldeta_i)\right)-\sup\limits_{\overline{\boldeta}'\in \Theta_{-l}}\sum\limits_{i=1}^K \lambda_i^* \left(\boldeta_i'^T\boldkappa_i-\mathcal{A}_i(\boldeta'_i)\right).
\end{eqnarray}
Let us now turn to the probability of interest. Observe that $Z_l(n) = \min_{m \neq l} Z_{lm}(n)$. Using \eqref{eqn:MGLR_statistic}, we obtain the following inequality:
\begin{eqnarray}
\lefteqn{P(Z_l(n) < \log((M-1)L))}\nonumber\\
&\leq& P\left(\log \mathcal{H}_l(\overline{\boldupsilon},{\bf{n_0}}) < -\epsilon'n\right) \nonumber \\
& &  + P\left(-\log\mathcal{H}_l({\bf{Y}}+\overline{\boldupsilon},{\bf{N}}+{\bf{n_0}})-n\sum\limits_{i=1}^K\lambda_i^* [\boldeta_i^T \boldkappa_i - \mathcal{A}_i(\boldeta_i)]< -\epsilon'n\right)\nonumber\\
                         & & +P\left(-\sup\limits_{\overline{\boldeta}' \in \Theta_{-l}} \sum\limits_{i=1}^K N_i^n\left(\boldeta_i'^T\hat{\boldkappa}_i - \mathcal{A}_i(\boldeta_i')\right)+\sup\limits_{\overline{\boldeta}' \in \Theta_{-l}} n \sum\limits_{i=1}^K\lambda_i^*\left(\boldeta_i'^T{\boldkappa}_i - \mathcal{A}_i(\boldeta_i')\right) < -\epsilon'n \right)\nonumber\\
                         & & +P\left(n D^*(\overline{\boldeta})-3\epsilon'n < \log((M-1)L)\right)\label{eqn:Etau_Zl};
\end{eqnarray}
the inequality in (\ref{eqn:Etau_Zl}) is obtained using union bound together with adding and subtracting $D^*(\overline{\boldeta})$. Our goal is now to show that each of the terms on the right-hand side above is either 0 or $O(n^{-3})$.

\vspace*{.3in}

\noindent (i) We begin with the last term in (\ref{eqn:Etau_Zl}). Let
\begin{equation}
\epsilon = \frac{D^*(\overline{\boldeta})}{D^*(\overline{\boldeta})-3\epsilon'} - 1,
\end{equation}
and
{\color{black}{\begin{equation}
n_0 = \frac{2\log((M-1)L)}{D^*(\overline{\boldeta})} > \frac{(1+\epsilon)\log((M-1)L)}{D^*(\overline{\boldeta})}.
\end{equation}}}
This is $n_0$ is one of the values that $n$ must exceed in the statement of the lemma. Then  for $n > n_0$, we have
\begin{equation}
n(D^*(\overline{\boldeta})-3\epsilon') > \frac{(1+\epsilon)\log((M-1)L)}{D^*(\overline{\boldeta})} [D^*(\overline{\boldeta})-3\epsilon'] = \log((M-1)L).
\end{equation}
Hence, we get that for $n >n_0$,
\begin{equation}
P\left(n D^*(\overline{\boldeta})-3\epsilon'n < \log((M-1)L)\right) = 0.
\end{equation}

\vspace*{.3in}

\noindent (ii) Consider next the first term in (\ref{eqn:Etau_Zl}):
\begin{equation}
P\left(\log \mathcal{H}_l(\overline{\boldupsilon},{\bf{n_0}}) < -\epsilon'n\right).
\end{equation}
The right-hand side inside the probability goes to negative infinity whereas, the left-hand side is a constant. Hence, the probability of the event under study is zero for all sufficiently large $n$ (independent of $L$).

\vspace*{.3in}

\noindent (iii) Next consider the second term in (\ref{eqn:Etau_Zl}). For convenience define $\mathcal{F}_i(\boldkappa_i) := \boldeta_i^T \boldkappa_i - \mathcal{A}_i(\boldeta_i)$, the Fenchel dual of $\mathcal{A}_i$ evaluated at $\boldkappa_i$. We then have
\begin{eqnarray}
\lefteqn{P\left(-\frac{1}{n}\log\mathcal{H}_l({\bf{Y}}+\overline{\boldupsilon},{\bf{N}}+{\bf{n_0}})-\sum\limits_{i=1}^K\lambda_i^*\mathcal{F}_i(\boldkappa_i)< -\epsilon'\right)}\nonumber\\
& \leq & P\left(-\frac{1}{n}\log\mathcal{H}_l({\bf{Y}}+\overline{\boldupsilon},{\bf{N}}+{\bf{n_0}})-\sum\limits_{i=1}^K\lambda_i^*\mathcal{F}_i(\boldkappa_i)< -\epsilon', \left|\frac{N_{i'}^n}{n}-\lambda_{i'}^*\right| \leq \epsilon_1, \left\| \frac{{\bf{Y}}_{i'}^n}{N_{i'}^n}-\boldkappa_{i'}\right\|_{\infty} \leq \epsilon_2, \forall i'\right)\nonumber\\
& & + \sum\limits_{i'} P\left(\left|\frac{N_{i'}^n}{n}-\lambda^*_{i'}\right| >\epsilon_1\right) + \sum\limits_{i'} P\left(\left\|\frac{{\bf{Y}}_{i'}^n}{N_{i'}^n}-\boldkappa_{i'}\right\|_{\infty} > \epsilon_2 \right), \label{eqn:Etau_ub2}
\end{eqnarray}
where $\epsilon_1, \epsilon_2$ are suitable constants that will be specified soon. Under the conditions
\[
  \left|\frac{N_{i'}^n}{n}-\lambda^*_{i'}\right| < \epsilon_1 \mbox{ and } \left\|\frac{{\bf{Y}}_{i'}^n}{N_{i'}^n}-\boldkappa_{i'}\right\|_{\infty} \leq \epsilon_2,
\]
we will follow the steps leading to (\ref{eqn:pre-laplace}) to lower bound $-\frac{1}{n}\log\mathcal{H}_l({\bf{Y}}+\overline{\boldupsilon},{\bf{N}}+{\bf{n_0}})$. First observe that
\begin{eqnarray}
\lefteqn{-\frac{1}{n}\log\mathcal{H}_l({\bf{Y}}+\overline{\boldupsilon},{\bf{N}}+{\bf{n_0}})}\nonumber\\
 &=& \frac{1}{n} \log \left\{\int\limits_{\overline{\boldeta}' \in \Theta_l}\exp\left\{\sum\limits_{i=1}^K \left( ({\boldeta}_i')^{T}({\bf{Y}}_i^n+\boldupsilon_i)-(N_i^n+n_{0i})\mathcal{A}_i(\boldeta'_i) \right) \right\}d\overline{\boldeta}'\right\}\nonumber\\
     &=& \frac{1}{n} \log \int\limits_{\overline{\boldeta}' \in \Theta_l}
     \exp\left\{n\sum\limits_{i=1}^K \left( \frac{N_i^n}{n} \left( (\boldeta_i')^T \frac{{\bf{Y}}_i^n}{N_i^n} - \mathcal{A}_i(\boldeta_i') \right)
     + (\boldeta_i')^T \frac{\boldupsilon_i}{n} - \frac{n_{0i}}{n}\mathcal{A}_i(\boldeta_i') \right) \right\}d\overline{\boldeta}'.\quad\quad\quad
     \label{eqn:Hl-lb}
\end{eqnarray}
Note that ${\boldeta}_i(\boldkappa_i)$  optimises the function $\boldeta'_i \mapsto (\boldeta')^T {\boldkappa}_i - \mathcal{A}_i(\boldeta'_i)$, for each $i = 1, \ldots, K$, over the set $\Theta_l$, since $\overline{\boldeta} \in \Theta_l$. As before, we leverage this.

Fix $\delta > 0$. Almost surely, there is a $C_{\delta} > 0$ such that for all sufficiently large $n$, $|| \hat{\boldkappa}_i - \boldkappa_i||_{\infty} \leq \epsilon_2$, and further, for all $\overline{\boldeta}' \in B_{\delta}(\overline{\boldeta})$, we have:
\begin{eqnarray*}
 \left| \sum_{i=1}^K \left( \frac{N_i^n}{n} - \lambda_i^* \right) ( (\boldeta_i')^T \boldkappa_i  - \mathcal{A}_i(\boldeta_i') ) \right| & \leq & C_{\delta} \epsilon_1 \\
 \left| (\boldeta_i')^T \boldupsilon_i - n_{0i} \mathcal{A}_i(\boldeta_i') \right| & \leq & C_{\delta} \\
 \left| \sum_{i=1}^K \frac{N_i^n}{n} (\boldeta_i')^T (\hat{\boldkappa}_i - \boldkappa_i ) \right| & \leq & C_{\delta} \epsilon_2 \\
 \left| \sum_{i=1}^K \left( (\boldeta_i')^T \boldkappa_i - \mathcal{A}_i(\boldeta_i') - \left( \boldeta_i^T \boldkappa_i - \mathcal{A}_i(\boldeta_i) \right)  \right) \right| & = & \left| \sum_{i=1}^K \left( (\boldeta_i')^T \boldkappa_i - \mathcal{A}_i(\boldeta_i') - \mathcal{F}(\boldkappa_i)  \right) \right| \leq \tau(\delta)
\end{eqnarray*}
where in the last inequality $\tau(\delta) \rightarrow 0$ as $\delta \rightarrow 0$ due to the continuity of $\mathcal{A}_i(\cdot)$. Further, we can lower bound the integral in (\ref{eqn:Hl-lb}) by restricting the integral to the set $\Theta_l \cap B_{\delta}(\overline{\boldeta})$. Putting these ideas together, we get that (\ref{eqn:Hl-lb}) is lower bounded by:
\begin{eqnarray}
  & \geq & \frac{1}{n} \log \int\limits_{\overline{\boldeta}' \in \Theta_l \cap B_{\delta}(\overline{\boldeta})}
     \exp\left\{n\sum\limits_{i=1}^K\frac{N_i^n}{n} \left( (\boldeta_i')^T \frac{{\bf{Y}}_i^n}{N_i^n} - \mathcal{A}_i(\boldeta_i') \right)
     - K C_{\delta} \right\}d\overline{\boldeta}' \\
  & = & \frac{1}{n} \log \int\limits_{\overline{\boldeta}' \in \Theta_l \cap B_{\delta}(\overline{\boldeta})}
     \exp\left\{n\sum\limits_{i=1}^K \lambda_i^* \left( (\boldeta_i')^T {\boldkappa}_i - \mathcal{A}_i(\boldeta_i') \right) \right. \nonumber \\
     & & \hspace*{2cm} + \left. n \sum_{i=1}^K \left( \frac{N_i^n}{n} - \lambda_i^* \right) ((\boldeta_i')^T \boldkappa_i - \mathcal{A}_i(\boldeta_i'))+ n \sum_{i=1}^K \frac{N_i^n}{n} (\boldeta_i')^T (\hat{\boldkappa}_i - \boldkappa_i ) - K C_{\delta} \right\} d\overline{\boldeta}' \nonumber \\
  & = & \frac{1}{n} \log \int\limits_{\overline{\boldeta}' \in \Theta_l \cap B_{\delta}(\overline{\boldeta})}
     \exp\left\{n\sum\limits_{i=1}^K \lambda_i^* \mathcal{F}(\boldkappa_i)
     + n\sum\limits_{i=1}^K \lambda_i^* \left( (\boldeta_i')^T {\boldkappa}_i - \mathcal{A}_i(\boldeta_i') - \mathcal{F}(\boldkappa_i)\right)
     \right. \nonumber \\
     & & \hspace*{2cm} + \left. n \sum_{i=1}^K \left( \frac{N_i^n}{n} - \lambda_i^* \right) ( (\boldeta_i')^T \boldkappa_i  - \mathcal{A}_i(\boldeta_i'))+ n \sum_{i=1}^K \frac{N_i^n}{n} (\boldeta_i')^T (\hat{\boldkappa}_i - \boldkappa_i ) - K C_{\delta} \right\} d\overline{\boldeta}' \nonumber \\
  & \geq & \sum\limits_{i=1}^K \lambda_i^* \mathcal{F}(\boldkappa_i)  + \frac{1}{n} \log \left( \textsf{Leb}\left(\overline{\boldeta}' \in \Theta_l \cap B_{\delta}(\overline{\boldeta}) \right) \right)
   - \tau(\delta) - C_{\delta} \epsilon_1 - C_{\delta} \epsilon_2 - \frac{K C_{\delta}}{n} \nonumber
\end{eqnarray}
Using this lower bound, we can now upper bound the first term in (\ref{eqn:Etau_ub2}) as
\begin{align}
P\Bigg(-\frac{1}{n}\log\mathcal{H}_l({\bf{Y}}+\overline{\boldupsilon},{\bf{N}}+{\bf{n_0}})- \sum\limits_{i=1}^K\lambda_i^*\mathcal{F}(\boldkappa_i)< -\epsilon' , \left|\frac{N_{i'}^n}{n}-\lambda_{i'}^*\right| \leq \epsilon_1,\left\|\frac{{\bf{Y}}_{i'}^n}{N_{i'}^n}-\boldkappa_{i'}\right\|_{\infty} \leq \epsilon_2, \forall i'\Bigg)\nonumber\\
\leq P\left(
\frac{1}{n} \log \left( \textsf{Leb}\left(\overline{\boldeta}' \in \Theta_l \cap B_{\delta}(\overline{\boldeta}(\overline{\boldkappa})) \right) \right)
   - \tau(\delta) - C_{\delta} \epsilon_1 - K C_{\delta} \epsilon_2 - \frac{K C_{\delta}}{n} < -\epsilon'\right);
   \label{eqn:prob-bound1}
\end{align}
the event inside the probability argument on the right-hand side of the above inequality will not occur, via suitable choices of $\delta$, $\epsilon_1$ and $\epsilon_2$, for all sufficiently large $n$ dependent on $\delta, \epsilon_1, \epsilon_2, \epsilon'$ but independent of $L$. For all such $n$, the probability on the left-hand side of \eqref{eqn:prob-bound1} is zero.

The third term in \eqref{eqn:Etau_ub2} is upper bounded by  $C/n^3$ for some constant $C$ by Lemma \ref{lem:concentration} (in fact it decays faster, $O(1/n^4)$). The constant $C$ is independent of $L$.

We now argue that there is an $\mathcal{N}_0$ independent of $L$ such that, for all $n \geq \mathcal{N}_0$,  the second term in \eqref{eqn:Etau_ub2} is upper bounded by $C_1/n^3$ for a suitable constant $C_1$ which is also independent of $L$.

Let us use the notation $n^a(m)$ to be the number of active samplings up to time $m$. First observe that, by triangle inequality,
\begin{equation}
  \label{eqn:Nin/n-violation}
  \left| \frac{N_{i'}^n}{n} - \lambda_{i'}^* \right| > \epsilon_1 \Rightarrow \left| \frac{N_{i'}^{n,a}}{n^a(n)} - \lambda_{i'}^* \right| > \frac{\epsilon_1}{2} \mbox{ or } \left| \frac{N_{i'}^n}{n} - \frac{N_{i'}^{n,a}}{n^a(n)} \right| > \frac{\epsilon_1}{2}.
\end{equation}
Choose a sufficiently small $\epsilon_3$, and a sufficiently small $\epsilon_2$ (a new $\epsilon_2$ that may depend on $\epsilon_3$), so that the following hold:
\begin{itemize}
  \item $3(K-1)\epsilon_3 < \epsilon_1/2$;
  \item for every $\overline{\boldkappa}'$ with $\max_i \|\boldkappa'_i - \boldkappa_i\|_{\infty} < \epsilon_2$, we have $\overline{\boldeta}(\boldkappa') \in \Theta_l$ (due to the openness of $\Theta_l$ and continuity of the $\overline{\boldeta}(\cdot)$ mapping);
  \item for every $\overline{\boldkappa}'$ with $\max_i \|\boldkappa'_i - \boldkappa_i\|_{\infty} < \epsilon_2$, we have $\max_i |\lambda_i^*(\overline{\boldeta}(\overline{\boldkappa})) - \lambda_i^*| < \epsilon_3$ (due to the continuities of the $\boldlambda^*(\cdot)$ and the $\overline{\boldeta}(\cdot)$ mappings).
\end{itemize}
Consider the conditions
\begin{equation}
  \label{eqn:Nin/n-violation-alternate}
  \sup_{m \geq n} \left| \frac{n^a(m)}{m} - \gamma \right| \leq \gamma \epsilon_1 \quad \mbox{ and }
  \sup_{m: n^a(m) \geq 3 \epsilon_3 \gamma (1-\epsilon_1) n} \max_{i'} \left\| \frac{{\bold{Y}^m_{i'}}}{N^m_{i'}} - \boldkappa_{i'} \right\|_{\infty} \leq \epsilon_2.
\end{equation}
Under these conditions, apply Lemma~\ref{lemma:lambda_conv} with $\epsilon = \epsilon_3$, $n_0 = 3 \epsilon_3 \gamma (1-\epsilon_1) n$, and $n_{\epsilon_3} = \gamma (1-\epsilon_1) n$ for $n \geq \mathcal{N}_0 := 1/(3\epsilon_3^2(1-\epsilon_1)\gamma)$, use the second and the third bullets above, and we get
\[
  \sup_{m: n^a(m) \geq \gamma (1-\epsilon_1) n} \left| \frac{N_{i'}^{m,a}}{n^a(m)} - \lambda_{i'}^* \right| \leq 3(K-1)\epsilon_3.
\]
In particular, since we are operating under $n^a(m)/m \geq \gamma (1 - \epsilon_1)$ for all $m \geq n$, taking $m = n$ in the above displayed equation, we get
\[
  \left| \frac{N_{i'}^{n,a}}{n^a(n)} - \lambda_{i'}^* \right| \leq 3(K-1)\epsilon_3,
\]
a condition which is incompatible with $\left| \frac{N_{i'}^{n,a}}{n^a(n)} - \lambda_{i'}^* \right| > \epsilon_1/2$ in view of the first bullet above. This contradiction, along with \eqref{eqn:Nin/n-violation}, \eqref{eqn:Nin/n-violation-alternate}, shows that, for $n \geq \mathcal{N}_0$,
\begin{eqnarray}
  \left| \frac{N_{i'}^n}{n} - \lambda_{i'}^* \right| > \epsilon_1 & \Rightarrow &
  \sup_{m \geq n} \left| \frac{n^a(m)}{m} - \gamma \right| > \gamma \epsilon_1 \nonumber \\
  & & \mbox{ or} \left[ \sup_{m \geq n} \frac{n^a(m)}{m} \leq \gamma(1+\epsilon_1) \quad \mbox{ and }
  \sup_{m: n^a(m) \geq 3 \epsilon_3 \gamma (1-\epsilon_1) n} \max_{i'} \left\| \frac{{\bold{Y}^m_{i'}}}{N^m_{i'}} - \boldkappa_{i'} \right\|_{\infty} > \epsilon_2 \right] \nonumber \\
  \label{eqn:Nin/n-violation-2}
  & & \mbox{ or } \left| \frac{N_{i'}^n}{n} - \frac{N_{i'}^{n,a}}{n^a(n)} \right| > \frac{\epsilon_1}{2}.
\end{eqnarray}

\noindent (a) By Bernstein inequality and the union bound, the first event has probability decaying exponentially in $n$.

\noindent (b) Using $n^a(m) \leq \gamma m(1+\epsilon_1)$, if the second event above holds, then we have
\[
  \sup_{m: m \geq 3 n \epsilon_3 (1-\epsilon_1) / (1+\epsilon_1) } \max_{i'} \left\| \frac{{\bold{Y}^m_{i'}}}{N^m_{i'}} - \boldkappa_{i'} \right\|_{\infty} > \epsilon_2;
\]
by Lemma~\ref{lem:concentration} and the union bound, its probability is upper bounded by
\[
  \sum_{m: m \geq 3 n \epsilon_3 (1-\epsilon_1) / (1+\epsilon_1) } \frac{C}{m^4} \quad \leq \quad \frac{C_2}{n^3}
\]
for some suitable constant $C_2$.

\noindent (c) Let us now address the third term in \eqref{eqn:Nin/n-violation-2}. Using \eqref{eqn:split_frac}, we get
\[
\left| \frac{N_{i'}^n}{n} - \frac{N_{i'}^{n,a}}{n^a(n)} \right| = \frac{N_{i'}^{n,a}}{n^a(n)} \left| \frac{n^a(n)}{n} \frac{1}{N_{i'}^{n,a}} \left( \sum_{k=1}^{N_{i'}^{n,a}} V_k^{(i')} + \overline{V}_i \right) - 1 \right|
\]
where $k$ runs over the indices involving the choice of $i'$ in an active slot. Using $\frac{N_{i'}^{n,a}}{n^a(n)} \leq 1$, we get
\begin{eqnarray*}
  \left| \frac{N_{i'}^n}{n} - \frac{N_{i'}^{n,a}}{n^a(n)} \right| > \frac{\epsilon_1}{2}
  & \Rightarrow & \left| \frac{n^a(n)}{n} \frac{1}{N_{i'}^{n,a}} \left( \sum_{k=1}^{N_{i'}^{n,a}} V_k^{(i')} + \overline{V}_i \right) - 1 \right| > \frac{\epsilon_1}{2} \\
  & \Rightarrow & \left| \frac{n^a(n)}{n}  - \gamma \right| > \gamma \delta \quad \mbox{(for a $\delta$ to be chosen soon)} \\
  & & \mbox{or } \left[ \left| \frac{n^a(n)}{n} - \gamma \right| \leq \gamma \delta \quad \mbox{ and } \quad \left| \frac{1}{N_{i'}^{n,a}} \sum_{k=1}^{N_{i'}^{n,a}} V_k^{(i')} - \frac{1}{\gamma} \right| > \frac{\delta}{\gamma} \right] \\
  & & \mbox{or } \left[ \left| \frac{n^a(n)}{n} - \gamma \right| \leq \gamma \delta \quad \mbox{ and } \quad \left| \frac{1}{N_{i'}^{n,a}} \sum_{k=1}^{N_{i'}^{n,a}} V_k^{(i')} - \frac{1}{\gamma} \right| \leq \frac{\delta}{\gamma} \right. \\
  & & \quad \quad
  \left. \mbox{ and } \quad
  \frac{n^a(n)}{n} \frac{\overline{V}_i}{N_{i'}^{n,a}} > \frac{\epsilon_1}{4} \right] \\
  & & \mbox{or } \left[ \left| \frac{n^a(n)}{n} - \gamma \right| \leq \gamma \delta \quad \mbox{ and } \quad \left| \frac{1}{N_{i'}^{n,a}} \sum_{k=1}^{N_{i'}^{n,a}} V_k^{(i')} - \frac{1}{\gamma} \right| \leq \frac{\delta}{\gamma} \right. \\
  & & \quad \quad
  \left. \mbox{ and } \quad
  \left| \frac{n^a(n)}{n} \frac{1}{N_{i'}^{n,a}} \sum_{k=1}^{N_{i'}^{n,a}} V_k^{(i')} - 1 \right| > \frac{\epsilon_1}{4} \right].
\end{eqnarray*}
Choose $\delta$ sufficiently small so that $(1+\delta)^2 < 1 + \epsilon_1/4$ and $(1-\delta)^2 > 1 - \epsilon_1/4$. The first of these events has exponentially (in $n$) small probability for all $n$ (Bernstein inequality). By Lemma~\ref{lemma:lambda_conv}, for all $n \geq \mathcal{N}_0$, we have $N_{i'}^{n,a} \geq (n^a(n))^{\beta}/2$. The Chernoff bound then gives that the second event too has exponentially (in $n$) small probability. The random variable $\overline{V}_i$ is stochastically dominated by a geometric random variable and hence the third event has exponentially small probability for all $n \geq \mathcal{N}_0$ and Lemma~\ref{lemma:lambda_conv}. Finally, by the choice of $\delta$, the fourth event cannot occur.

The above arguments (a)-(c) establish that the probability of the event $\left| \frac{N_{i'}^n}{n} - \lambda_{i'}^* \right| > \epsilon_1$, i.e., the second term in \eqref{eqn:Etau_ub2}, is also upper bounded by $C_1/n^3$ for some constant $C_1$ independent of $L$, for all $n \geq \mathcal{N}_0$.

Note that, with the above, we have also established that the second term in (\ref{eqn:Etau_Zl}) is upper bounded by $C/n^3$, for some $C_3$, for all $n \geq \mathcal{N}_0$.

\vspace*{.3cm}

\noindent (iv) Finally, consider the third term in (\ref{eqn:Etau_Zl}). The following chain of inequalities are self-evident:
\begin{eqnarray}
\lefteqn{P\left(-\sup\limits_{\overline{\boldeta}' \in \Theta_{-l}}
\sum\limits_{i=1}^K N_i^n\left(\boldeta_i'^T\hat{\boldkappa}_i - \mathcal{A}_i(\boldeta_i')\right)+\sup\limits_{\overline{\boldeta}' \in \Theta_{-l}} n \sum\limits_{i=1}^K\lambda_i^*\left(\boldeta_i'^T{\boldkappa}_i - \mathcal{A}_i(\boldeta_i')\right) < -\epsilon'n \right)}\nonumber\\
&=& P\left(\sup\limits_{\overline{\boldeta}' \in \Theta_{-l}} \sum\limits_{i=1}^K \frac{N_i^n}{n}\left(\boldeta_i'^T\hat{\boldkappa}_i - \mathcal{A}_i(\boldeta_i')\right)- \sup\limits_{\overline{\boldeta}' \in \Theta_{-l}} \sum\limits_{i=1}^K\lambda_i^*\left(\boldeta_i'^T{\boldkappa}_i - \mathcal{A}_i(\boldeta_i')\right) >\epsilon' \right)\nonumber\\
& = & P\left(\sup\limits_{\overline{\boldeta}' \in \Theta_{-l}} \sum\limits_{i=1}^K \frac{N_i^n}{n}\left( -D(\hat{\boldeta}_i \parallel \boldeta_i') + \hat{\boldeta}_i^T \hat{\boldkappa}_i - \mathcal{A}_i(\hat{\boldeta}_i) \right) \right. \nonumber \\
& & \hspace*{1cm} \left. -
\sup\limits_{\overline{\boldeta}' \in \Theta_{-l}} \sum\limits_{i=1}^K \lambda_i^* \left( -D(\boldeta_i \parallel \boldeta_i') + \boldeta_i^T \boldkappa_i - \mathcal{A}_i(\boldeta_i) \right) >\epsilon'\right)\label{eqn:sup_bounds}\\
& = & P\left( \sum\limits_{i=1}^K \frac{N_i^n}{n}\left( - \inf\limits_{\overline{\boldeta}' \in \Theta_{-l}} D(\hat{\boldeta}_i \parallel \boldeta_i') + \hat{\boldeta}_i^T \hat{\boldkappa}_i - \mathcal{A}_i(\hat{\boldeta}_i) \right) \right. \nonumber \\
& & \hspace*{1cm} \left.
- \sum\limits_{i=1}^K \lambda_i^* \left( - \inf\limits_{\overline{\boldeta}' \in \Theta_{-l}} D(\boldeta_i \parallel \boldeta_i') + \boldeta_i^T \boldkappa_i - \mathcal{A}_i(\boldeta_i) \right) >\epsilon'\right)\label{eqn:sup_bounds-2}\\
& = & P\left( \sum\limits_{i=1}^K \left( \frac{N_i^n}{n} - \lambda_i^* \right) \left( - \inf\limits_{\overline{\boldeta}' \in \Theta_{-l}} D(\hat{\boldeta}_i \parallel \boldeta_i') + \hat{\boldeta}_i^T \hat{\boldkappa}_i - \mathcal{A}_i(\hat{\boldeta}_i) \right) \right. \nonumber \\
& & \hspace*{1cm} \left.
- \sum\limits_{i=1}^K \lambda_i^* \left( \inf\limits_{\overline{\boldeta}' \in \Theta_{-l}} D(\hat{\boldeta}_i \parallel \boldeta_i')  - \inf\limits_{\overline{\boldeta}' \in \Theta_{-l}} D(\boldeta_i \parallel \boldeta_i') \right) \right. \nonumber \\
& & \hspace*{1cm} \left. + \sum\limits_{i=1}^K \lambda_i^*
\left( \hat{\boldeta}_i^T \hat{\boldkappa}_i - \mathcal{A}_i(\hat{\boldeta}_i)
- \boldeta_i^T \boldkappa_i + \mathcal{A}_i(\boldeta_i) \right) >\epsilon'\right)\label{eqn:sup_bounds-3}\\
&\leq& P\Bigg(
\sum\limits_{i=1}^K \left( \frac{N_i^n}{n} - \lambda_i^* \right) \left( - \inf\limits_{\overline{\boldeta}' \in \Theta_{-l}} D(\hat{\boldeta}_i \parallel \boldeta_i') + \hat{\boldeta}_i^T \hat{\boldkappa}_i - \mathcal{A}_i(\hat{\boldeta}_i) \right) \nonumber \\
& & \hspace*{1cm} \left.
- \sum\limits_{i=1}^K \lambda_i^* \left( \inf\limits_{\overline{\boldeta}' \in \Theta_{-l}} D(\hat{\boldeta}_i \parallel \boldeta_i')  - \inf\limits_{\overline{\boldeta}' \in \Theta_{-l}} D(\boldeta_i \parallel \boldeta_i') \right) \right. \nonumber \\
& & \hspace*{1cm} + \sum\limits_{i=1}^K \lambda_i^*
\left( \hat{\boldeta}_i^T \hat{\boldkappa}_i - \mathcal{A}_i(\hat{\boldeta}_i)
- \boldeta_i^T \boldkappa_i + \mathcal{A}_i(\boldeta_i) \right)
>\epsilon',\nonumber\\
& & \hspace*{4cm}  \left|\frac{N_{i'}^n}{n}-\lambda_{i'}^*\right| \leq \epsilon_1,\left\|\frac{{\bf{Y}}_{i'}^n}{N_{i'}^n}-\boldkappa_{i'}\right\|_{\infty} \leq \epsilon_2 , \forall i' \Bigg) \nonumber \\
& & \qquad + \sum\limits_{i'} P\left(\left|\frac{N_{i'}^n}{n}-\lambda^*_{i'}\right| >\epsilon_1\right) + \sum\limits_{i'} P\left(\left\|\frac{{\bf{Y}}_{i'}^n}{N_{i'}^n}-\boldkappa_{i'}\right\|_{\infty} > \epsilon_2 \right). \label{eqn:convergences}
\end{eqnarray}
Following the approach that led to the bound for (\ref{eqn:Etau_ub2}), since
$
  \boldeta_i \mapsto \inf\limits_{\overline{\boldeta}' \in \Theta_{-l}} D(\boldeta_i \parallel \boldeta_i')
$
is a continuous function by Lemma~\ref{lem:Fl-continuous}, we obtain that (\ref{eqn:convergences}) is also bounded by $C'/n^3$. This establishes Lemma \ref{lemma:exp_bounds_Zl}.
}}
\end{IEEEproof}

Proof of result in (\ref{eqn:upper_bound_Etau}): A sufficient condition to establish the convergence of expected stopping time is to show the second moment condition:
\begin{equation}
\limsup\limits_{L \rightarrow \infty} E\left[\left(\frac{\tau(\pi_{SMF}(L,\gamma,\beta))}{\log(L)}\right)^2\right] < \infty.
\end{equation}
We now proceed to establish this. Define
\begin{equation}
u(L) := \left(\frac{(1+\epsilon)\log((M-1)L)}{D^*(\overline{\boldeta})\log(L)}+\frac{1}{\log(L)}\right)^2.
\end{equation}
We then have
\begin{eqnarray}
\lefteqn{\limsup\limits_{L \rightarrow \infty} E\left[\left(\frac{\tau(\pi_{SMF}(L,\gamma,\beta))}{\log(L)}\right)^2\right]}\nonumber\\
 &=& \limsup\limits_{L \rightarrow \infty} \int\limits_{x \geq 0} P\left(\frac{\tau(\pi_{SMF}(L,\gamma,\beta))}{\log(L)} > \sqrt{x}\right)dx \nonumber\\
 &\leq& \limsup\limits_{L \rightarrow \infty} \int\limits_{x \geq 0} P\left(\tau^l(\pi_{SMF}(L,\gamma,\beta)) > \lfloor \sqrt{x}\log(L)\rfloor\right) dx \nonumber\\
 &\leq& \limsup\limits_{L \rightarrow \infty} \left[u(L)+\int\limits_{x \geq u(L)} P\left(\tau^l(\pi_{SMF}(L,\gamma,\beta)) > \lfloor \sqrt{x}\log(L)\rfloor\right) dx\right] \nonumber\\
&\leq& \left(\frac{1+2\epsilon}{D^*(\overline{\boldeta})}\right)^2 + \limsup\limits_{L \rightarrow \infty} \sum\limits_{n \geq \lfloor \sqrt{u(L)}\log(L)\rfloor}
\left( \left(\frac{n+1}{\log(L)}\right)^2 - \left(\frac{n}{\log(L)}\right)^2 \right) P\left(\tau^l(\pi_{SMF}(L,\gamma,\beta)) > n \right)\nonumber\\
& & \label{eqn:int_sum}\\
&\leq& \left(\frac{1+2\epsilon}{D^*(\overline{\boldeta})}\right)^2 + \limsup\limits_{L \rightarrow \infty} \sum\limits_{n \geq \lfloor \sqrt{u(L)}\log(L)\rfloor} \left(\frac{2n+1}{(\log(L))^2}\right) P\left(Z_l(n) < \log((M-1)L) \right) \nonumber\\
&\leq& \left(\frac{1+2\epsilon}{D^*(\overline{\boldeta})}\right)^2 + \limsup\limits_{L \rightarrow \infty} \sum\limits_{n \geq \lfloor \sqrt{u(L)}\log(L)\rfloor} \left(\frac{2n+1}{(\log(L))^2}\right) \frac{B}{n^3} \quad \mbox{ (for $L$ sufficiently large)} \label{eqn:use_n^3}\\
&<& \infty; \nonumber
\end{eqnarray}
inequality in (\ref{eqn:int_sum}) is obtained using the fact that $P\left(\tau^l(\pi_{SMF}(L,\gamma,\beta)) > \lfloor \sqrt{x}\log(L)\rfloor\right)$ is constant in the interval
\begin{equation*}
x \in \left[\left(\frac{n}{\log(L)}\right)^2, \left(\frac{n+1}{\log(L)}\right)^2 \right];
\end{equation*}
for inequality (\ref{eqn:use_n^3}), we have from Lemma \ref{lemma:exp_bounds_Zl} that
\begin{equation}
\text{for all } n \geq \frac{(1+\epsilon)\log((M-1)L)}{D^*(\overline{\boldeta})},
\end{equation}
$P(Z_l(n) < \log((M-1)L)) < B/n^3$. This completes the proof.

\subsection{Proof of (\ref{eqn:upper_bound_cost})}
To prove this, observe that
\begin{eqnarray}
E[C\left(\pi_{SMF}\left(L,\gamma\right)\right)] &=& E\Big[\tau\left(\pi_{SMF}\left(L,\gamma\right)\right) + \sum\limits_{l=1}^{\tau\left(\pi_{SMF}\left(L,\gamma\right)\right)-1} g\left(A_l,A_{l+1}\right)\Big]\nonumber\\
							 &\leq & E[\tau\left(\pi_{SMF}\left(L,\gamma\right)\right)]+ g_{\max}E\Big[\sum\limits_{l=1}^{\tau\left(\pi_{SMF}\left(L,\gamma\right)\right)-1} 1_{\{A_l \neq A_{l+1}\}}\Big]\nonumber\\
							 &\leq & E[\tau\left(\pi_{SMF}\left(L,\gamma\right)\right)] + g_{\max}E\Big[\sum\limits_{l=1}^{\tau\left(\pi_{SMF}\left(L,\gamma\right)\right)} U_{l+1}\Big]\nonumber\\
							 &=& E[\tau\left(\pi_{SMF}\left(L,\gamma\right)\right)] + g_{\max} \gamma E[\tau\left(\pi_{SMF}\left(L,\gamma\right)\right)]\nonumber\\
							 & = & E[\tau\left(\pi_{SMF}\left(L,\gamma\right)\right)]\left(1+g_{\max} \gamma\right),\nonumber
\end{eqnarray}
where in the penultimate equality, we have used Doob's optional stopping theorem. Divide by $\log L$ and let $L \rightarrow \infty$ to get the required result. This completes the proof of (\ref{eqn:upper_bound_cost}), completes the proof of all three results in the proposition, and thus finishes the proof of Proposition~\ref{prop:achievability}.

\section{Verification of Continuous Selection in Some Examples}\label{app:verifyCS}

In this section we shall show that the odd arm identification problem and the best arm identification problem admit continuous selections. 

\subsection{Odd Arm Identification Problem}
In order to show that the correspondence $\boldlambda \mapsto \boldlambda^*(\overline{\boldeta})$ admits a continuous selection, we show that the function $F(\cdot, \boldeta)$ defined in (\ref{eqn:F-definition}) is strictly concave for each $\boldeta \in \Theta_{l}$. We begin with
\begin{equation}\label{eqn:lambda*_def}
\boldlambda^*(\overline{\boldeta}) = \argmax\limits_{\boldlambda \in \mathcal{P}(K)} F(\boldlambda, \boldeta) = \argmax\limits_{\boldlambda \in \mathcal{P}(K)} \inf\limits_{\overline{\boldeta}'\in \Theta_{-l}} \sum\limits_{i=1}^K \lambda_i D(\boldeta_i \parallel \boldeta_i').
\end{equation}

Recall the example discussed in Section (\ref{eg:odd_arm}). We observe from \cite{8755796} that $\boldlambda^*(\overline{\boldeta})$ is of the form
\begin{equation*}
\left[\frac{1-\lambda_l}{K-1}, \cdots \frac{1-\lambda_l}{K-1}, \lambda_l,\frac{1-\lambda_l}{K-1}, \cdots \frac{1-\lambda_l}{K-1} \right]
\end{equation*}
and the expression for $D^*(\cdot)$ in (\ref{eqn:D*}) can be reduced to
\begin{equation}\label{eqn:D_odd}
D^*(\overline{\boldeta}) = \max\limits_{\lambda_l \in [0,1]} \lambda_lD(\boldeta_1||\tilde{\boldeta}) + (1-\lambda_l)\frac{K-2}{K-1} D(\boldeta_2||\tilde{\boldeta})
\end{equation}
where $\tilde{\boldeta} = \boldeta(\tilde{\boldkappa})$ and
\begin{equation}
\tilde{\boldkappa} = \frac{\lambda_l\boldkappa_1 + (1-\lambda_l)\frac{K-2}{K-1}\boldkappa_2}{\lambda_l + (1-\lambda_l)\frac{K-2}{K-1}}.
\end{equation}

To establish the  strict concavity, we show that the second derivative of the objective function in (\ref{eqn:lambda*_def}) is strictly negative for all values of $\lambda_l$. Using the result in (\ref{eqn:D_odd}), we redefine the objective function in (\ref{eqn:lambda*_def}) as
\begin{equation}
\Phi(\lambda_l) = \lambda_lD(\boldeta_1||\tilde{\boldeta}) + (1-\lambda_l)\frac{K-2}{K-1} D(\boldeta_2||\tilde{\boldeta}).
\end{equation}
Taking the first derivative,
\begin{eqnarray}
\frac{d\Phi}{d\lambda_l} &=& D\left(\boldeta_1||\tilde{\boldeta}\right) - \frac{K-2}{K-1}D\left(\boldeta_2||\tilde{\boldeta}\right)+ \left[\lambda_l \nabla_{\tilde{\boldeta}}D\left(\boldeta_1||\tilde{\boldeta}\right)+\left(1-\lambda_l\right)\frac{K-2}{K-1}\nabla_{\tilde{\boldeta}}D\left(\boldeta_2||\tilde{\boldeta}\right)\right]^T\frac{d\tilde{\boldeta}}{d\lambda_l}\nonumber\\
					   &=& D\left(\boldeta_1||\tilde{\boldeta}\right) - \frac{K-2}{K-1}D\left(\boldeta_2||\tilde{\boldeta}\right). \label{eqn:Phi_first_deriv}
\end{eqnarray}
The equality in (\ref{eqn:Phi_first_deriv}) follows from the fact that the $
\boldeta'$ that attains the infimum in (\ref{eqn:lambda*_def}) is $\tilde{\boldeta}$. Differentiating again by applying chain rule and using the result that $\nabla_{\boldeta_2} D(\boldeta_1||\boldeta_2) = \boldkappa_2-\boldkappa_1$ we get
\begin{equation}\label{eqn:sec_deriv}
\frac{d^2\Phi}{d\lambda_l^2} = \left[(\tilde{\boldkappa}-\boldkappa_1) - \frac{K-2}{K-1}(\tilde{\boldkappa}-\boldkappa_2)\right]^T \frac{d\tilde{\boldeta}}{d\lambda_l}.
\end{equation}
Observe that
\begin{eqnarray}
\frac{d\tilde{\boldeta}}{d\lambda_l} &=& D_{\tilde{\boldkappa}}\tilde{\boldeta} \cdot \frac{d\tilde{\boldkappa}}{d\lambda_l}\label{eqn:eta_lambda}\\
                                     &=& \textsf{Hess}(\mathcal{F}(\tilde{\boldkappa})) \cdot \frac{-1}{\lambda_l + \frac{K-2}{K-1}(1-\lambda_l)}\left((\tilde{\boldkappa}-\boldkappa_1) - \frac{K-2}{K-1}(\tilde{\boldkappa}-\boldkappa_2)\right)\label{eqn:hess}.
\end{eqnarray}
Equality in (\ref{eqn:eta_lambda}) is obtained using chain rule for differentiation and $D_{\tilde{\boldkappa}} \tilde{\boldeta}$ is the matrix $\left(\frac{\partial}{\partial \tilde{\boldkappa}_j} \tilde{\boldeta}_i\right)_{1 \leq i,j \leq d}$. From (\ref{dual}), we recognise that $D_{\tilde{\boldkappa}} \tilde{\boldeta} = \textsf{Hess}(F(\tilde{\boldkappa}))$, the Hessian of the function $F(\boldkappa)$ with respect to $\boldkappa$ evaluated at $\tilde{\boldkappa}$. Using this and a straightforward calculation of the derivative $d\tilde{\boldkappa}/d\lambda_l$, we get (\ref{eqn:hess}).

Substituting (\ref{eqn:hess}) in (\ref{eqn:sec_deriv}) and using the fact that the Hessian of the strictly convex function $\mathcal{F}(\cdot)$ is positive definite, we get the required inequality as
\begin{equation}
\frac{d^2\Phi}{d\lambda_l^2} < 0
\end{equation}
thereby completing the proof of strict concavity.  Using this and the first sufficient condition for Assumption A to hold (see the first bullet after Assumption A), the correspondence $\boldlambda \mapsto \boldlambda^*(\overline{\boldeta})$ admits a continuous selection.

\subsection{Best Arm Identification Problem}

That the continuous selection assumption, Assumption A, holds for this problem has been proved by Garivier and Kaufmann \cite[Prop.~6.2]{garivier16}.

\bibliographystyle{IEEEtran}
\bibliography{ref}

\end{document}